\documentclass[aps,prb,groupedaddress,showpacs,twocolumn]{revtex4}
\usepackage{graphicx}
\usepackage{amsmath}
\usepackage{amssymb}
\usepackage{bbm}
\usepackage{xspace}
\usepackage{color}
\usepackage{footnote}
\usepackage[caption=false]{subfig}
\usepackage{hyperref}

\newcommand{\fig}[1]{Fig.\thinspace{}\ref{#1}}

\newcommand{\eq}[1]{Eq.\thinspace{}(\ref{#1})}
\newcommand{\Eq}[1]{Eq.\thinspace{}(\ref{#1})}
\newcommand{\eqs}[1]{Eqs.\thinspace{}(\ref{#1})}

\newcommand{\se}{Sec.\@\xspace}
\newcommand{\Se}{Sec.\@\xspace}

\newcommand{\App}{App.\@\xspace}

\newcommand{\etal}[0]{\textit{et al.}}
\newcommand{\tcite}[1]{Ref.~\onlinecite{#1}}

\newcommand{\un}{{\cal U}}
\newcommand{\teu}[1]{{}_\un(#1)}
\newcommand{\noteu}[1]{(#1)}
\newcommand{\vd}{\vv d} 
\newcommand{\aux}{\text{aux}}
\newcommand{\beq}{\begin{equation}}
\newcommand{\eeq}{\end{equation}}
\newcommand{\und}{\underline}
\newcommand{\iim}{\Im{m}\,}
\newcommand{\li}{\mathcal L}
\newcommand{\hhat}[1]{\hat{\hat{#1}}}
\newcommand{\up}{\ensuremath{\uparrow}}
\newcommand{\dw}{\ensuremath{\downarrow}}
\newcommand{\omegap}{\omega}
\newcommand{\vv}[1]{\boldsymbol{#1}}
\newcommand{\ga}{\varsigma}  

\newcommand{\rtime}[2]{_{#1,#2}} 
\newcommand{\ee}{\varepsilon}

\newcommand{\ze}{0}
\newcommand{\LLO}{{\hat L}}
\newcommand{\LL}{\ell}

\newcommand{\tr}[1]
{
\text{tr}\,#1 
}

\newcommand{\uu}{1\hspace{-3pt}1}

\def\bra#1{\mathinner{\langle{#1}|}}
\def\ket#1{\mathinner{|{#1}\rangle}}
\def\braket#1{\mathinner{\langle{#1}\rangle}}

\newcommand{\nag}{{\phantom{\dag}}}

\newcommand{\diffd}{D}
\newcommand{\UU}{U}

\begin{document}

\title{Auxiliary master equation approach to non-equilibrium correlated impurities}

\author{Antonius Dorda}
\email[]{dorda@tugraz.at}
\affiliation{Institute of Theoretical and Computational Physics, Graz University of Technology, 8010 Graz, Austria}
\author{Martin Nuss}
\affiliation{Institute of Theoretical and Computational Physics, Graz University of Technology, 8010 Graz, Austria}
\author{Wolfgang von der Linden}
\affiliation{Institute of Theoretical and Computational Physics, Graz University of Technology, 8010 Graz, Austria}
\author{Enrico Arrigoni}
\affiliation{Institute of Theoretical and Computational Physics, Graz University of Technology, 8010 Graz, Austria}

\date{\today}

\begin{abstract}
We present a numerical method for the study of correlated quantum impurity problems
out of equilibrium, which is particularly suited to address 
steady state properties within Dynamical Mean Field Theory. The approach, 
recently introduced 
 in [Arrigoni \etal{}, Phys. Rev. Lett. \textbf{110},
086403 (2013)], 
is based upon a mapping of the original impurity 
problem onto an auxiliary open quantum system, consisting of the interacting 
impurity coupled to bath sites as well as to a Markovian environment. The 
dynamics of the auxiliary system is governed by a Lindblad master equation 
whose parameters are used to optimize the mapping. 
The accuracy of the results can be readily estimated and systematically improved by increasing the number 
of auxiliary bath sites, or by introducing a linear correction. 
Here, we focus on a detailed discussion of the proposed approach including technical remarks.
To solve for the Green's functions of the auxiliary impurity problem, a non-hermitian Lanczos diagonalization is 
applied. As a benchmark, results for the steady state current-voltage characteristics of the 
single impurity Anderson model are presented. Furthermore, the bias dependence of the 
single particle spectral function and the splitting of the Kondo resonance 
are discussed. 
In its present form the method is fast, efficient and features a controlled accuracy.
\end{abstract}


\maketitle

\section{Introduction}\label{sec:introduction}

Correlated systems out of
equilibrium have recently 
attracted increasing interest 
due to the significant progress in a number of related experimental fields.
 Advances in microscopic control and manipulation of quantum
mechanical many-body systems within quantum optics~\cite{ha.br.08} and
ultra cold quantum gases, for example in optical
lattices,~\cite{ra.sa.97,ja.br.98,gr.ma.02, tr.ch.08, sc.ha.12} have long reached high accuracy and
versatility. 
Ultrafast laser
spectroscopy~\cite{iw.on.03,ca.de.04} offers
the possibility to explore and understand electronic dynamics in
unprecedented detail. Experiments in condensed matter
nano-technology,~\cite{bo.gr.05} spintronics,~\cite{zu.fa.04}
molecular junctions~\cite{cu.fa.05,sm.no.02,pa.ab.02,li.sh.02,ag.ye.03,ve.la.06}
and quantum wires or quantum dots,~\cite{go.go.98, kr.sh.12} are  
able to reveal effects of  the interference of few microscopic quantum states.
The non-equilibrium nature of such
experiments does not only offer a new route to explore fundamental
aspects of quantum physics, such as 
non-equilibrium quantum phase
transitions,~\cite{mi.ta.06},  the interplay between
quantum entanglement, dissipation and decoherence~\cite{le.ch.87}, or 
the pathway to thermalization,~\cite{caza.06, ri.du.08}, but also suggests the possibility of exciting
future applications.~\cite{cu.fa.05,ni.ra.03}

Addressing the 
dynamics of correlated quantum systems 
poses a major challenge to theoretical
endeavors. In this respect, quantum impurity models  help improving our
understanding of fermionic many-body systems. 
 In particular, the
single-impurity Anderson model (SIAM),~\cite{ande.61}
which 
was originally
devised to study magnetic impurities in
metallic hosts,~\cite{frie.56, cl.ma.62}
has become an important tool in many areas of condensed
matter physics.~\cite{br.sc.74,ge.ko.96} Most
prominently, it features non-perturbative many-body physics which
manifest in the Kondo effect.~\cite{hews.97} It provides the backbone for all
calculations within dynamical mean field theory
(DMFT),~\cite{ge.ko.96,voll.10} a technique which allows
to understand the properties of  a broad range of correlated systems  
and becomes exact in the limit of infinite 
dimensions.~\cite{me.vo.89} The basic physical properties of
the SIAM in equilibrium are quite well
understood~\cite{hews.97} thanks to the pioneering work from
Kondo,~\cite{kond.64}  renormalization
group~\cite{ande.70} as well as perturbation
theory (PT)~\cite{yo.ya.70,yama.75,yo.ya.75,yama.75b}
and the mapping to its low energy realization, the Kondo
model.~\cite{sc.wo.66}

The SIAM out of equilibrium provides a description for several
physical processes such as,
 for example, nonlinear transport through quantum
dots,~\cite{go.go.98,mu.pl.13} correlated
molecules~\cite{bo.sc.12,pa.ab.02,li.sh.02,yu.ke.05,to.ro.12}
or the influence of 
adsorbed atoms
 on surfaces or
bulk transport.~\cite{pr.we.11} 
As in the equilibrium case, the solution of the SIAM constitutes the
bottleneck 
 of  non-equilibrium
DMFT~\cite{ao.ts.13u,sc.mo.02u,fr.tu.06,free.08,jo.fr.08,ec.ko.09,okam.07,ar.kn.13}
calculations.
Therefore,
accurate and efficient methods to obtain dynamical correlation
functions of  impurity models out of equilibrium 
are required in order to describe
time resolved  
experiments on strongly correlated compounds~\cite{iw.on.03,ca.de.04}
ant to understand their steady state transport characteristics.~\cite{ni.ra.03}

However, nonequilibrium correlated impurity models  still pose
an exciting challenge to theory. Our work addresses this issue
  with special emphasis on the
steady state. But before introducing the present work in Sec.~\ref{pres}, we
briefly review previous approaches.
In recent times a number
of computational techniques have been devised to handle the SIAM out
of equilibrium. Among them are scattering-state
BA,~\cite{me.an.06} scattering-state
NRG (SNRG),~\cite{ande.08,an.sc.10,rosc.12}
non-crossing approximation studies,~\cite{me.wi.93,wi.me.94} fourth order Keldysh PT,~\cite{fu.ue.03} other perturbative
methods~\cite{sc.sc.94, he.da.91} in combination
with the renormalization group
(RG),~\cite{scho.09,ro.pa.05,an.sc.06,ro.we.08,do.an.06} iterative
summation of real-time path integrals,~\cite{we.ec.08} time dependent
NRG,~\cite{an.sc.05} flow equation techniques,~\cite{mo.ke.08,kehr.05}
the time dependent density matrix RG
(DMRG)~\cite{vida.04,whit.93,da.ko.04,wh.fe.04,scho.11,schm.04}  
applied to the SIAM,~\cite{he.fe.09,nu.ga.13} non-equilibrium
cluster PT (CPT),~\cite{nu.he.12} the
non-equilibrium variational cluster approach
(VCA),~\cite{kn.li.11,ho.ec.13} dual
fermions,~\cite{ju.li.12} the functional RG
(fRG),~\cite{ge.pr.07,ja.me.07}
diagrammatic QMC,~\cite{we.ok.10,co.gu.13u} continuous time
QMC (CT-QMC) calculations on an auxiliary system with an imaginary
bias,~\cite{han.06, ha.he.07, di.we.10, ha.di.12, ha.di.12b} super
operator techniques,~\cite{du.ko.11, mu.bo.13} many-body PT and time-dependent density functional theory,~\cite{ui.kh.11} generalized slave-boson methods~\cite{sm.gr.11},  real-time RG (rtRG),~\cite{so.ko.00}, time dependent Gutzwiller mean-field calculations~\cite{sc.fa.10} and generalized master equation approaches.~\cite{timm.08} Comparisons of the results of some of these methods are available in literature~\cite{ec.he.10,nu.ga.13,an.me.10} and time scales have been discussed in \tcite{pu.sp.12}.

Despite this large number of approaches, only a limited number of them
is applicable to non-equilibrium DMFT, 
and very few are still accurate
for large times in steady state. 
Beyond the quadratic
action for the Falicoff Kimball
model,~\cite{fa.ki.69,fr.tu.06,ec.ko.08}
iterated PT (IPT),~\cite{sc.mo.02u} numerical
renormalization group~\cite{jo.fr.08}, real time
QMC,~\cite{jo.fr.08,ec.ko.10} the noncrossing
approximation (NCA)~\cite{okam.08,ar.ko.12} and recently Hamiltonian based impurity
solvers~\cite{gr.ba.13} have been applied in the time dependent
case. Some of the above approaches, such as QMC~\cite{ec.ko.09} and DMRG~\cite{wh.fe.04} are very accurate in addressing the short
and medium-time dynamics, but in some cases the accuracy 
decreases at long times and a steady state cannot be reliably identified. Some
other methods are perturbative and/or  valid only in certain parameter
regions or for restricted models. RG approaches (e.g., \tcite{scho.09}) are certainly
more appropriate to identify the low-energy behavior.

\subsection{Present work}
\label{pres}
In this paper we discuss a method, first proposed in
 \tcite{ar.kn.13}, which addresses the correlated impurity problem out
 of equilibrium,
and is particularly efficient for the steady state.
 The accuracy of the results is {\it controlled} as it
can be directly estimated by analyzing the 
 bath hybridization function (details below).
Here, 
 we extend, test and provide details of this approach
and its implementation. The basic idea is to map the impurity
problem onto an auxiliary open system, consisting of a small number of bath
sites coupled to the interacting impurity and additionally, to a so-called
Markovian environment.~\cite{carmichael1} The parameters of this auxiliary open quantum system are obtained by optimization
in order to represent the original impurity problem as accurately as possible. The auxiliary system dynamics are governed by a Lindblad
Master equation which is solved exactly with the non-hermitian Lanczos method. The crucial point is, that 
the overall accuracy of the method is thus solely determined by how
well the auxiliary system reproduces the original one. This can be, in
principle, improved by increasing the number of auxiliary bath
sites.
 
In the present study we provide convincing benchmarks for the steady state properties of the SIAM coupled to two metallic
leads under bias voltage. We include a discussion of convergence as a
function of the number of bath sites and present a scheme to estimate the
error and partially correct for it. In its presented form the method is
fast, efficient and is directly applicable to steady state
dynamical mean field theory~\cite{ar.kn.13} for which previously suggested methods are
less reliable. Extending the method to treat time dependent properties and multi-orbital systems is possible, in principle, 
however with a much heavier computational effort.

The paper is organized as follows: In \se~\ref{ssec:SIAM} the SIAM under bias voltage is introduced. In \se~\ref{ssec:ssnegf} we introduce non-equilibrium Green's functions and in  \se~\ref{ssec:mapping} and \ref{ssec:delta}
we outline the auxiliary master equation approach where we also focus on 
details of our particular implementation. Results
for the steady state, including the equilibrium situation are
presented in \se~\ref{sec:results}. This includes the steady state
current-voltage characteristics which we compare with exact results from
Matrix Product State (MPS) time evolution~\cite{nu.ga.13} as
well as data for the spectral function under bias which we compare with
non-equilibrium NRG.~\cite{an.sc.10} We conclude and give an outlook in \se~\ref{sec:conclusion}. 

\section{Auxiliary master equation approach}\label{ssec:method}
As discussed above, the method is  particularly suited to deal with
non-equilibrium steady state properties caused by  different
temperatures and/or chemical potential in the leads of a correlated
quantum impurity system. As such, it can be readily used as impurity
solver for non-equilibrium DMFT.~\cite{fr.tu.06,ar.kn.13} Here we
illustrate its application to the fermionic SIAM with two leads 
having different chemical potentials, and, in principle, different temperatures.

\subsection{Non-equilibrium single impurity Anderson model}\label{ssec:SIAM}
 \begin{figure}
 \subfloat{
\includegraphics[width=0.5\textwidth]{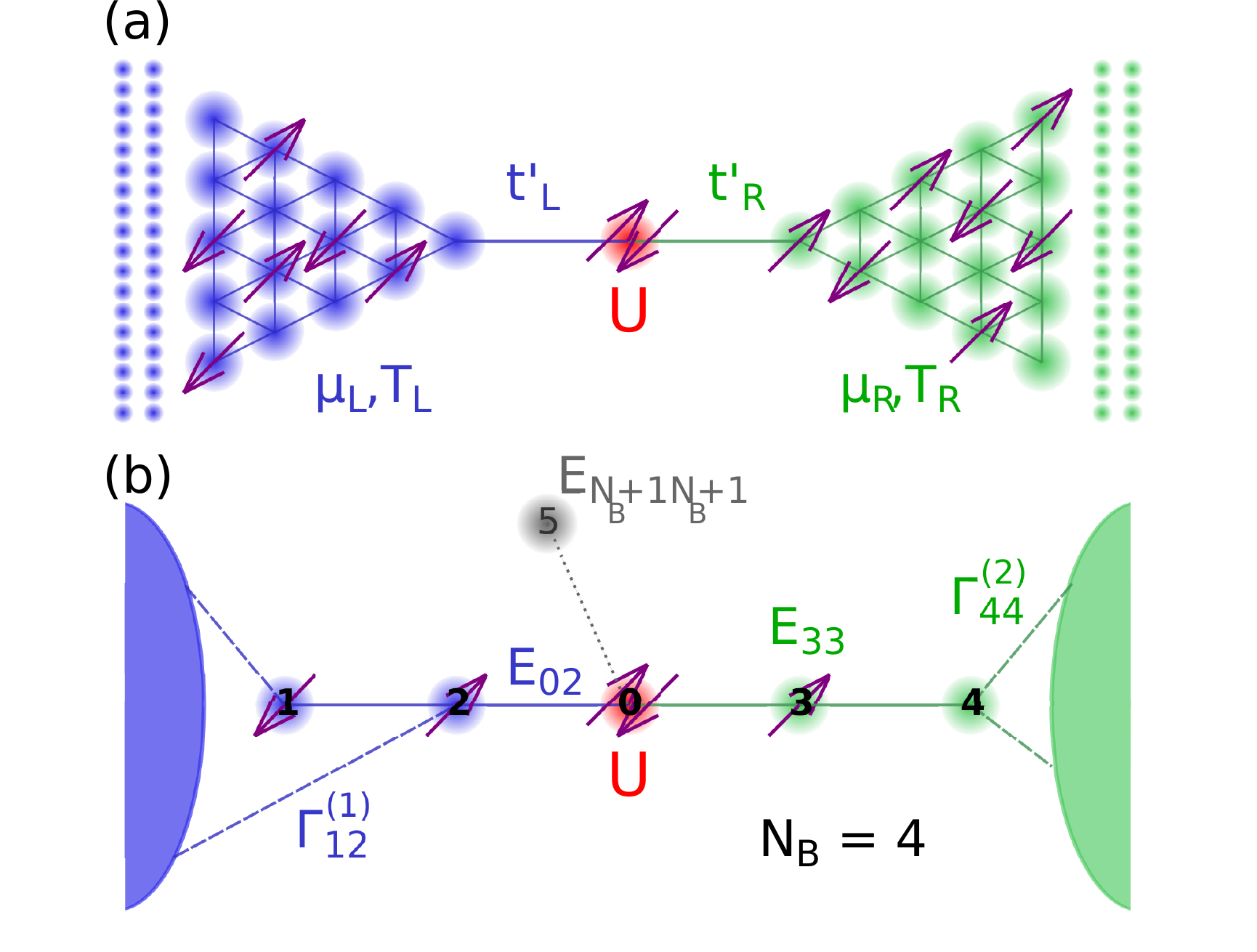}}
\caption{(Color online) (a) Sketch of the quantum impurity model \eq{eq:H} consisting of an
  impurity with interaction $U$ coupled via hybridizations
  $t'_{\lambda}$ to noninteracting leads at chemical potential
  $\mu_\lambda$ and temperature $T_\lambda$, $\lambda\in\{L,R\}$. (b)
Illustration of the auxiliary open quantum system \eq{li} with single particle parameters
  $E_{\mu\nu}$ and Lindblad dissipators $\Gamma^{\kappa}_{\mu\nu}$
  consisting of the impurity at site $f=0$, $N_B$ bath sites ($N_B=4$ in the plot) as well as a Markovian
  environment (shaded areas). When evaluating linear corrections (see
  Sec.~\ref{ssec:error}),  an additional site $N_B+1$ is used.}
\label{fig:model}
\end{figure}
We consider a single Anderson impurity coupled to electronic leads under bias voltage (see \fig{fig:model} (a))
\begin{align}
\hat{\mathcal{H}} &= \hat{\mathcal{H}}_{\text{imp}} + \hat{\mathcal{H}}_{\text{res}} + \hat{\mathcal{H}}_{\text{coup}}\,\mbox{.}\label{eq:H}
\end{align}
The impurity orbital features charge as well as spin degrees of freedom and is subject to a local Coulomb repulsion $U$
\begin{align*}
\hat{\mathcal{H}}_{\text{imp}} &= \epsilon_f \, \sum\limits_{\sigma}  \, f_{\sigma}^\dagger \, f_{\sigma}^\nag + U \, \hat{n}^{f}_{\uparrow} \, \hat{n}^{f}_{\downarrow}\,\mbox{.}
\end{align*}
Here $f_{\sigma}^\dag/f_{\sigma}^\nag$ denote fermionic
creation/annihilation operators for the impurity orbital with spin
$\sigma\in\{\uparrow,\downarrow\}$ respectively. The particle number
operator is defined in the usual way
$\hat{n}^{f}_{\sigma}=f_{\sigma}^\dag f_{\sigma}^\nag$ and the
impurity on-site potential is $\epsilon_f=(V_G-\frac{U}{2})$, with
gate voltage $V_G=0$ at particle hole symmetry. The impurity is
coupled to two noninteracting electronic leads $\lambda\in\{L,R\}$
with dispersion $\epsilon_{\lambda k}$ 
\begin{align*}
\hat{\mathcal{H}}_{\text{res}} &= \sum\limits_{\lambda k \sigma}
\left(\epsilon_\lambda + \epsilon_{\lambda k}\right) 
 \, c_{k\lambda\sigma}^{\dagger} \, c_{k\lambda\sigma}^{\nag}\,\mbox{.}
\end{align*}
The effect of a bias voltage $\phi$ is to shift 
the chemical potential and the on-site energies  of the two leads
by $\epsilon_{\lambda}=\pm\frac{\phi}{2}$, respectively. 
For the energies $\epsilon_{\lambda k}$ of the 
leads we will consider 
two cases: (i)  Two tight-binding semi-infinite chains 
with nearest-neighbor hopping $t$,
corresponding to a 
semi-circular electronic density of states (DOS).
In this case, the boundary retarded single particle Green's
function  of the two uncoupled leads
is given by~\cite{footnote1, footnote14, economou}
\begin{align}
\label{gsc}
g_{\lambda}^R(\omega) &=
g_{\text{SC},\lambda}^R(\omega)=
\frac{\omega-\epsilon_\lambda}{2t^2}-i\frac{\sqrt{4t^2-(\omegap-\epsilon_\lambda)^2}}{2t^2}\,\mbox{,}
\end{align}
with a bandwidth of $D_{SC}^\nag=4\,t$. (ii) A constant DOS with a bandwidth $D_{\text{WB}}^\nag=\pi\,t$, resulting in boundary Green's functions~\cite{footnote14}
\begin{align}
\label{gwb}
g_{\lambda}^R(\omega)&=
g_{\text{WB},\lambda}^R(\omega)=
-\frac{1}{D_{WB}^\nag}
\text{ln}\left(\frac{\omegap-\epsilon_\lambda-\frac{D_{\text{WB}}^\nag}{2}}{\omegap-\epsilon_\lambda+\frac{D_{\text{WB}}^\nag}{2}}\right)
\,\mbox{.}
\end{align}
The choice $D_{\text{WB}}^\nag=\pi\,t$ makes sure that the DOS at $\omega=0$ of both lead types coincide. The leads are
coupled to the impurity orbital by
\begin{align*}
\hat{\mathcal{H}}_{\text{coup}} &= \sum\limits_{\lambda\sigma}
t_{\lambda}'\frac{1}{\sqrt{N_k}}\sum_k
\left(c_{k\lambda\sigma}^{\dagger} \, f_{\sigma}^\nag +
  f_{\sigma}^\dagger \, c_{k\lambda\sigma}^{\nag} \right)\,\mbox{,} 
\end{align*}
where we take the same hybridization $t'_\lambda=-0.3162\,t$  for both
leads, and $N_k\to\infty$ is the number of $k$ points. Expressions presented below are valid for arbitrary temperatures, although we will show results for zero temperature only, which is numerically the most unfavorable case.~\cite{footnote13} The setup chosen here represents by no means a limitation of the method and extensions to more complicated situations, such as  non-symmetric couplings, off particle-hole symmetry, etc. are straightforward. 

\subsection{Steady state non-equilibrium Green's functions}\label{ssec:ssnegf}
We are interested in the steady state behavior under bias voltage of the model described
by \eq{eq:H}. We assume that such a steady state
exists and is unique.~\cite{footnote2}
We denote the single particle Green's function of the
impurity in the non-equilibrium Green's function (Keldysh) formalism by~\cite{kad.baym,schw.61,keld.65,ha.ja,ra.sm.86} 
\begin{align}
\label{gund}
  \und{G}(\omega) &= \begin{pmatrix} G^R(\omega) & G^K(\omega) \\ 0 & G^A(\omega) \end{pmatrix}\,\mbox{.}
\end{align}
Fourier transformation to energy $\omega$ is possible 
since in the steady state the system becomes time translationally
invariant. In that case, the memory of the initial
condition has been fully washed away, so there is no contribution from
the Matsubara branch.~\cite{kame.11} We will use an underline $\und{\cdots}$ to denote
two-point functions with the Keldysh matrix structure as in \eq{gund}.

The Green's function of the correlated impurity can be expressed via
Dyson's equation
\begin{align}
\underline{G}^{-1}(\omega)&=\underline{G}^{-1}_{0}(\omega)-\underline{\Sigma}(\omega)\,\mbox{,}\label{dys}
\end{align}
where $\underline{\Sigma}(\omega)$ is the impurity self-energy. 
The noninteracting impurity Green's function $\und G_0(\omega)$ can be written in the form
\begin{align}
\underline{G}^{-1}_{0}(\omega)&= \underline{g}^{-1}_{0}(\omega) - \underline{\Delta}(\omega)\,\mbox{,} \label{g0}
\end{align}
$\underline{g}_{0}^\nag(\omega)$ being the noninteracting Green's function of the disconnected impurity,~\cite{footnote1}
and
\begin{align}
\und{\Delta}(\omega) &= \sum\limits_\lambda t_{\lambda}^{\prime\ 2}\,\underline{g}_{\lambda}^\nag(\omega)\,\mbox{,}\label{deltaw}
\end{align}
is the hybridization function of the leads (a $2\times2$ Keldysh object, in contrast to the equilibrium case, where it is convenient to work in Matsubara space). We define an equilibrium Anderson width~\cite{hews.97} for each lead $\Delta_0 \equiv -\frac{1}{2}\iim(\und{\Delta}^R(\omega=0))=\frac{t'^2_{\lambda}}{t} \approx 0.1\,t$.  Below, we will use $\Delta_0$ as a unit of energy and in addition we choose $\hbar=e=1$.

The boundary Green's functions $\und{g}_{\lambda}$ of each disconnected lead is determined by (a) its retarded component $g_{\lambda}^R$ 
(either \eqref{gsc} or \eqref{gwb}), (b) its advanced component $g_{\lambda}^A= g_{\lambda}^{R*}$, and (c) its Keldysh component, which satisfies the fluctuation dissipation theorem
\begin{align}
\label{fd}
g_{\lambda}^K(\omega)&=2 i \left(1-2 p_{\text{F}}(\omega-\mu_\lambda)\right) \iim (g_{\lambda}^R(\omega))\,\mbox{,}
\end{align}
since the disconnected leads are in equilibrium. Here, $p_{\text{F}}(\omega-\mu_\lambda)$ is the Fermi distribution with chemical potential $\mu_\lambda$.
For the noninteracting isolated impurity one can take $\left(g_{0}^{-1}\right)^R=\omega -\epsilon_f$, and $\left(g_{0}^{-1}\right)^K=0$, since infinitesimals $0^+$ can be neglected after coupling to the leads (unless there are bound states).

As usual, the presence of the interaction $U$ makes the solution of the problem impurity plus leads a major challenge both in equilibrium as well as out of equilibrium, which we plan to address in the present paper.

Similarly to the equilibrium case, the action of the leads on the impurity is completely determined by the hybridization function
$\und\Delta(\omega)$, {\em independently of how the leads are represented in detail}. In other words, if one constructs a different configuration
of leads (e.g. with more leads with different temperatures, DOS, etc.), which has the same $\und \Delta(\omega)$, i.e. the same $\Delta^R(\omega)$ {\em and} $\Delta^K(\omega)$ as \eq{deltaw}, then the resulting local properties of the interacting impurity, e.g. the Green's function $\und G(\omega)$ are the same. This holds provided the leads contain {\em noninteracting} fermions only.

The approach we suggested in \tcite{ar.kn.13} precisely exploits this
property. The idea is to replace the impurity plus leads system (\eq{eq:H}) by
an auxiliary one which reproduces $\und \Delta(\omega)$ as accurately as possible, and at the
same time can be solved exactly by numerical methods, such as Lanczos exact diagonalization. Details on the construction of the auxiliary impurity system are given below.

The self energy $\und\Sigma_\aux(\omega)$
of the auxiliary system, obtained by exact diagonalization, is used in analogy to 
DMFT~\cite{ca.kr.94,ge.ko.96} as an approximation to the physical self energy of the 
original impurity system. Inserting $\und\Sigma(\omega) \approx \und\Sigma_\aux(\omega)$ into \eqs{dys},\thinspace(\ref{g0}), 
together with the exact hybridization function $\und\Delta(\omega)$ yields an approximation
for the physical Green's function. From this, observables such as the current or the spectral function are then calculated.
We emphasize that the accuracy of this approximation can be controlled by
the difference between the $\und\Delta_\aux(\omega)$ of the auxiliary system and
the physical one $\und\Delta(\omega)$, and that this can be, in
principle, systematically improved, as discussed below.

\subsection{Auxiliary open quantum system}
\label{ssec:mapping}
The idea presented here  is strongly related to the ED approach for the DMFT impurity problem in {\em equilibrium}.~\cite{ca.kr.94,ge.ko.96} Here, the
infinite leads are replaced by a small number of bath
sites, whose parameters are optimized by fitting
the hybridization function in Matsubara space.
The reduced system of bath sites plus impurity is then solved
by Lanczos ED.~\cite{lanc.51} This approach cannot be straightforwardly  extended to the
non-equilibrium steady state case for several reasons: (i) since the small bath is finite,
its time dependence is (quasi) periodic, i.e. no steady state is
reached, (ii) there is no Matsubara representation out of
equilibrium,~\cite{footnote3} thus, one is forced to use real
energies but (iii) in this case $\iim(\Delta^R_{\text{aux}}(\omega))$ of the small bath consists
of $\delta$-peaks and can hardly be fitted to a smooth
$\Delta^R(\omega)$. The solution we suggested in \tcite{ar.kn.13}
consists in additionally coupling the small bath to a Markovian
environment, which makes it effectively ``infinitely large'', and solves problems (i) and (iii) above.
Specifically, we replace the impurity plus leads model (\eq{eq:H}) by an auxiliary {\em open} quantum system consisting of the impurity plus a small number of bath sites, which in turn are coupled to a Markovian environment.

The dynamics of the system (consisting of bath sites and impurity), 
including the effect of the Markovian environment
is expressed in terms of the Lindblad quantum master equation which controls the time dependence
of its reduced density operator $\hat\rho$:~\cite{br.pe,carmichael1}
\begin{align}
  \dot{\hat{\rho}} &=  \hat{\hat{\li}}\hat{\rho}\,\mbox{.} \label{eq:lindbladmeq}
\end{align}
\begin{subequations}
\label{eq:L}
The Lindblad super-operator~\cite{footnote4}
\begin{align}
\label{li}
\hhat\li &= \hhat\li_{H} + \hhat\li_D\,\mbox{,}
\end{align}
consists of a unitary contribution
\begin{align*}
\hhat\li_{H} \hat{\rho} &= 
-i [\hat{\mathcal{H}}_{\text{aux}},\hat{\rho}]\,\mbox{,}
\end{align*}
as well as a non-unitary, dissipative term originating from the coupling to the Markovian environment
\begin{align}
\nonumber
&\hhat\li_D\hat\rho&\equiv2\sum\limits_{\mu\nu = 0}^{N_B}\sum\limits_\sigma\Bigg(\Gamma_{\nu\mu}^{(1)}\left(d_{\mu\sigma}^\nag\hat\rho\  d_{\nu\sigma}^{\dag}-\frac{1}{2}\{\hat\rho,d_{\nu\sigma}^{\dag}d_{\mu\sigma}^\nag\}\right)+\\
\label{ld}
&&+\Gamma_{\nu\mu}^{(2)}\left(d_{\nu\sigma}^{\dag}\hat\rho\  d_{\mu\sigma}^\nag-\frac{1}{2}\{\hat\rho,d_{\mu\sigma}^\nag d_{\nu\sigma}^{\dag}\}\right)\Bigg)\,\mbox{,}
\end{align}
\end{subequations}
where $[\hat{A},\hat{B}]$ and $\{\hat{A},\hat{B}\}$ denote the commutator and anti-commutator respectively. The unitary time evolution is generated by the Hamiltonian 
\begin{align}
\hat{\mathcal{H}}_{\text{aux}} &= \sum\limits_{\mu\nu = 0}^{N_B}\sum\limits_\sigma E_{\mu\nu}d_{\mu\sigma}^{\dag}d_{\nu\sigma}^\nag+U d_{f\up}^{\dag}d_{f\up}^\nag d_{f\dw}^{\dag}d_{f\dw}^\nag\,\mbox{,}
\label{eq:Haux}
\end{align}
describing a fermionic ``chain'' ($E_{\mu\nu}$ is non-zero only for on-site and nearest neighbor terms). It is convenient to choose
the interacting impurity at site $f=0$ and $N_B$ auxiliary bath sites at $\mu,\nu=1,\cdots,N_B$ (see \fig{fig:model}). 
As usual, $d_{\mu \sigma}^\dag/d_{\mu\sigma}^\nag$ create/annihilate
the corresponding auxiliary particles.
The quadratic form of the dissipator (\eq{ld}) corresponds to 
a noninteracting Markovian environment. The dissipation matrices $\Gamma_{\mu\nu}^{(\kappa)}$, $\kappa\in\{1,2\}$ are hermitian and positive
semidefinite.~\cite{br.pe} The advantage of replacing the impurity problem by the auxiliary one
described by \eqs{eq:lindbladmeq}-(\ref{eq:Haux}), is that for a small number of bath sites the
dynamics of the interacting auxiliary system can be solved exactly
by diagonalization of the super-operator $\hhat \li$ in the space of
many-body density operators (see \se~\ref{ssec:sigma}).

Intuitively, one can consider the effective system as a truncation of
the original chain described by \eq{eq:H}, whereby the
Markovian environment compensates for the missing ``pieces''. However, this would still
be a crude approximation, and in addition, it would
not be clear how to introduce the chemical potential in the Markovian
environment (except for weak coupling).
Our strategy, similarly to the equilibrium case,
 consists in simply using the parameters of the auxiliary 
system in order to provide an optimal fit to the bath spectral function $\und
\Delta(\omega)$.
The parameters for the fit are, in principle, $E_{\mu\nu}$ and 
$\Gamma_{\mu\nu}^{(\kappa)}$.
 However, one should consider that there is a certain redundancy. In
 other words several combinations of parameters lead to the same $\und
 \Delta(\omega)$. For example, it is well known in equilibrium  that
 in the case of the  $E_{\mu\nu}$  one can restrict to diagonal and nearest neighbor
terms only.~\cite{footnote5} 

The  accuracy of the results will be directly related to the accuracy
of the fit to $\und \Delta(\omega)$, and this is expected to
increase rapidly with the number of fit parameters, which obviously increases with $N_B$.
On the other hand, also the computational complexity necessary to
exactly diagonalize the interacting auxiliary system increases
exponentially with $N_B$.
The fit does not present a major numerical difficulty, as the determination of
the hybridization functions of both the original model (\eq{deltaw}), as well as the one of the auxiliary system  $\und\Delta_{\text{aux}}(\omega)$ described by the Lindblad equation \eqref{eq:L} require the  evaluation of  $\und G_0$ (cf. \eqref{g0}), i.e. the solution  of a {\em noninteracting} problem.

The fit is obtained by minimizing the cost function
\begin{align}
  \nonumber \chi(E_{\mu\nu},\Gamma_{\mu\nu}^{(\kappa)})
= &\sum\limits_{\alpha\in\{R,K\}}\int\limits_{-\infty}^{\infty} d \omega \,W^\alpha(\omega) \\
&\times\left|\Delta^\alpha(\omega)-\Delta^\alpha_{\text{aux}}(\omega;E_{\mu\nu },\Gamma_{\mu\nu}^{(\kappa)})\right|^n\,\mbox{.}
\label{eq:SC}
\end{align} 
with respect to the parameters of the auxiliary system. The advanced component does not need to be considered as $\Delta^A=\Delta^{R*}$. Of course, like in ED based DMFT, there exists an ambiguity which is related to the choice of the weight function $W^\alpha(\omega)$, which also sets the integral boundaries. This uncertainty is clearly reduced upon increasing $N_B$. 

Depending on the expected physics, it might be useful to adopt a energy-dependent weight function. This could be used for example to describe the physics around the chemical potentials more accurately. 

Once the auxiliary system is defined in terms of $E_{\mu\nu} \mbox{ and }
\Gamma_{\mu\nu}^{(\kappa)}$, the corresponding interacting non-equilibrium problem
\eq{eq:L} can  be solved by an exact diagonalization of the non-hermitian
super-operator $\hhat \li$ within the space of many-body density
operators. The dimension of this space is equal to the square of the dimension of
the many-body Hilbert space, and thus it grows exponentially as a
function of $N_B$. Therefore, for $N_B\geq 4$ a non-hermitian Lanczos
treatment must be used. The solution of the noninteracting Lindblad problem is non standard (see e.g.~\tcite{pros.08}), and a method particularly suited for the present approach is discussed in \se~\ref{nonint}.

\subsection{Green's functions of the auxiliary Lindblad problem}
\label{ssec:delta}
In this section we present expressions for the Green's functions
of the auxiliary system. Specifically, we will  derive an analytic
expression for the noninteracting Green's functions in
\se~\ref{nonint}, and illustrate the numerical procedure to determine
the interacting ones in \se~\ref{ssec:sigma}. The derivations make
largely use of the formalism of \tcite{dz.ko.11}, see also
\tcite{schm.78}. For an alternative appealing approach to the
noninteracting case, see also \tcite{pros.08}. 
All Green's functions discussed in Sec.~\ref{ssec:delta} are the ones
of the auxiliary system, which are different from the physical ones
for  $N_B<\infty$. 

The dynamics of the auxiliary open quantum system described by the super-operator
$\hat{\hat{\mathcal{L}}}$ (\eq{eq:L}) can be recast in an elegant way
as a standard operator problem in an 
augmented fermion Fock space with twice as
many sites.~\cite{dz.ko.11,schm.78,ha.mu.08,pros.08}
Specifically, one introduces ``tilde'' operators 
$\tilde{d}_\mu^\nag/\tilde{d}_\mu^\dag$ together with the original
ones ${d}_\mu^\nag/{d}_\mu^\dag$.~\cite{footnote12}
Introducing the so-called left-vacuum 
\begin{align}
\label{lvac}
\ket{I}&=\sum\limits_{S} (-i)^{N_S}
  \ket{S}\otimes\ket{\tilde{S}}\,\mbox{,}
\end{align}
where $\ket{S}$ are many-body
states of the original Fock space,  $\ket{\tilde{S}}$ the
corresponding ones of the tilde space~\cite{dz.ko.11} and $N_S$
the number of particles in $S$. The non-equilibrium density operator can be written as a state vector in this augmented space
\beq
\label{rhoi}
\ket{\rho(t)} \equiv \hat{\rho}(t)\ket{I}\,.
\eeq 
The Lindblad equation is rewritten in a Schr\"odinger like fashion~\cite{dz.ko.11,footnote4} 
\begin{align}
\label{drhot}
\frac{d}{dt}\ket{\rho(t)}&= \hat{\mathcal{L}}\ket{\rho(t)}\,\mbox{,}
\end{align}
where now $\hat\li$ is an ordinary {\em operator} in the augmented
space. $\hat\li=\hat\li_0+\hat\li_{I}$ is conveniently represented in terms of 
 the operators of the augmented space in a vector notation:~\cite{footnote12}
\begin{align*}
 \vd^\dag &= \left( d_{0}^\dag, \hdots, d_{N_B}^\dag,\tilde{d}_{0},
\hdots \tilde{d}_{N_B}\right)\,\mbox{.}
\end{align*}
Its noninteracting part $\li_0$ reads in the augmented space~\cite{dz.ko.11,footnote4}
\begin{align}
i\hat{\li}_0&=\sum\limits_\sigma\left(\vd^\dag \vv h \vd - \,\text{Tr}\left(\vv E+i\vv \Lambda\right)\right)\,\mbox{,}
\label{eq:L0}
\end{align}
where $\text{Tr}$ denotes the matrix trace and the matrix $\vv h$ is given by
\begin{align}
\label{eq:L02}
\vv h &=
\begin{pmatrix}
\vv E+i\vv \Omega & 2 \vv \Gamma^{(2)}\\
 -2 \vv \Gamma^{(1)} &  \vv E-i \vv \Omega
\end{pmatrix} \,\mbox{,}
\end{align}
with 
\begin{align*}
\vv \Lambda = \left(\vv \Gamma^{(2)}+\vv \Gamma^{(1)}\right)\,\mbox{,}
&\quad\quad
\vv \Omega = \left(\vv \Gamma^{(2)}-\vv \Gamma^{(1)}\right)\,\mbox{.}
\end{align*}
Its  interacting part has the form~\cite{dz.ko.11}
\begin{align*}
i\li_I &= U d^\dag_{f\up}d^\nag_{f\up}d^\dag_{f\dw}d^\nag_{f\dw}-U \tilde d^\dag_{f\up}\tilde d^\nag_{f\up}\tilde d^\dag_{f\dw}\tilde d^\nag_{f\dw}\,\mbox{.}
\end{align*}

In this auxiliary open system,
dynamic two-time correlation functions for two operators
$\hat A$ and $\hat B$ of the system can be expressed as
\begin{align}
\label{gab}
i G_{BA}(t_2,t_1) & \equiv
\langle\hat{B}\teu{t_2}
\hat{A}\teu{t_1}\rangle = \tr_{\un}\left(\hat{B}\teu{t_2}\hat{A}\teu{t_1}\hat{\rho}_\un\right)
\nonumber \\
& = \tr\left(\hat{B}\hat{A}\rtime{t_1}{t_2-t_1}\right)\,\mbox{,}
\end{align}
where $\hat{\rho}_\un$ is the density operator of the ``universe'' $\un$
composed of the system and Markovian environment, $\text{tr}$ is the
trace over the system degrees of freedom,  $\tr_E$ the one over the
environment,  $\tr_{\un}=\tr \otimes \tr_E$ the one over the
universe, $\hat O\teu{\cdots}$ denotes the unitary time
evolution of an  operator $\hat O$ according to the Hamiltonian
of the universe
$\hat{\mathcal{H}}_\un$.
Here,~\cite{carmichael1}
\begin{align}
\label{rtime}
\hat A\rtime{t_1}{t}&\equiv\tr_{E}\left(e^{-i\hat{\mathcal{H}}_\un t}\hat{A} \hat{\rho}_{\un}(t_1)e^{+i\hat{\mathcal{H}}_\un t}\right)\,\mbox{.}
\end{align}
Notice that the time evolution of $\hat{\rho}_\un(t)$, as well as the one
in \eq{rtime} are {\em opposite with respect to the Heisenberg time evolution of operators}. This is
the convention  for  density operators.
For $t=t_2-t_1>0$ one can use the quantum regression theorem~\cite{carmichael1} which holds under the
same assumptions as for \eq{eq:lindbladmeq}. It states that
\begin{align}
\label{dadt}
\frac{d}{d t} \hat{A}\rtime{t_1}{t}&=\hhat\li \hat{A}\rtime{t_1}{t}\,\mbox{.}
\end{align}
In the augmented space,  in the same way as for \eqref{rhoi}-\eqref{drhot},
one can associate
the operator \eqref{rtime}
with the state vector
$\ket{A\rtime{t_1}{t}} = \hat{A}\rtime{t_1}{t} \ket{I}$. For this vector, \eqref{dadt} translates into
\begin{align}
\label{dadtv}
\frac{d}{d t} \ket{A\rtime{t_1}{t}} &=\hat\li \ket{A\rtime{t_1}{t}} \,\mbox{.}
\end{align}
Considering its initial value (time $t=0$)
\begin{align*}
\ket{A\rtime{t_1}{0}} &= \hat A \ket{\rho(t_1)}\,\mbox{,}
\end{align*}
the solution of \eqref{dadtv} reads
\begin{align}
\label{at}
\ket{A\rtime{t_1}{t}} &= e^{\hat \li t} \hat A \ket{\rho(t_1)}\,\mbox{.}
\end{align}

Therefore, we have for the correlation function \eqref{gab} for $t_2>t_1$, which we denote as $G_{BA}^+(t_2,t_1)$
\begin{align*}
i G_{BA}^+(t_2,t_1) &
= \bra{I} \hat B e^{\hat \li (t_2-t_1)} \hat A \ket{\rho(t_1)}\\
&= \bra{I} \hat B(t_2-t_1) \hat A \ket{\rho(t_1)}\,\mbox{,}
\end{align*}
where
\begin{align}
\label{heis}
\hat B(t) &:= e^{-\hat \li t} \hat B e^{\hat \li t}\,\mbox{,}
\end{align}
is the non-hermitian  time evolution of the operator $\hat{B}$,
and we have exploited the relation~\cite{dz.ko.11} \mbox{$\bra{I}\hat{\li}=0$}.
For the steady state correlation function, which depends on $t=t_2-t_1$, we have
\begin{align}
\label{gba} i G_{BA}^+(t)&= \bra{I} \hat B(t) \hat A\ket{\rho_\infty}\,\mbox{,}
\end{align}
where $\hat\rho_\infty$ is the steady state density operator. 
Since the quantum regression theorem only propagates forward in time,
for $t<0$ one has to take the complex conjugate of \eq{gab}, which
gives for the $t<0$ steady state correlation function denoted as 
$G_{BA}^-$:
\begin{align}
\label{gbam}
i G_{BA}^-(t) &= -i G_{A^\dag B^\dag}^+(-t)^*  =
\bra{I} \hat A^\dag(-t) \hat B^\dag
\ket{\rho_\infty}^*\,\mbox{.}
\end{align}
Using \eqref{gba}, the steady state greater Green's function for times $t>0$ reads~\cite{footnote9}
\begin{align}
\nonumber G_{\mu\nu}^{>+}(t)&\equiv
 -i \theta(t) \  \left<
   d_\mu^\nag\noteu{t+t_1}d^\dag_\nu\noteu{t_1}\right>_{t1\to\infty}
\\
\label{eq:ssGtg0}&=-i \theta(t) \ \bra{I}d_\mu^\nag(t)
d^\dag_\nu\ket{\rho_\infty} \,\mbox{.}
\end{align}
We can use \eqref{gba} also for the
lesser Green's function, however for~\cite{footnote9}  $t<0$
\begin{align*}
G_{\mu\nu}^{<+}(t)& \equiv
 i  \theta(-t)\ \left<
   d_\nu^\dag\noteu{t_1}d^\nag_\mu\noteu{t+t_1}\right>_{t1\to\infty} 
\nonumber \\ & 
= i \theta(-t)\ \bra{I}d_\nu^\dag(-t) d^\nag_\mu\ket{\rho_\infty}\,\mbox{.}
\end{align*}
For the opposite sign of $t$, we can use  \eqref{gbam}, so that for both Green's functions one has~\cite{footnote4,footnote9}
\begin{align}
\label{gcc}
\vv G^{\stackrel{>}{<}-}(t) &=- \vv G^{\stackrel{>}{<}+}(-t)^\dag\,\mbox{.}
\end{align}
For the Fourier transformed Green's function, defined, with abuse of notation as
\begin{align}
\label{four}
\vv G^{\stackrel{>}{<}\pm}(\omega) &= \int dt \ e^{i\omega t} \vv G^{\stackrel{>}{<}\pm}(t)\,\mbox{,}
\end{align}
relation \eqref{gcc} translates into
\begin{align}
\label{gccw}
\vv G^{\stackrel{>}{<}-}(\omega) &= - \vv G^{\stackrel{>}{<}+}(\omega)^\dag\,\mbox{.}
\end{align}
We need the retarded and the Keldysh Green's functions
\begin{align}
\label{grgk}
& \vv G^R = \vv G^{>+} - \vv G^{<-} =  \vv G^{>+} + \vv G^{<+\dag}\,\mbox{,}
\\
\nonumber
&\vv G^K = \vv G^{>+}+ \vv G^{<-}+\vv G^{>-}+ \vv G^{<+} =
\vv G^{>+}+ \vv G^{<+} - h.c.\,\mbox{,}
\end{align}
whereby both relations hold for the time-dependent as well as for the Fourier transformed ones.

\subsubsection{Noninteracting case}\label{nonint}
To solve the noninteracting Lindblad problem described by \eqref{eq:L0},
one first diagonalizes the non-hermitian matrix~\cite{dz.ko.11} $\vv h$ in \eqref{eq:L02}:
\begin{align}
\label{ee}
 \vv \ee &={\vv  V}^{-1} \vv h {\vv V}\,\mbox{,}
\end{align}
where $\vv \ee$ is a diagonal matrix of eigenvalues $\ee_\mu$. The
noninteracting Lindbladian \eq{eq:L0} can then be written as
\begin{align*}
i\hat\li_0 &=  \boldsymbol{\bar{\xi}} \ \vv \ee\ \boldsymbol{\xi} + \eta\,\mbox{,}
\end{align*}
in terms of the normal modes
\begin{align}
\label{xig}
\boldsymbol{\xi} &={\vv V}^{-1}\vv{d}\,\mbox{,} \quad\quad \boldsymbol{\bar{\xi}}=\vv{d}^\dag {\vv V}\,\mbox{,}
\end{align}
and a constant $\eta$. The normal modes still obey canonical anticommutation rules
\begin{align}
\label{acr}
\{\xi_\mu,\bar\xi_\nu\}&=\delta_{\mu\nu}\,\mbox{,}
\end{align}
but are not mutually hermitian conjugate.

The steady state $\ket{\rho_\infty}$ obeys the equation
\begin{align*}
\hat \li \ket{\rho_\infty} &=0\,\mbox{.}
\end{align*}
Let us now consider the time evolution
\eqref{at} of a state initially consisting of 
the normal mode operators applied to the steady state density matrix:
\begin{align*}
e^{\hat \li_0 t} \xi_\mu \ket{\rho_\infty} &=
e^{\hat \li_0 t} \xi_\mu e^{-\hat \li_0 t}  \ket{\rho_\infty} =
e^{i\ee_\mu t} \xi_\mu\ket{\rho_\infty}\,\mbox{.}
\end{align*}
If $\iim(\ee_\mu)<0$ this term diverges exponentially in the long-time
limit, which would be in contradiction to the fact that
$\ket{\rho_\infty}$ is a steady state,
 unless the state created by $\xi_\mu$ is zero. 
Therefore, we must have
\begin{subequations}
\label{vac}
\begin{align}
\xi_\mu \ket{\rho_\infty} &=0 \quad \quad\hbox{ for } \iim(\ee_\mu)<0\,\mbox{.}
\end{align}
Similarly, we must have
\begin{align}
\label{vacb}
\bar\xi_\mu \ket{\rho_\infty} &=0
\quad \quad\hbox{ for } \iim(\ee_\mu)>0\,\mbox{.}
\end{align}
These  equations, thus, define the steady state as a kind of ``Fermi sea''. In addition, by requiring that expectation values of the form
\begin{align*}
\bra{I} \xi_\mu(t) \bar \xi_\nu \ket{\rho}\,\mbox{,}
\end{align*}
do not diverge for large $t$, we obtain that
\begin{align}
\bra{I} \xi_\mu  &= 0
\quad \quad\hbox{ for } \iim(\ee_\mu)>0\,\mbox{,}
\\
\label{vacd}
\bra{I} \bar\xi_\mu  &= 0
\quad \quad\hbox{ for } \iim(\ee_\mu)<0\,\mbox{.}
\end{align}
\end{subequations}
From \eqref{vacd} it follows  that an expectation value of the form 
$\bra{I} \bar\xi_\mu \xi_\nu \ket{\rho_\infty}$ vanishes for the case $\iim(\ee_\mu)<0$. For $\iim(\ee_\mu)>0$ 
we make use of the anticommutation rules \eqref{acr} together with
\eqref{vacb} and the fact that~\cite{dz.ko.11} 
$\braket{I | \rho_\infty}=\tr\rho_\infty =1$
and arrive at
\begin{align*}
\bra{I} \bar\xi_\mu \xi_\nu \ket{\rho_\infty}&= D_{\mu\nu}\,\mbox{,}
\end{align*}
where the matrix
\begin{align*}
 D_{\mu\nu} &= \delta_{\mu\nu} \
 \theta({\iim(\ee_{\mu}}))\,\mbox{.}
\end{align*}
Similarly,
\begin{align*}
\bra{I} \xi_\mu \bar\xi_\nu \ket{\rho_\infty}
&= \bar D_{\mu\nu} \equiv
\delta_{\mu\nu}- D_{\mu\nu}\,\mbox{.}
\end{align*}
The expression for the steady state correlation functions of the eigenmodes $\vv \xi$ of $\hat\li_0$ can be now  evaluated by considering that, due to the anticommutation rules the Heisenberg time evolution \eqref{heis} gives
\begin{align*}
\xi_\mu(t) &= e^{-i\ee_\mu t} \ \xi_\mu\,\mbox{,}\quad\quad \bar\xi_\mu(t) = e^{i\ee_{\mu}t} \ \bar\xi_\mu\,\mbox{.}
\end{align*}
Thus,
\begin{align*}
 \bra{I}\xi_\mu^\nag(t)
\bar{\xi}^\nag_\nu\ket{\rho_\infty}&=
e^{-i\ee_{\mu} t}
\bra{I}\xi_\mu^\nag
\bar{\xi}^\nag_\nu\ket{\rho_\infty}
= e^{-i\ee_{\mu} t} D_{\mu\nu}
\,\mbox{.}
\end{align*}
In this way, the greater Green's function for $t>0$ becomes
\begin{align}
\label{ggr}
& iG_{\ze\mu\nu}^{>+}(t)=
\bra{I} d_\mu^\nag(t)d^\dag_\nu\ket{\rho_\infty}
\\ \nonumber
& =\sum\limits_{\ga}V_{\mu\ga}e^{-i\ee_{\ga} t}\bar D_{\ga\ga}(V^{-1})_{\ga\nu} = \left({\vv V} e^{-i\vv \ee t} \vv {\bar D} {\vv V}^{-1} \right)_{\mu\nu}\,\mbox{,}
\end{align}
where we have used \eqref{xig}. The Green's functions are defined with operators $d_\mu^\nag/d^\dag_\mu$ 
in the original Fock space, so that it is sufficient to know the first $N_B+1$ 
rows (columns) of ${\vv V}$ (${\vv V}^{-1}$).
For this purpose we introduce 
\begin{align*}
 \vv \UU &=  \vv T {\vv V}\,,\\
 \vv \UU^{(-1)} &=  {\vv V}^{-1} \vv T^{\dagger}\,\mbox{,}
\end{align*}
whereby $\vv T$ is a $(N_B+1)\times(2N_B+2)$ matrix, which in block
form reads \mbox{$\vv T =  (\uu \hspace{1em}\vv 0)$}. Notice that \mbox{$\vv \UU^{-1} \neq \vv \UU^{(-1)}$}.
With this, the Fourier transform \eqref{four} of \eqref{ggr} is given by~\cite{footnote7}
\begin{align}
\label{ggw}
\vv G^{>+}_\ze(\omega) = \left(\vv
   \UU\frac{\vv{\bar D}}{\omega -\vv\ee}\vv
   \UU^{(-1)}\right)\,\mbox{,}
\end{align}
and $G_{\ze\mu\nu}^{>-}(\omega)$ is obtained with the help of \eqref{gccw}. 
Similarly, the lesser Green's function for $t<0$
\begin{align*}
& iG_{\ze\mu\nu}^{<+}(t)=
 -\bra{I} d_\nu^\dag d_\mu(t)\ket{\rho_\infty}
\\ \nonumber
& =- \sum\limits_{\ga}V_{\mu\ga}
e^{-i\ee_{\ga}   t} D_{\ga\ga} 
(V^{-1})_{\ga\nu}
 = - \left(\vv \UU e^{-i\vv \ee t} \vv D \vv \UU^{(-1)} \right)_{\mu\nu}\,\mbox{,}
\end{align*}
with the Fourier transform 
\begin{align}
\label{glw}
\vv G^{<+}_\ze(\omega) &= \left(\vv \UU\frac{\vv{D}}{\omega - \vv\ee }\vv \UU^{(-1)}\right) \,\mbox{,}
\end{align}
and $ \vv G^{<-}_\ze(\omega)$ is obtained from \eqref{gccw}. 
Using \eqref{grgk} together with \eqref{ggw} and \eqref{glw}, we get
\begin{align}
\label{grw}
\vv G^R_\ze(\omega) &= 
\vv \UU\frac{\vv{\bar D}}{\omega - \vv\ee }\vv \UU^{(-1)}+\left(\vv \UU\frac{\vv D }{\omega -\vv\ee }\vv \UU^{(-1)}\right)^\dag\,\mbox{,}
\end{align}
and for the Keldysh Green's function using also \eqref{gccw}
\begin{align}
\label{gkw}
\vv G^K_\ze(\omega) &=\vv \UU \left(\frac{\vv{\bar D}}{\omega - \vv\ee }+\frac{\vv D }{\omega -\vv\ee }\right) \vv \UU^{(-1)} - \hbox{h.c.} \nonumber \\
&=\vv \UU \left(\frac{1}{\omega - \vv\ee }\right) \vv \UU^{(-1)} - \hbox{h.c.}\,\mbox{.}
\end{align}

In principle, one could just carry out the diagonalization \eqref{ee} and then
evaluate \eqref{grw} and \eqref{gkw} numerically, which is a rather lightweight task.
However, it is possible to obtain a (partially) analytical 
expression for the Green's functions.
Indeed, a lengthy but straightforward calculation yields
 for the retarded one
\begin{align}
\label{gro}
\vv G^R_\ze(\omega) &= \left(\omega -\vv E+i\vv\Lambda\right)^{-1}\,\mbox{.}
\end{align}
Similarly, for   the Keldysh component of the inverse Green's function we obtain
\begin{align}
\label{gko}
\left(\vv{\und G^{-1}_\ze}\right)^K &\equiv - \vv G^{R-1}_\ze \vv G^K_\ze \vv G^{A-1}_\ze = -2i  \vv \Omega \,\mbox{.}
\end{align}
To sum up, 
\eqref{gro} and \eqref{gko} are the main results of this
subsection. To evaluate $\und\Delta_{\text{aux}}(\omega)$, one then uses \eqref{g0},
whereby one should consider that the matrix $\und G_0$ in Keldysh
space is just the {\em local one}, i.e., in terms of the  components 
local at the impurity $G^R_{\ze ff}$ and $G^K_{\ze ff}$: 
\[
\und G_0 \equiv \left(\begin{array}{cc}
              G^R_{\ze ff} & G^K_{\ze ff}   \\[4pt]
                0          & G^A_{\ze ff} 
\end{array}\right) \,\mbox{.}
\]
In turn, $G^K_{\ze ff}$, the $ff$ component of $\vv G^K_\ze$ has to be
obtained from \eqref{gko} by the well-known expression~\cite{ha.ja} $\vv G^K_0 = -\vv G^R_0 \left(\vv{\und G^{-1}_\ze}\right)^K \vv G^A_0 \,\mbox{.}$

\subsubsection{Interacting case}
\label{ssec:sigma}
The next step consists in solving
the interacting auxiliary Lindblad problem described by \eqref{li}
in order to determine the Green's function and the self energy at the
impurity site. This is done by Lanczos exact diagonalization within
the many-body augmented Fock space.

First, the steady state $\ket{\rho_{\infty}}$ has to be determined as the right-sided eigenstate of the Lindblad operator $\hat\li$ with eigenvalue $l_0=0$. For convenience we introduce
\begin{align}
\LLO &= i \hat\li \,\mbox{,} \label{eq:LLO}
\end{align}
which is a kind of non-hermitian Hamiltonian with complex eigenvalues $\LL$.
The dimension of the Hilbert space can be reduced by exploiting 
symmetries similar to the equilibrium case. 
The conservation of the particle number per spin $\hat N_\sigma$ is
replaced here by the conservation of $\hat{N}_\sigma-\hat{\tilde{N}}_\sigma$.~\cite{ar.kn.13} The steady state lies in the sector \mbox{$N_\sigma-\tilde N_\sigma=0$}. 

Starting from \eq{eq:ssGtg0}, the steady state greater Green's function of the impurity reads in a non-hermitian Lehmann representation, for $t>0$
\begin{align*}
 G_{\mu\nu}^{>+}(t)&= -i \sum_n e^{-i\LL_n^{(+1)}t} \bra{I}d_\mu^\nag \ket{R_n^{(+1)}} \bra{L_n^{(+1)}} d^\dag_\nu\ket{\rho_{\infty}} \,\mbox{,}
\end{align*}
where the identity $\sum_n\ket{R_n^{(+1)}} \bra{L_n^{(+1)}}$
in the sector $N_\sigma-\tilde N_\sigma=+1$ has been 
inserted, in terms of right ($\ket{R_n^{(+1)}}$) and left
($\bra{L_n^{(+1)}}$) eigenstates of $\LLO$ with eigenvalues $\LL_n^{(+1)}$, 
and $\ket{I}$ is the left vacuum \eqref{lvac}. 
Its Fourier transform reads
\begin{align}
 G_{\mu\nu}^>(\omega)&= \sum\limits_n \frac{1}{\omega-\LL_n^{(+1)}} \bra{I}d_\mu^\nag \ket{R_n^{(+1)}} \bra{L_n^{(+1)}} d^\dag_\nu\ket{\rho_{\infty}} \nonumber\\
 & - \sum_n \frac{1}{\omega-\LL_n^{(+1)*}} \left(\bra{I}d_\nu^\nag \ket{R_n^{(+1)}} \bra{L_n^{(+1)}} d^\dag_\mu\ket{\rho_{\infty}}\right)^* \,\mbox{.}
 \label{eq:mbGFgreater}
\end{align}
The analogous expression for the lesser Green's function $G_{\mu\nu}^<(\omega)$ is obtained by inserting a complete set of eigenstates in the $N_\sigma-\tilde N_\sigma=-1$ sector and exchanging the elementary operators accordingly. $G_{\mu\nu}^K(\omega)$ and $G_{\mu\nu}^R(\omega)$ are obtained using \eq{grgk}, see also \eqref{gccw}.

For a small number of bath sites $N_B\leq 3$ the dimension of the
augmented Fock space is still moderate, and eigenvalues and eigenvectors
can be determined by full diagonalization. For $N_B\geq 4$ a
non-hermitian Lanczos procedure has to be carried out. Especially
extracting the steady state is not an easy task, since it lies in the
center of the spectrum. Details of our numerical procedure are given in \App~\ref{calcself}.

Once the interacting and non-interacting Green's functions of the
auxiliary system at the
impurity site $\underline{G}(\omega)$ and $\underline{G}_0(\omega)$, respectively,
are determined, the corresponding self energy 
is obtained via Dyson's equation in Keldysh space \eq{dys}.
The individual components are explicitly~\cite{ar.kn.13}
\begin{subequations}
\begin{align*}
 \Sigma^R(\omega) &= 1/G_0^R(\omega) - 1/G^R(\omega) \\
 \Sigma^K(\omega) &= -G_0^K(\omega)/|G_0^R(\omega)|^2 + G^K(\omega)/|G^R(\omega)|^2\,\mbox{.}
\end{align*}
\end{subequations}
As discussed in Sec.~\ref{ssec:ssnegf}, this is used
in the Dyson equation \eqref{dys} for 
the physical Green's function.

\begin{center}
 \begin{figure*}
\includegraphics[width=1\textwidth]{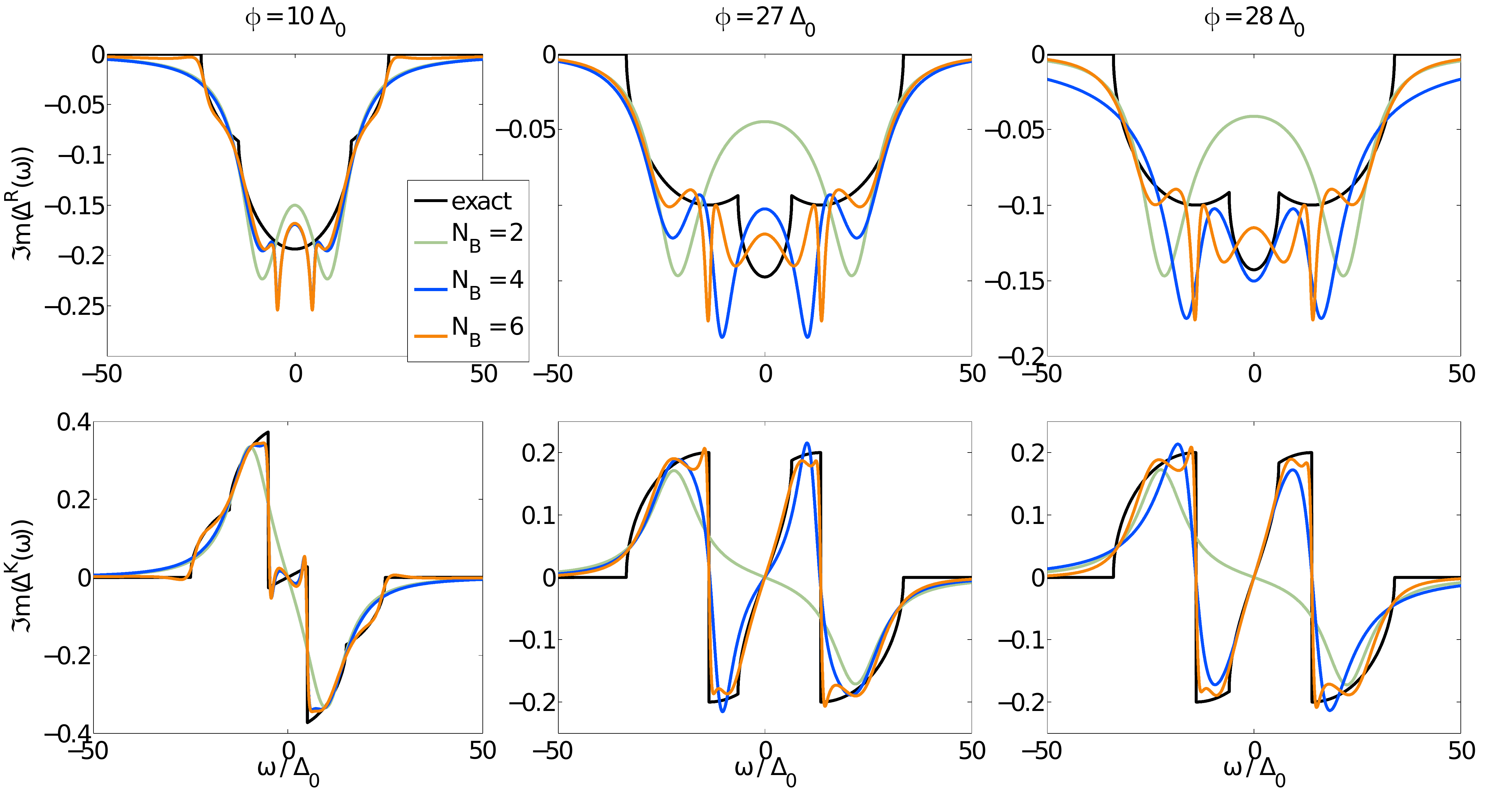}
\captionsetup{justification=raggedright,singlelinecheck=false}
\caption{(Color online) Comparison of $\iim(\Delta^\alpha(\omega))$ from
  \eqref{deltaw}
  (black) with $\iim(\Delta^\alpha_{\text{aux}}(\omega))$ at the
  absolute minimum of the cost function \eq{eq:SC} for auxiliary
  system sizes  $N_B = 2,4,\mbox{ and }6$ bath sites (green, blue
  and orange, respectively), and $\alpha=R$ (top) and $K$ (bottom). 
Results are shown for tight-binding leads \eq{gsc} and
  \eq{fd} with $t=10\,\Delta_0$, and three different bias
  voltages $\phi \in \{10,27,28\}\,\Delta_0$ from left to right.} 
\label{fig:fits}
\end{figure*}
\end{center}
\section{Results}\label{sec:results}
In this section, results for the steady state properties of a symmetric, correlated Anderson impurity coupled to two metallic leads 
under bias voltage are provided. We assess the validity of the proposed method by discussing the fit of the hybridization 
function and outline how uncertainties are estimated. Results for the current voltage characteristics and the non 
equilibrium spectral function are presented and compared with data from TEBD~\cite{nu.ga.13} and SNRG~\cite{an.sc.10} 
calculations, respectively. The  effect of a linear correction of the calculated Green's functions is 
illustrated.

\subsection{Hybridization functions}\label{ssec:resultsFit}

The optimal representation of the exact bath $\underline{\Delta}(\omega)$ by the auxiliary one 
$\underline{\Delta}_{\text{aux}}(\omega)$ is obtained by minimizing the cost function \eq{eq:SC}. 
In practice this is done by employing a quasi-Newton line search.~\cite{shan.70,pr.te.07} 
In particular, we chose an equal weighting of the retarded and the Keldysh component 
$W^R(\omega)=W^K(\omega)=\Theta(\omega_c- |\omega|)$.
After finding our results to be robust upon different values for the cut-off $\omega_c$, as well as upon using 
different norms ($n=1,2$) in \eq{eq:SC}, we finally choose $\omega_c = 50\, \Delta_0$ and consider imaginary parts 
$\left(\iim(\Delta^\alpha(\omega)-\Delta^\alpha_{\text{aux}}(\omega))\right)^2$ in the cost function only. This is justified 
since $\Delta_{\text{aux}}^K(\omega)$ is purely imaginary and the real part of $\Delta_{\text{aux}}^R(\omega)$ 
is connected to its imaginary part via the Kramers-Kronig relations.~\cite{jack.75} The asymptotic behavior of $\Delta^R_{\text{aux}}(\omega)$ is determined by $\Lambda_{ff}$ whereas the one of $\Delta^K_{\text{aux}}(\omega)$ by $\Omega_{ff}$. Therefore, the correct asymptotic limit 
$\lim\limits_{\omega\rightarrow\pm\infty}\underline{\Delta}_{\text{aux}}(\omega) = \und 0$ is guaranteed by taking 
$\Gamma^{(1)}_{ff}=\Gamma^{(2)}_{ff}=0$, which results in
$\Gamma^{(\kappa)}_{\mu f}=\Gamma^{(\kappa)}_{f\mu}=0$ due to the
requirement of semi-positive definiteness of $\Gamma^{(\kappa)}_{\mu\nu}$. Particle-hole symmetry allows for a further reduction of the auxiliary system parameters.~\cite{footnote8}

In this work we use an even number of auxiliary bath sites $N_B = 2,4, \mbox{ and }6$ in a 
linear setup (see \fig{fig:model} (b)) with an equal number to the left and to the right of the impurity 
(only \fig{fig:lincorr} displays one calculation for an odd number of bath sites). In \fig{fig:fits}, 
the obtained auxiliary hybridization functions are compared with the exact ones for various bias voltages. We find a quick 
convergence as a function of $N_B$, which degrades
for large
 bias voltage $\phi$. The Fermi steps at the chemical 
potentials in $\Delta_{\text{aux}}^K(\omega)$ cannot be properly resolved in the case of $N_B = 2$. Especially in the case of 
$\phi = 10 \,\Delta_0$ the auxiliary hybridization functions for $N_B = 6$ as well as for $N_B = 4$ agree fairly well with the 
exact one and capture all essential features, in particular the Fermi steps. The auxiliary bath develops spurious oscillations in 
$\Delta_{\text{aux}}^R(\omega)$ at the energies of the Fermi levels of the contacts. Here the discrepancy with 
$\Delta^R(\omega)$ is considerable in magnitude, but extends over small $\omega$-intervals, thus inducing only small errors 
in the self energies.

\begin{figure*}
\includegraphics[width=1\textwidth]{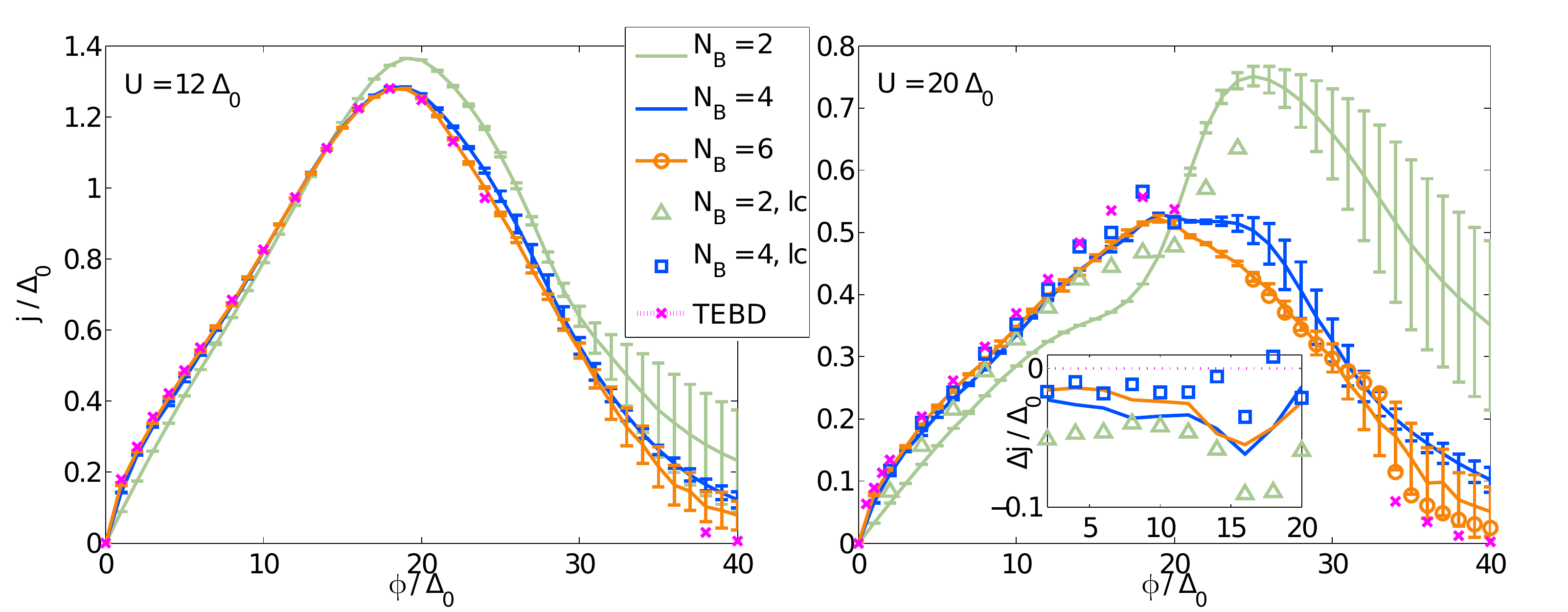}
\captionsetup{justification=raggedright,singlelinecheck=false}
\caption{(Color online) Current $j$ vs. voltage $\phi$ 
 for the model \eqref{eq:H} with tight-binding leads 
and  on-site interaction
   $U = 12\,\Delta_0$ (left) and $U = 20\,\Delta_0$ (right). 
Results
  for three different auxiliary  systems with $N_B \in
  \{2,4,6\}$  are displayed 
  and compared with reference data from TEBD (magenta  dotted and $\times$).~\cite{nu.ga.13}
 We plot the
  averaged mean values connected by lines together with error bars determined according to
  Sec. \ref{ssec:resultsFit} and \App~\ref{app:Averaging}. The
  additional data marks for $U = 20\,\Delta_0$ are as
  follows: The circles for $N_B=6$ display $j(\phi)$ 
when considering
   the absolute minimum in the fit \eqref{eq:SC}. $N_B=2, \mathrm{lc}$
   and $N_B=4, \mathrm{lc}$  present the results
  of a linear correction of the current values of the absolute minima
  as described in \App~\ref{ssec:error}. The inset displays the
  difference $\Delta j$ of
  the calculated currents to the TEBD results.
} 
\label{fig:jV}
\end{figure*}

When following the absolute minimum of the cost function \eq{eq:SC} as a function of some external parameter, such as, e.g., the bias voltage $\phi$, spurious discontinuities appear due to the
fact that local minima cross each other. This occurs for large bias
voltages and large $U$, and/or small $N_B$, for which the approach is more challenging.
An example for such a situation is shown in \fig{fig:fits} for the case $N_B = 4$, when comparing the hybridization functions 
just before and after such a crossing, i.e. for $\phi = 27\,\Delta_0$ and $\phi = 28\,\Delta_0$. Even though the changes in 
the exact hybridization function are only minor, $\underline{\Delta}_{\text{aux}}(\omega)$ displays a considerable difference. 
The influence of this spurious effect on observable quantities is shown in 
\fig{fig:jV} (right panel, orange circles) for a different parameter set of 
$N_B=6$ at around $\phi_c=33\,\Delta_0$. The artificial discontinuity in the current is caused by the shift of spectral 
weight in $\underline{\Delta}_{\text{aux}}(\omega)$.

To deal with these discontinuities,
 we adopt a scheme which is suitable for obtaining a continuous dependence of observables on external 
parameters and in addition, allows to estimate their uncertainties
(see \fig{fig:jV}).
We first identify a set of local minima 
of the cost function \eq{eq:SC}, obtained by a series of minimum
searches starting with random initial values. 
 These local minima are then used to calculate an average and 
variance of physical quantities, such as the current. We consider the
distribution of local minima with a Boltzmann weight 
associated with an artificial ``temperature'', whereby the value of the cost function \eq{eq:SC} is the associated ``energy''. 
This artificial temperature for the Boltzmann weight is chosen in such a way, that the averaged spectral weight of the 
hybridization function as a function of $\phi$ is as smooth as possible. Details are outlined in \App~\ref{app:Averaging}. 
A possible pitfall however is, that physical discontinuities, i.e., real phase transitions could be overlooked. It is thus compulsory to additionally investigate the results for the absolute minima and for different bath setups carefully. This approach has a certain degree of arbitrariness. However, we point
out that it only affects regions with large error bars in
Fig.~\ref{fig:jV}, i.e. large $\phi$ and large $U$ for which also
other techniques are less accurate.

\subsection{Current voltage characteristics}\label{ssec:resultsjV}

After evaluating the interacting impurity Green's function of the
physical system according to \eqref{dys} 
with the self energy evaluated in Sec.~\ref{ssec:delta}, we are able to determine the steady-state current.
This is done with the help
of the Meir-Wingreen expression~\cite{me.wi.92,ha.ja,jauh} in its
symmetrized form, where we have already summed over spin 
\begin{align} j&=i\int\limits_{-\infty}^\infty\frac{d\omega}{2\pi}\left(\left(\gamma_L(\omega)-\gamma_R(\omega)\right)G^{<}(\omega) \right.\label{eq:MW}\\
\nonumber&+\left.\left(p_{\text{F},L}(\omega)\gamma_L(\omega)-p_{\text{F},R}(\omega)\gamma_R(\omega)\right)\left(G^{R}(\omega)-G^{A}(\omega)\right)\right)\,\mbox{,}
\end{align}$\gamma_\lambda(\omega)=-2|t'_\lambda|^2\iim(g_\lambda^{R}(\omega))$ are the ``lead self-energies`` and 
$p_{\text{F},\lambda}(\omega) = p_{\text{F}}(\omega-\mu_\lambda)$ denotes the Fermi distribution of lead $\lambda$ with 
chemical potential $\mu_\lambda$.

To quantify the accuracy of the method we compare the results for the current voltage characteristics with quasi exact reference 
data from TEBD.~\cite{nu.ga.13} We find very good agreement for
interaction strength $U < 12 \,\Delta_0$. Since in this paper we want to benchmark
the approach in ``difficult'' parameter regimes, in the following, we will discuss $U\gtrsim
12 \Delta_0$ only.
 In \fig{fig:jV} we 
display data for  $U = 12\,\Delta_0$ and $U = 20\,\Delta_0$. 
The data points and error bars shown are obtained by using the averaging scheme as described 
in \App~\ref{app:Averaging}. For the universal physics at small and medium bias voltages $\phi \lesssim 20\,\Delta_0$, 
the current as a function of the auxiliary system size ($N_B \in \{2,4,6\}$) converges rapidly to the expected result. The convergence is even monotonic in a broad region of the parameter space.
The zero bias response is linear for all $N_B$ and approaches the
results expected from the Friedel sum rule~\cite{hews.97}
 $j(\phi=0^+)=2\frac{e^2}{h}\phi$ 
quickly for increasing $N_B$. For $U = 12\,\Delta_0$ already the $N_B\gtrsim 4$ results yield a good reproduction of 
the current in this bias regime. For $U = 20\,\Delta_0$ and $\phi \gtrsim 20\,\Delta_0$ a larger difference between the 
$N_B = 4$ and $N_B = 6$ results is observed. Notice that also other available methods do not yield a satisfactory result in this parameter regime.
In the lead dependent high bias regime the fit becomes more challenging and 
large variances appear in the calculated quantities. This indicates the presence of many competing local minima with similar 
values for the cost function whose value tends to increase with increasing $\phi$. For $\phi \geq 40 \, \Delta_0$ the 
densities of states of the left and the right contact do not overlap anymore and the current has to vanish. This limit 
cannot be exactly reproduced by the proposed approach due to spurious long-range Lorentzian tails present in the auxiliary 
Markovian environment. Nevertheless, $j(\phi = 40 \, \Delta_0)$ approaches zero as one increases the number of bath sites. 
This holds true for quantities obtained at the absolute minimum of the cost function as well as for averaged ones.

To extrapolate our results to larger $N_B$, a scheme for linear corrections is discussed in \App~\ref{ssec:error}. 
Data for $N_B=2, \mathrm{lc}$ and $N_B=4, \mathrm{lc}$, whereby ``lc'' denotes  ``linear correction'', is shown in \fig{fig:jV}. 
For large $U=20\,\Delta_0$ and small to medium bias voltages $\phi \lesssim 20\,\Delta_0$, a solid improvement towards the 
TEBD reference values is observed (see inset \fig{fig:jV}). Correction ratios $r$ (see \App~\ref{ssec:error}) close to one 
indicate a good applicability of the linear correction scheme. We find on average 
$r\approx0.75$ for $\phi \lesssim 20\,\Delta_0$ ($N_B=2, \mathrm{lc}$ and $N_B=4, \mathrm{lc}$). 
In the high bias regime, however, the linear correction cannot be applied with large magnitude and $r$ drops below $0.5$ for 
$N_B=2, \mathrm{lc}$. Nevertheless, the calculation of the effective, auxiliary hybridization function $\und{\Delta}_{\aux,r}(\omega)$
as described in \App \ref{ssec:error}, successfully avoids an ''over-correction'' of the current values and automatically allows
one to estimate the reliability of the results.

Judging from the larger uncertainty from the averaging procedure and the strong effects of the linear corrections, we conclude that the 
high bias regime is more sensitive to the details of the fitted, auxiliary hybridization function. The universal low and 
medium bias regime are however very well reproduced even with a small number of auxiliary bath sites.

\subsection{Non-equilibrium spectral function}\label{ssec:resultsAwV}

\begin{center}
 \begin{figure*}
 \subfloat{
\includegraphics[width=0.45\textwidth]{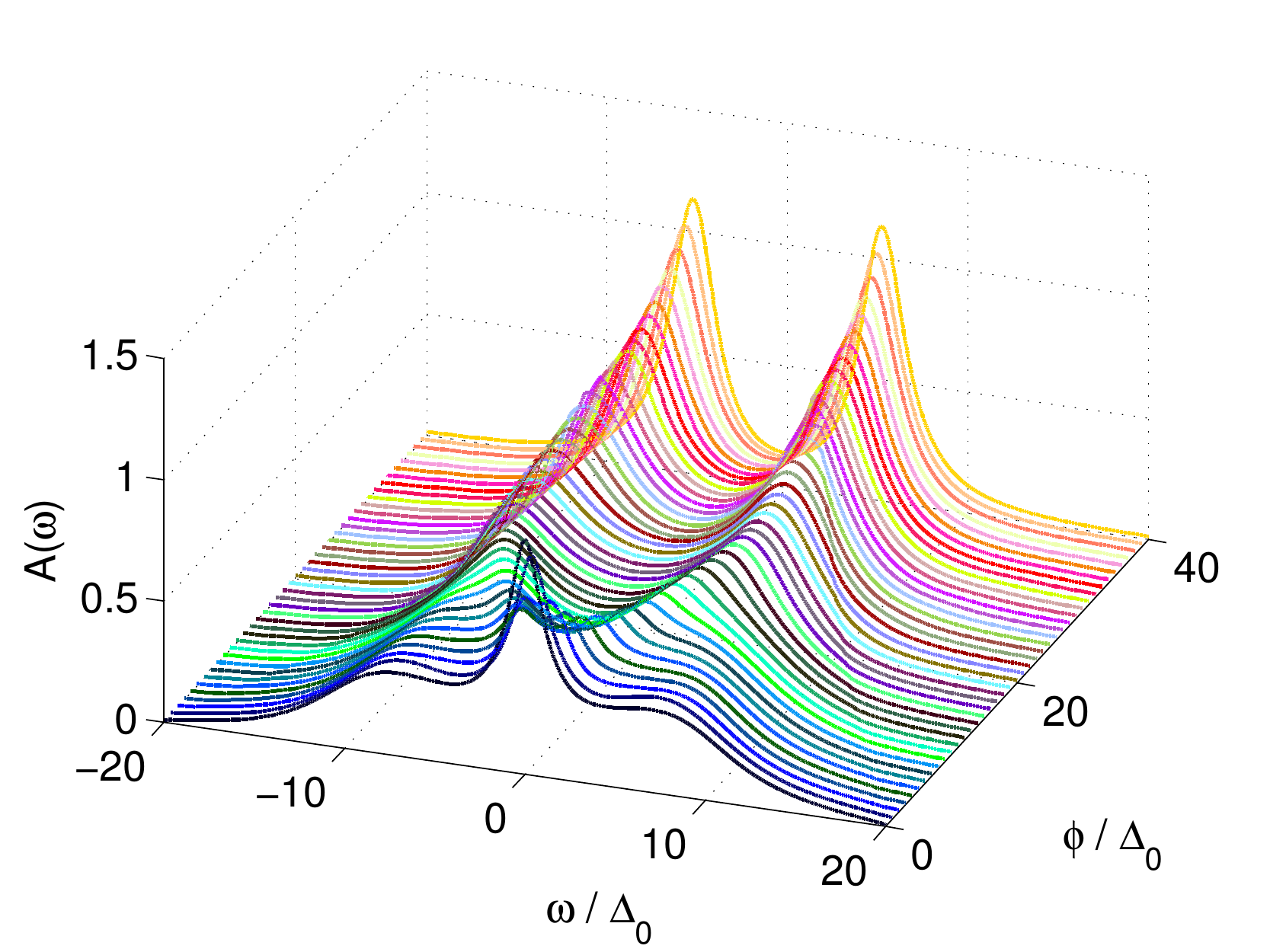}}  \hspace{-0.025\textwidth}
\subfloat{
\includegraphics[width=0.45\textwidth]{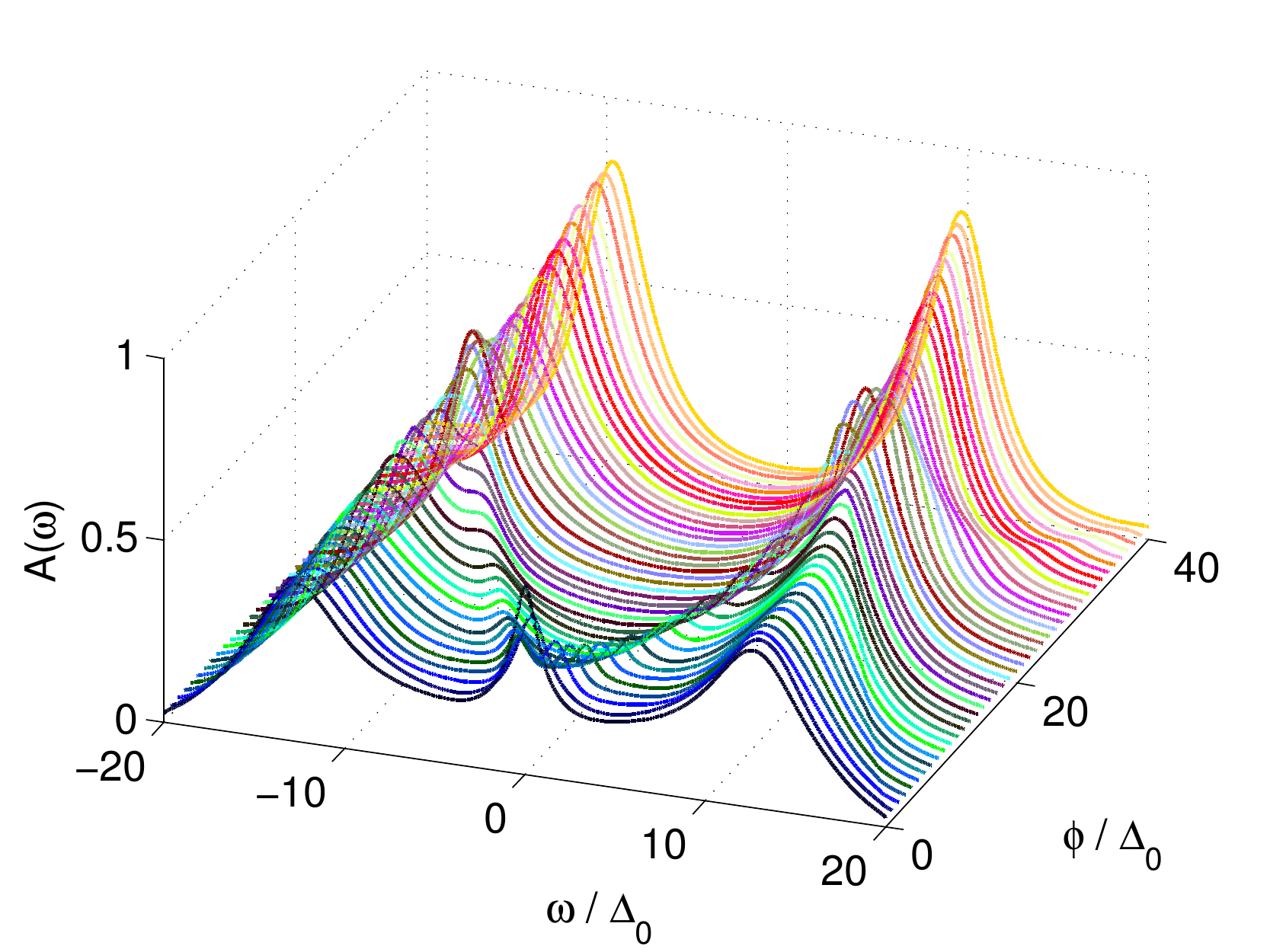}}  \\ \vspace{-0.025\textwidth}
\subfloat{
\includegraphics[natwidth=2000bp,natheight=1500bp,width=0.45\textwidth]{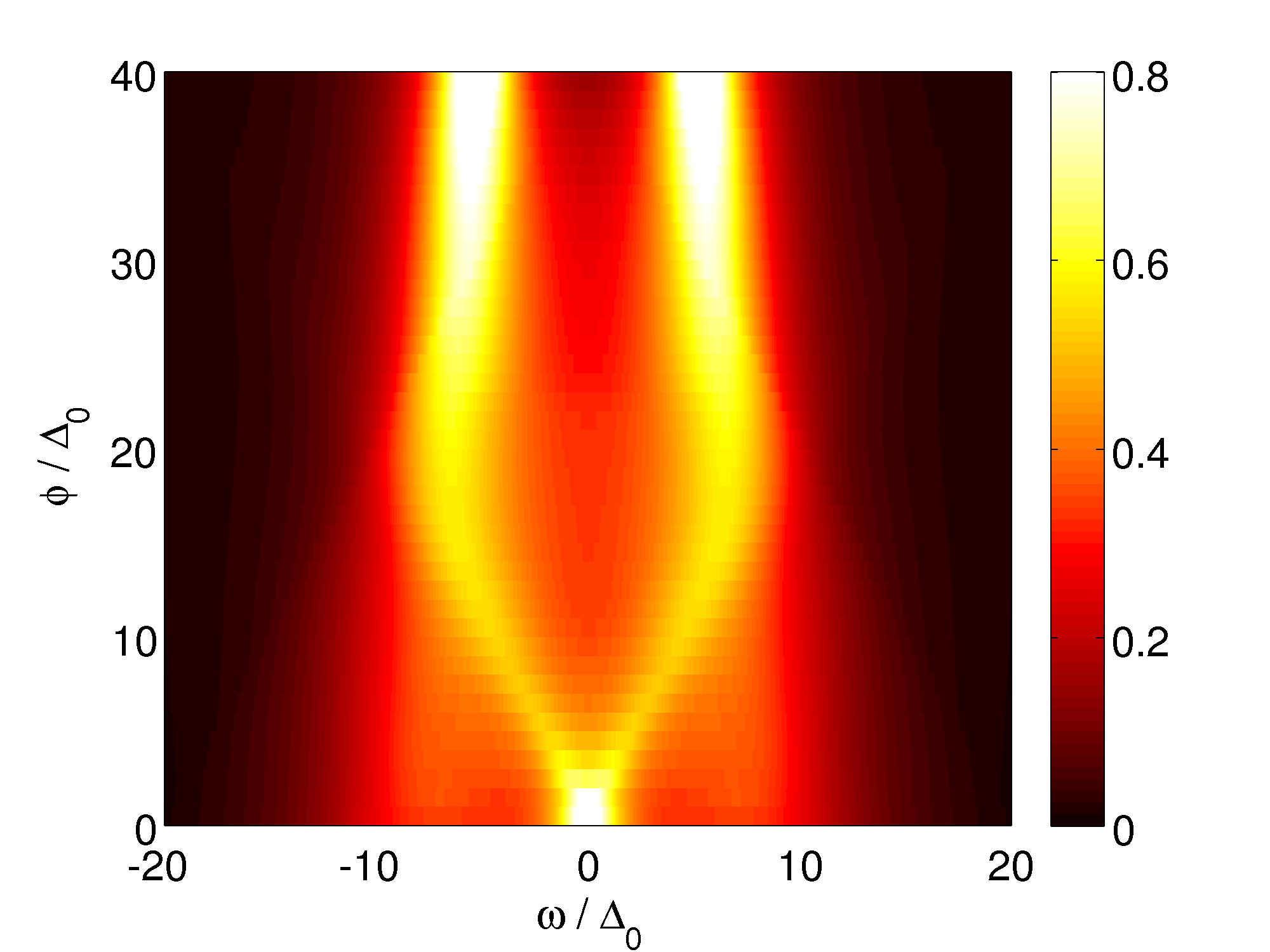}} \hspace{-0.02\textwidth}
\subfloat{
\includegraphics[natwidth=2000bp,natheight=1500bp,width=0.45\textwidth]{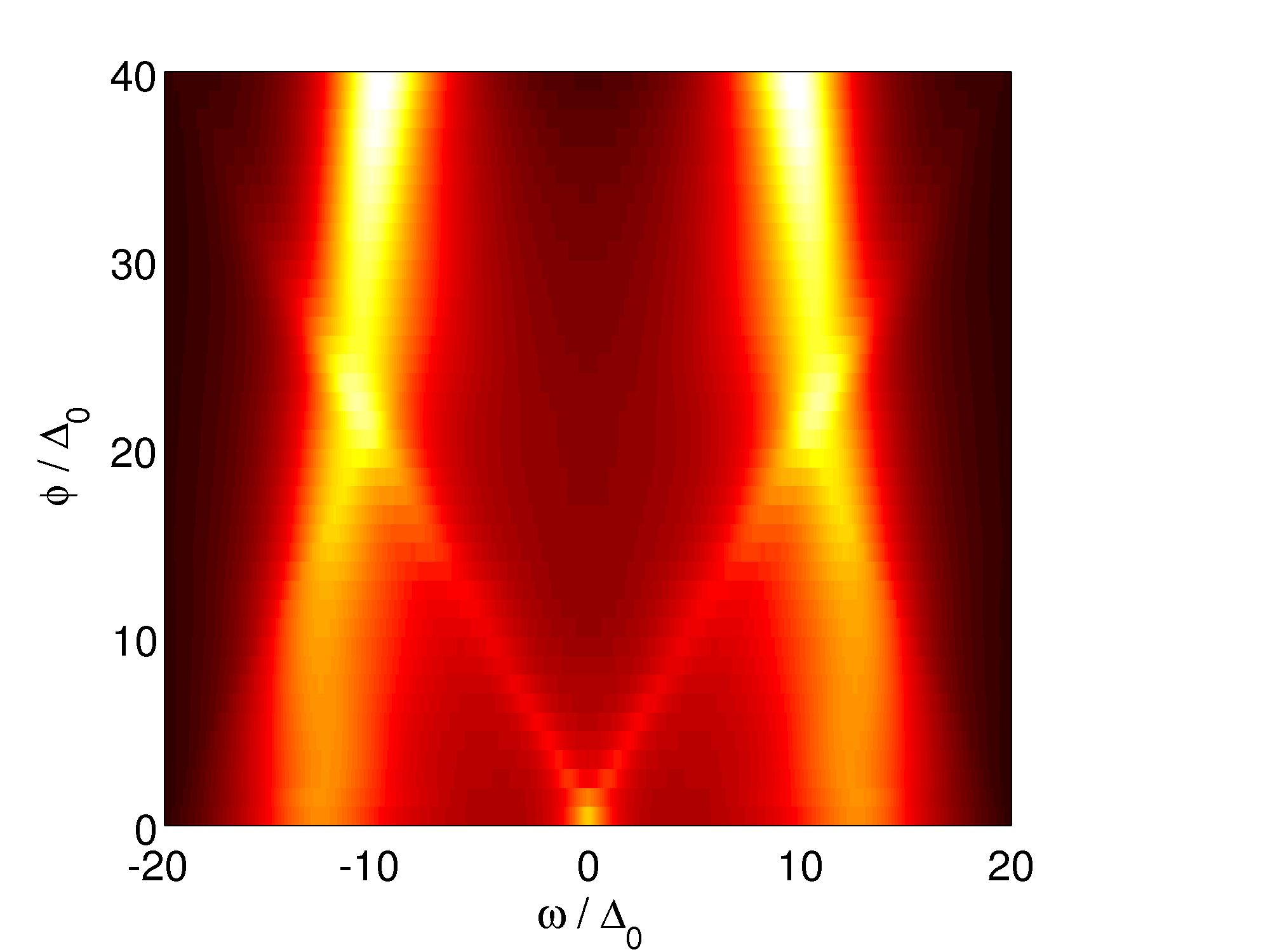}} 
\captionsetup{justification=raggedright,
singlelinecheck=false}
\caption{(Color online) Single particle spectral function at the
  impurity evaluated for $N_B = 6$, different bias voltages $\phi$ and $U = 12\,\Delta_0$ (left)
and $U = 20\,\Delta_0$ (right). Data is obtained according to
\se~\ref{ssec:resultsFit} and \App~\ref{app:Averaging}. Other
parameters are as in Fig.~\ref{fig:jV}. }
\label{fig:Aw}
\end{figure*}
\end{center}

The bias-dependent single particle spectral function is evaluated from the physical steady state Green's function of the 
impurity $A(\omega)=-\frac{1}{\pi}\iim(G^{R}(\omega))$. Results obtained using $N_B = 6$ for $U = 12\,\Delta_0$ and 
$U = 20\,\Delta_0$ are presented for the whole bias range of interest in \fig{fig:Aw}. Data for $N_B = 4$ are  
similar but here the Kondo physics cannot be reproduced as accurately as in the case of $N_B = 6$. Our approach does preserve 
the local charge density $\langle n_f\rangle = \sum\limits_\sigma \frac{1}{2}+\frac{1}{2}\int\limits_{-\infty}^\infty\frac{d\omega}{2\pi}\iim(G^{K}(\omega))=1$ 
and magnetization $\langle m_f\rangle=0$ as well as the spectral sum rule.~\cite{negele.orland}

The presented method reproduces qualitatively correctly also the equilibrium physics at $\phi = 0 $, since $A(\omega)$ 
displays a Kondo resonance at $\omega = 0$ and two Hubbard satellites at the approximate positions $\omega \approx \pm U/2$. 
This renders the application to equilibrium DMFT problems an interesting perspective. The width and magnitude of the Kondo 
resonance is discussed in comparison with (S)NRG data in \se~\ref{ssec:resultsSNRGlincorr}.

Upon increasing the bias voltage, the Kondo resonance splits up and two excitations are observed at the energies of the 
Fermi levels of the leads.~\cite{fr.ha.02,le.sc.05, nu.he.12} For $U = 12\,\Delta_0$, the splitted resonances merge into the 
Hubbard bands at approximately $\phi \approx 15 \, \Delta_0$ and cannot be clearly identified thereafter. In contrast, in the 
case of $U = 20\,\Delta_0$  the resonances overlap with the Hubbard satellites and can still be observed in the spectrum 
$A(\omega)$ at higher voltages. Calculations with increasing $U$ in the high bias regime $\phi \approx 40 \, \Delta_0$ have 
shown the consistency of this effect and that a minimum value of $U \approx 15\,\Delta_0$ is needed in order for the 
resonances at the Fermi energies to be perceptible after having crossed the Hubbard bands.

\subsubsection{Comparison with scattering states numerical renormalization group}\label{ssec:resultsSNRGlincorr}
\begin{center}
 \begin{figure}
\includegraphics[width=0.45\textwidth]{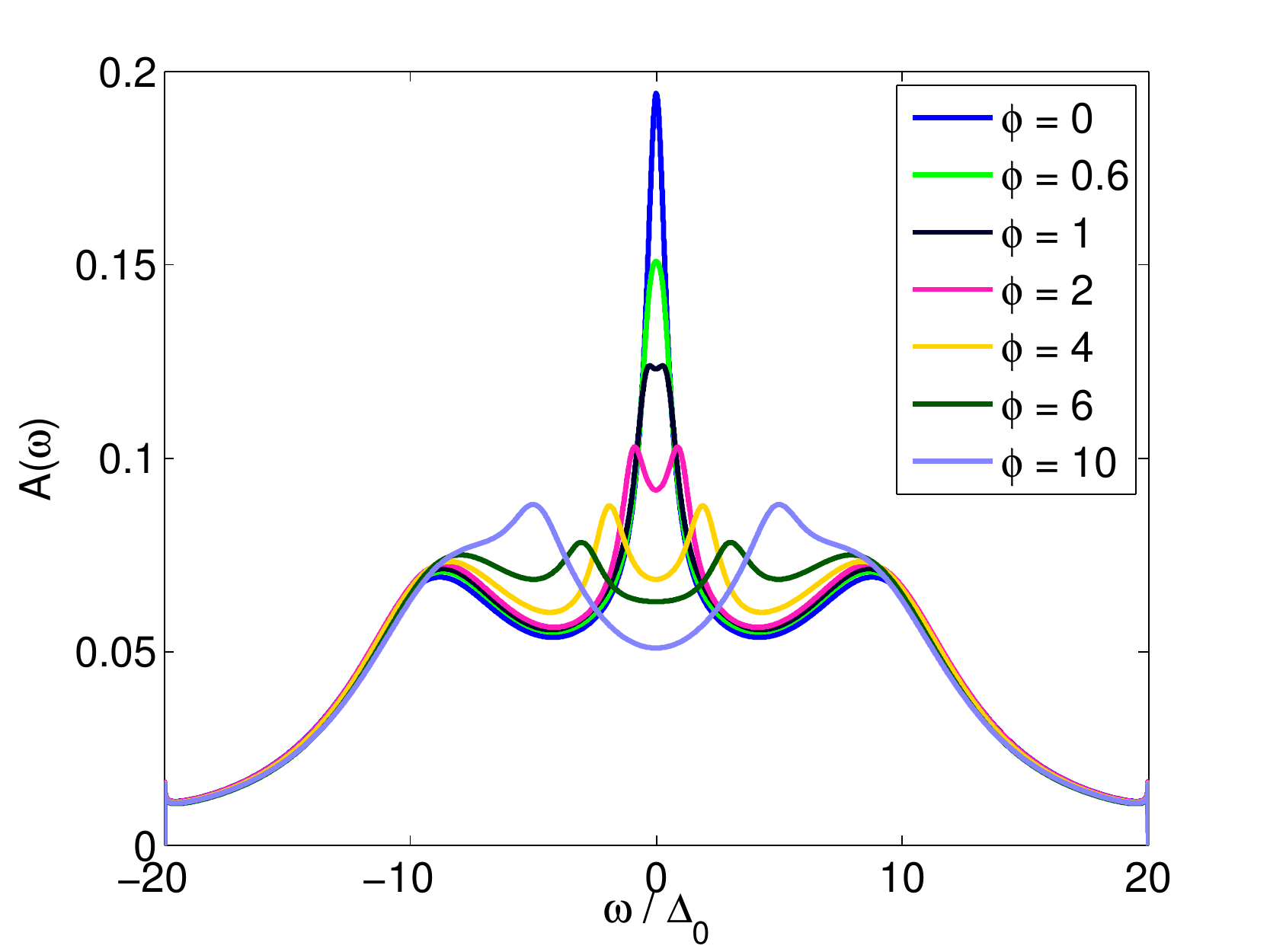}
\captionsetup{justification=raggedright,singlelinecheck=false}
\caption{(Color online) Single particle spectral
  function for 
a constant DOS of the leads \eq{gwb}, with $D_\text{WB} =
20\,\Delta_0$ and $U = 16\,\Delta_0$
and different bias voltages $\phi$ (in units of $\Delta_0$).
Results are obtained for $N_B = 6$ and at the absolute minimum of \eq{eq:SC}.  
 For a comparison with SNRG  \cite{ande.08} (Fig. 2a therein), note that
 their $\Gamma = 2\,\Delta_0$.
}
\label{fig:Aw_compSNRG}
\end{figure}
\end{center}

We compare the computed spectral functions with results obtained by means of SNRG.~\cite{ande.08} For this purpose, we use a 
flat DOS \eq{gwb} for the leads, as in \tcite{ande.08}. Focusing on the low bias regime and $N_B = 6$, the obtained spectral 
functions are depicted in \fig{fig:Aw_compSNRG}. Compared with SNRG, our results do not achieve the same accuracy in the low 
energy domain, i.e. in the vicinity of $\omega \approx 0$. However, our data provides a better resolution at higher energies. 
When inspecting the Kondo peak in the equilibrium case $\phi = 0$, our results do not fully fulfill the Friedel 
sum rule.~\cite{hews.97,lang.61,lang.66} Depending on parameters the height of the Kondo resonance is underestimated. 
This is due to the fact that the imaginary part of the self-energy at $\omega=0$ has a small finite value which is due to the 
Lorentzian tails of the Markovian environment. 

The resolution does not suffice to tell whether a two or a three peak structure is present for very low bias voltages  $\phi \lesssim 2\,\Delta_0$. Nevertheless, one can say that the higher bias regime $\phi > 4\,\Delta_0$ is resolved more accurately and one is able to clearly distinguish the excitations at the Fermi energies of the contacts from the Hubbard satellites. The observed linear splitting is consistent with experiments on nanodevices.~\cite{le.sc.05,fr.ha.02} Within second-order Keldysh PT~\cite{fu.ue.03} and QMC results~\cite{mu.ur.11} the resonance does not split but is suppressed only. In fourth-order and in NCA it splits into two, which are located near the chemical potentials of the two leads.~\cite{fu.ue.03} Other methods yield a splitting with features slightly different in details: real-time diagrammatics,~\cite{ko.ju.96} VCA,~\cite{nu.he.12} imaginary potential QMC~\cite{di.ha.13} or scaling methods.~\cite{ro.pa.03}

Overall, a good qualitative agreement with the SNRG results is achieved which underlines the reliability of the calculated spectral functions.

\subsubsection{Linear correction of Green's functions}
\begin{center}
 \begin{figure*}
 \subfloat{
\includegraphics[width=0.35\textwidth]{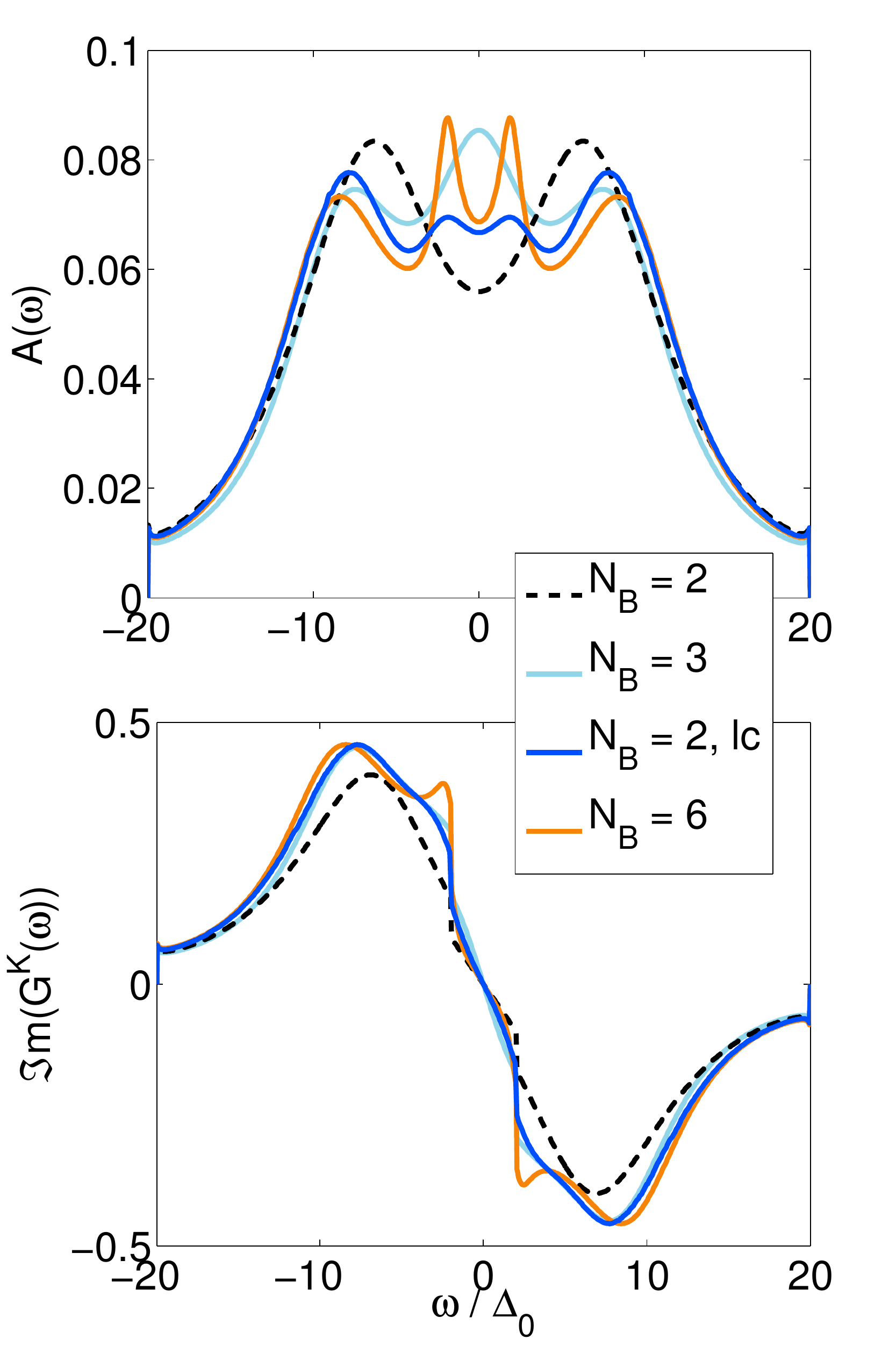}
} \hspace{-0.035\textwidth}
 \subfloat{
\includegraphics[width=0.35\textwidth]{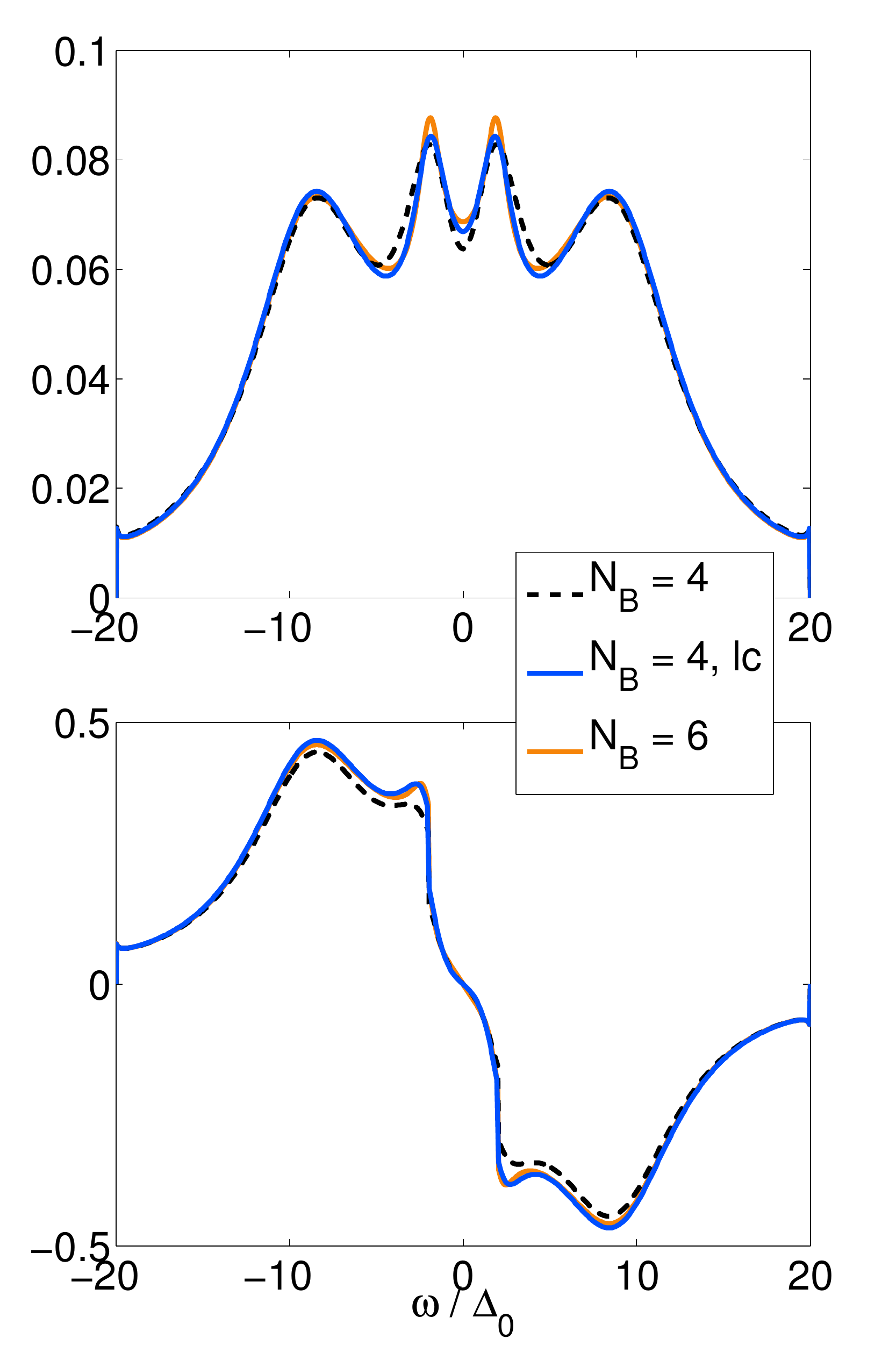}
} 
\captionsetup{justification=raggedright, singlelinecheck=false}
\caption{(Color online) 
 Effects of the linear corrections
 of the  Green's functions according to \se~\ref{ssec:error}
(solid blue lines). 
 The dashed lines indicate the uncorrected
  $\underline{G}(\omega)$ with the same $N_B$,
while solid orange and light blue  lines display results for larger $N_B$ for comparison.
Results are shown for a constant lead DOS  \eq{gwb} with
  $D_\text{WB} = 20\,\Delta_0$, $U = 16\,\Delta_0$ and $\phi = 4\,
  \Delta_0$.}
\label{fig:lincorr}
\end{figure*}
\end{center}

Here we consider the effect of a linear correction of the Green's functions, as outlined in \App~\ref{ssec:error}. 
In the left panels (right panels) of \fig{fig:lincorr}, we show data for $N_B = 2$ ($N_B=4$) including linear corrections 
($r=1$) for a high interaction strength in the low bias regime. We benchmark to data obtained using $N_B=6$ without corrections.

For $N_B=2$ without linear corrections, the spectral function of the auxiliary system does not feature excitations at the 
Fermi energies of the contacts ($\omega=\pm2\,\Delta_0$), which are present in the $N_B=6$ data. Also the spectra appear 
washed out. The linearly corrected result, however, features not only the two resonances at the appropriate energies but also 
the shoulders present in the reference data. Again in the Keldysh Green's function a large correction towards the more 
accurate $N_B = 6$ results is observed.
To highlight the fact that the improvement of the linear correction is not only due to the inclusion of one additional bath site,
also a calculation for an auxiliary system with $N_B=3$ is shown. Evidently, the $N_B=3$ spectral function exhibits a large
weight at low frequencies, but, the resolution is rather low and only a single, smeared out peak at $\omega = 0\,\Delta_0$ is observed. It clearly does not account for the splitting of the Kondo resonance.

For $N_B = 4$, a similar enhancement is found. Clearly the size of the corrections is much smaller. 
Especially in the Keldysh component, the Green's function for $N_B = 6$ and for the corrected $N_B = 4$ system nearly coincide. 
In general, the difference between the $N_B = 6$ and the $N_B = 4$ calculations (raw and corrected) are quite small, 
so that the presented spectral functions in \fig{fig:Aw_compSNRG} for larger values of $\phi \lesssim 12\,\Delta_0$ can be 
assumed to be quite accurate.

Overall, the linear correction enables a vast improvement in the universal low and medium bias regime for all $U$, 
which becomes especially important for large $U$. For large bias voltages, when lead band effects become prominent, the linear correction 
is more challenging (see also \se~\ref{ssec:resultsjV}).

\section{Conclusions}\label{sec:conclusion}
We have presented a numerical approach to study correlated quantum
impurity problems out of equilibrium.~\cite{ar.kn.13} The auxiliary master equation
approach presented here is based on a mapping of the original
Hamiltonian to an auxiliary open quantum system consisting of the
interacting impurity coupled to bath sites as well as to a
Markovian environment. The dynamics of the auxiliary open system are
controlled by a Lindblad master equation. Its  parameters  are determined by a fit to the impurity-environment
hybridization function. This has many similarities to the procedure used for the exact-diagonalization dynamical mean field theory
impurity solver, but has the advantage that one can work directly with
real frequencies, which is mandatory for non-equilibrium systems.

We have illustrated how the accuracy of the results can be 
estimated, and systematically improved by increasing the number of
auxiliary bath sites. A scheme to introduce linear corrections has
been devised. We presented in detail how the non-equilibrium Green's
functions of the correlated open quantum system are obtained by making
use of non-hermitian Lanczos diagonalization in a super-operator
space. These techniques make the whole method fast and  
efficient as well as particularly suited as an impurity solver for
steady state dynamical mean field theory.~\cite{ar.kn.13} 

In this work we have applied the approach to the single impurity Anderson model, which is one of the paradigmatic quantum impurity models. We have analyzed in detail the systematic improvement of the current-voltage characteristics as a function of the number of auxiliary bath sites. Already for four auxiliary bath sites, results show a rather good agreement with quasi exact data from time evolving block decimation~\cite{nu.ga.13} in the low and medium bias regime. In the high bias regime, the current deviates from
the expected result with increasing interaction strength.  However, we have shown how to estimate the reliability of the data from the
deviation of the hybridization functions and how results can 
be corrected to linear order in this deviation. 
The impurity spectral function obtained in our calculation
features
 a linear splitting of the Kondo resonance as a function of bias
 voltage. Good agreement with data from scattering state
 numerical renormalization group~\cite{ande.08} was found.   

Applications of the presented method to multi-orbital correlated impurities or
correlated clusters is in principle straightforward, although
numerically more demanding. Such systems are themselves of interest  as
models for transport through molecular or nanoscopic objects and as
solvers for non-equilibrium cluster dynamical mean field
theory. In this case, a larger number of auxiliary sites might be
necessary to obtain a good representation of the various hybridization
functions. For this situation, one should use numerically more efficient methods to
solve for larger correlated open quantum systems, such as 
matrix product states and density matrix renormalization group, 
possibly combined with  stochastic wave-function
approaches,~\cite{da.ca.92,da.ta.09,pr.zn.09}, sparse polynomial space,~\cite{al.fe.09,we.we.06} or configuration interaction approaches.~\cite{lo.sa.99}
A more accurate determination of low-energy, and possibly critical
properties might be achieved by a combination with renormalization group iteration schemes,
 similar to the numerical renormalization group.
Work along these lines is in progress.

Although we have presented results for the steady state, where the
method is most efficient, also extensions to time dependent phenomena
provide an interesting and feasible perspective.
 While other approaches, such as time-dependent density matrix renormalization group,~\cite{wh.fe.04} or quantum Monte Carlo,~\cite{ec.ko.09} are certainly more accurate at short times, the present approach could be used to estimate directly slowly-decaying modes by inspecting the behavior of the low-lying spectrum of
the Lindblad operator.

\begin{acknowledgments}
We acknowledge
discussions with A. Rosch, W. Hofstetter, S. Diehl, M. Knap, D. Rost and F. Schwarz.
AD and MN thank the
Forschungszentrum J{\"u}lich, in particular the autumn school on
correlated electrons, for hospitality and support. 
This work was supported by the Austrian Science Fund (FWF) P24081-N16, 
P26508-N20, as well as SfB-ViCoM project F04103. 
\end{acknowledgments}

\appendix

\section{Numerical calculation of the auxiliary interacting Green's function}\label{calcself}
In this section we present details of the numerical evaluation of the auxiliary Green's function, as described in \se~\ref{ssec:sigma}. We focus on large Hilbert spaces for which a sparse-matrix approach is mandatory.

To determine the steady state, which is the right-sided eigenstate of $\LLO$ with eigenvalue zero, one can make use of a shift-and-invert Arnoldi procedure.~\cite{saad.11,arbe.12u,ba.de.87,kn.ar.11.ec} The spectrum of $\LLO$ \eq{eq:LLO} has the property that $\iim(\LL_n)<0$ for all eigenvalues $\LL_n$ (except the steady state $\LL_0=0$). Therefore, given a small shift $s>0$, the eigenvector of $(\LLO - is \hat{\uu})^{-1}$ with the largest eigenvalue is the steady state. Since $\LLO$ is non-hermitian, the three term recurrence of the ordinary Lanczos scheme~\cite{lanc.51} does not apply, and one  
has to resort to an Arnoldi scheme instead. To construct the corresponding Krylov space, a system of equations
$(\LLO - is \hat{\uu})\ket{\tilde\phi_{n+1}} = \ket{\phi_n}$ has to be solved in each step. For the problem at hand we found that this can be done most efficiently by combining a stabilized biconjugate gradient method with an incomplete LU-decomposition as preconditioner.~\cite{saad.03,ba.be.94} Despite using sparse matrix methods the memory requirements of this approach are rather 
high compared to the schemes presented below.

A second possible route to determine the steady state $\ket{\rho_{\infty}}$ is to perform an explicit time evolution. For unitary 
time evolutions a well-established method relies on the Lanczos scheme to construct an approximate time evolution 
operator.~\cite{pa.ju.86} Such an approach can be adapted to the non-unitary case by using a two-sided Lanczos scheme (see below),
or also by employing an Arnoldi procedure.~\cite{kn.ar.11.ec}
Since $\LLO$ is non-hermitian, one can equally well use a simpler backward or forward Euler scheme~\cite{pr.te.07} to discretize the non-unitary 
time evolution operator. These approaches may not yield a highly accurate time evolution of $\ket{\rho(t)}$, but can nevertheless
determine the steady state within a moderate number of steps.
As for the shift-and-invert approach above, to solve the implicit update of $\ket{\rho(t_{n+1})}$ at time step $t_n$ in the case 
of the backward Euler, a biconjugate gradient routine has proven to be effective. 
For the forward time-integration a Runge Kutta method of second order is sufficient, with the great
advantage that only matrix-vector multiplications are needed, which
reduces memory requirements. In practice, for the considered cases   
it was found that for not too large systems ($N_B<6$) the shift-and-invert Arnoldi procedure is best suited, whereas a 
forward time-integration is advantageous for the case $N_B=6$.

Once the steady state is determined, Green's functions can be effectively calculated by employing a two-sided Lanczos scheme.~\cite{saad.11,ba.de.97,gutk.90,gutk.99,fe.gu.93,be.pa.85}
We therefore express the right- and left-sided eigenvectors of $\LLO$ in \eq{eq:mbGFgreater} in a Krylov space basis
\begin{align*}
 \ket{R_n} &= \sum_k U_{kn} \ket{\phi_R^k} \,\mbox{,} \quad\bra{L_n} = \sum_k \left(U^{-1}\right)_{nk} \bra{\phi_L^k} \,\mbox{.}
\end{align*} 
Here, we have 
omitted the $N_\sigma-\tilde N_\sigma$ symmetry sector index for the sake of clarity.
The biorthogonal Lanczos vectors
\begin{align*}
 \braket{\phi_L^k | \phi_R^{k'}} &= \delta_{kk'}\,\mbox{,}
\end{align*}
are determined by the three term recurrence
\begin{align*}
\ket{\phi_R^{n+1}} &= \frac{1}{c_{n+1}} \left( \LLO \ket{\phi_R^{n}} - e_n \ket{\phi_R^{n}} - k_n \ket{\phi_R^{n-1}} \right) \,\mbox{,} \\
\bra{\phi_L^{n+1}} &= \frac{1}{c_{n+1}^*} \left( \bra{\phi_L^{n}} \LLO - e_n \bra{\phi_L^{n}} - k_n^* \bra{\phi_L^{n-1}} \right) \,\mbox{,}
\end{align*}
with
\begin{align*}
 e_n &= \bra{\phi_L^{n}} \LLO \ket{\phi_R^{n}} \,\mbox{,} \\
 k_n &= \bra{\phi_L^{n-1}} \LLO \ket{\phi_R^{n}} = \left( \bra{\phi_L^{n}} \LLO \ket{\phi_R^{n-1}} \right)^*  \,\mbox{,}
\end{align*}
and a normalization constant $c_n$ such that $\braket{\phi_L^{n} | \phi_R^{n}} = 1$. One has a certain degree of freedom in the
choice of $c_n$ and $k_n$ due to the relation $k_n^* = c_n$, which is fulfilled for example by $k_n = k_n^* = c_n$.

In the Krylov basis, $\LLO$ takes on a tridiagonal form $T_{nm} = \bra{\phi_L^{n}} \LLO \ket{\phi_R^{m}}$ with the matrix elements
$T_{nn} = e_n$, $T_{n-1n} = k_n$ and $T_{nn-1} = k_n^*$. When $n+1$ becomes as large as the degree of the minimal
polynomial of $\LLO$, the eigenvalues and eigenvectors of $T$ represent those of $\LLO$.~\cite{gutk.90,saad.11} If one truncates the
Krylov basis, this statement holds still approximately true, especially for the largest eigenvalues in magnitude. Analogous to the
hermitian case,~\cite{we.me.96} an exponential convergence of the eigenspectrum of $T$ towards the one of $\LLO$ is observed, 
which is of particular importance for the calculation of Green's functions. A peculiarity of the two-sided Lanczos scheme is,
that not  every Krylov subspace  guarantees that $\iim(\LL_n)<0$ for all eigenvalues $\LL_n$ of $T$. In order to obtain
the appropriate pole structure for the estimated Green's functions, one has to check $\iim(\LL_n)<0$ together with 
convergence criteria. In cases in which $\iim(\LL_n)<0$ cannot be fulfilled exactly, it has to be ensured at least that 
the corresponding weights of these eigenvalues are negligible.
 
For the calculation of the Green's functions needed here it is convenient to choose appropriate initial vectors, which are in the case of the
greater Green's function~\eqref{eq:mbGFgreater}
\begin{align*}
 \ket{\phi_R^{0}} &= \frac{1}{c_{0}} \left( d^\dag_f\ket{\rho_{\infty}} \right) \,\mbox{,}\quad \bra{\phi_L^{0}} = \frac{1}{c_{0}^*} \left(\bra{I}d_f^\nag  \right) \,\mbox{.}
\end{align*}
When denoting by $\LL_n$ and $U_{k,n}$ the eigenvalues and right-sided eigenvectors of $T$, respectively, \eq{eq:mbGFgreater} can
be cast into the form
\begin{align*}
 G^>(\omega)&= \sum_{n,k,k'} \frac{U_{kn}{U}^{-1}_{nk'}}{\omega-\LL_n}  \bra{I}d_f^\nag \ket{\phi_R^{k}} \bra{\phi_L^{k'}} d^\dag_f\ket{\rho_{\infty}} \nonumber\\
 & - \sum_{n,k,k'} \frac{\left(U_{kn}{U}^{-1}_{nk'} \right)^*}{\omega-\LL_n^*}   \left(\bra{I}d_f^\nag \ket{\phi_R^{k}} \bra{\phi_L^{k'}} d^\dag_f\ket{\rho_{\infty}}\right)^* \nonumber\\
 & = \left|c_0\right|^2 \sum_{n} \frac{U_{0n}{U}^{-1}_{n0}}{\omega-\LL_n}  - \left|c_0\right|^2 \sum_{n}\frac{\left(U_{0n}{U}^{-1}_{n0} \right)^*}{\omega-\LL_n^*}  \,\mbox{.} 
\end{align*}

\section{Averaging scheme for multiple local minima}\label{app:Averaging}
This section contains details on the approach we used to determine the artificial ''temperature`` for the Boltzmann weights as described in \Se \ref{ssec:resultsFit}. We consider the situation that a set of local minima for which \Eq{eq:SC} becomes stationary is known. Let us specify by $\mathbf{a}_y(\phi)$ the vector of parameters $\{E_{\mu\nu },\Gamma_{\mu\nu}^{(\kappa)}\}_y$ corresponding to one certain local minimum for a set of model parameters, labeled by $y$. In order to quantify the spectral weight distribution of the corresponding hybridization function $\underline{\Delta}_{\text{aux}}(\omega;\mathbf{a}_y(\phi))$, we define
\begin{align*}
 m_2^R(\mathbf{a}_y(\phi)) & = \int\limits_{-\omega_c}^{\omega_c} \iim( \Delta^R_{\text{aux}}(\omega;\mathbf{a}_y(\phi)) )\,\omega^2 d\omega \,\mbox{,}\\
 m_3^K(\mathbf{a}_y(\phi)) & = \int\limits_{-\omega_c}^{\omega_c} \iim( \Delta^K_{\text{aux}}(\omega;\mathbf{a}_y(\phi)) )\,\omega^3 d\omega \,\mbox{,}
\end{align*}
which are similar
 to the second and third moment of
$\Delta^R_\aux$ and $\Delta^K_\aux$, respectively. For the Keldysh component a definition analogous to the first moment would
yield the desired information as well but the choice above has been found to be more sensitive to details of $\Delta^K_\aux$.
The value of the corresponding cost function $\chi(\mathbf{a}_y(\phi))$ 
of the $y$-th minimum  is used as an artificial ''energy`` and
enables one to define weights when making use of Boltzmann's statistic
\begin{align*}
 P_y(\phi,\beta) &= \frac{1}{Z} e^{-\beta \chi(\mathbf{a}_y(\phi))}\,\mbox{,}
\end{align*}
where we introduced an artificial ''temperature`` $\beta^{-1}$. For each bias voltage separately, we are then able to calculate averaged quantities
\begin{align*}
 \overline{m_2^R}(\phi,\beta) & = \sum\limits_{y} P_y(\phi,\beta) m_2^R(\mathbf{a}_y(\phi)) \,\mbox{,}
\end{align*}
as well as $\overline{m_3^K}(\phi,\beta)$ and $\overline{\chi}(\phi,\beta)$ in an analogous manner. The quantities 
$\overline{m_2^R}(\phi,\beta)$ and $\overline{m_3^K}(\phi,\beta)$ 
provide an estimate of the center of the spectral weight for the averaged set 
of hybridization functions for each bias voltage $\phi$.

Our goal is that these quantities vary in a smooth way when changing the bias voltage. To achieve this, we employ a minimum curvature scheme,~\cite{pr.te.07} meaning that we optimize the function
\begin{align*}
 v_c(\beta) &= \int\limits_0^{\phi_{max}} \left\{  w^R\left|\frac{\partial^2}{\partial \phi^2} \overline{m_2^R}(\phi,\beta) \right|^2  
              + w^K\left|\frac{\partial^2}{\partial \phi^2} \overline{m_3^K}(\phi,\beta) \right|^2  \right. \nonumber \\
           &  \left. + w^\chi\left|\frac{\partial^2}{\partial \phi^2} \overline{\chi}(\phi,\beta) \right|^2  
 \right\} d\phi\,\mbox{,}
\end{align*}
with respect to $\beta$.
 This determines the optimal artificial ''temperature``, which ensures that the averaged cost function 
as well as the averaged spectral weight are as smooth functions of $\phi$ as possible, 
given the set of calculated minima $\{\mathbf{a}_y(\phi)\}$.
As in many optimization problems, an arbitrariness exists in the definition of the quantities $\overline{m_2^R}(\phi,\beta)$ and $\overline{m_3^K}(\phi,\beta)$, as well as in choosing the values of the weights $w^R$, $w^K$ and $w^\chi$. In our case all of the weights were chosen to be equal to one in units of $t$.

An improvement of the results, to a certain degree at least, could be expected when making use of extensions like a bias-dependent $\beta(\phi)$. This has not been considered in the present work since already a single variable $\beta$ provided quite smooth observables. As mentioned in the main text, in any case, it is obligatory to examine besides the averaged results also the ones for the absolute minima and/or for different averaging schemes, in order to avoid that physical discontinuities are averaged out.
We stress that this approach has to be taken with due care, since it
is in some aspects arbitrary. However, it is useful to give an
estimate of the error of the calculation, and can certainly identify
regions in parameter space where the error is negligible small.

\section{Linear corrections}\label{ssec:error}
In this section, we present a scheme to correct physical quantities up
to linear order in the difference~\cite{footnote11}
\begin{align*} 
\und\diffd(\omega)&=\und\Delta_{ex}(\omega)-\und\Delta_\aux(\omega)\,\mbox{,}
\end{align*} 
between the auxiliary and the exact hybridization
functions.
Although $\und \diffd(\omega)$
 decreases rapidly with increasing number
of auxiliary bath sites $N_B$, the size of the Hilbert space  also
increases exponentially with $N_B$. This poses a clear limit to the
maximum value of $N_B$. 

The idea is based on the fact that each physical quantity
$O[\und\Delta]$ is a functional of $\und\Delta(\omega)$. Its  exact
value is, thus, obtained as $O[\und\Delta_{ex}]$. For a finite $N_B$
there will always be a nonzero value of $\und\diffd(\omega)$ at some
energies, so we will always obtain an approximate value
$O[\und\Delta_\aux]$. A linear correction can be obtained by
evaluating numerically the functional derivative of $O[\und\Delta]$. 
Strictly speaking,  considering that only $\iim(\Delta^R(\omega))$ and $\iim(\Delta^K(\omega))$ are independent functions,  $O$ is a functional $O[\iim(\Delta^R),\iim(\Delta^K)]$. Suppose one knows the functional derivatives
\begin{align*}
\frac{\delta O[\und\Delta]}{\delta \iim(\Delta^{\alpha}(\omega))}\,\mbox{,}
\quad\quad&\alpha\in\{R,K\}\,\mbox{,}
\end{align*}
then to linear order in $\und \diffd(\omega)$
\begin{align}
\label{lincorr}
O[\und \Delta_{ex}] & \approx
O[\und \Delta_\aux] 
\\ \nonumber &\hspace{-3em}
+ r \sum_{\alpha\in\{R,K\}}\int 
\left.\frac{\delta O[\und\Delta]}{\delta \iim(\Delta^{\alpha}(\omega_0))} \right|_{\und\Delta = \und\Delta_\aux} \hspace{-2.8em}
 \iim(\diffd^\alpha(\omega_0)) \ d \omega_0 + \mathcal{O}(\und\diffd^2)\,\mbox{,}
\end{align}
with $r=1$.

We evaluate the functional derivative  numerically in the following way. One first evaluates $O[\und \Delta_\aux]$ at the optimum $\und \Delta_\aux(\omega)$. Then $O$ is evaluated at a ``shifted'' $\iim(\Delta^\alpha(\omega))$, obtained by adding a delta function peaked around a certain energy $\omega_0$
\begin{align*}
\delta_{\omega_0}(\omega) &\equiv \delta(\omega-\omega_0)\,\mbox{,}
\end{align*}
multiplied by  a small coefficient $\epsilon$. The functional derivatives are then approximated linearly, by making use of the equations
\begin{align}
\label{fder} \hspace{-0.3em}\frac{\delta O[\und\Delta]}{\delta \iim(\Delta^{R}(\omega_0))} \pm 2 \frac{\delta O[\und\Delta]}{\delta \iim(\Delta^{K}(\omega_0))}\\
\nonumber & \hspace{-10em} \approx \frac{1}{\epsilon}\left( O[\iim(\Delta^R),\iim(\Delta^K)]   \right. \nonumber \\
\nonumber & \hspace{-10em} - \left. O[\iim(\Delta^R)-\epsilon \delta_{\omega_0}, \iim(\Delta^K) \mp 2 \epsilon \delta_{\omega_0}] \right)\,\mbox{,}
\end{align}
which become exact in the $\epsilon\to 0$ limit.

A (quasi) delta-peak correction $\epsilon \delta_{\omega_0}$ to $\Delta^\alpha(\omega)$ can be obtained by attaching an additional bath site ($N_B+1$) with on-site energy $E_{N_B+1,N_B+1}=\omega_0$ directly to the impurity site with a hopping $E_{N_B+1,f}=\sqrt{\epsilon/\pi}$. The sum of  $\Gamma_{N_B+1,N_B+1}^{(1)}$ and $\Gamma_{N_B+1,N_B+1}^{(2)}$ is proportional to the width of $\delta_{\omega_0}$ and thus, should be taken as small as possible. In practice, one uses a discretization of the integration over $\omega_0$ in \eq{lincorr} and the width of the delta peaks has to be adjusted accordingly. Setting one of the components $\Gamma_{N_B+1,N_B+1}^{\kappa}$ to zero yields a peak in the Keldysh component with a coefficient $\pm 2\epsilon$, respectively, as used in \eq{fder}.

Notice that the functional derivative \eq{fder} amounts to carrying out two many-body calculations for each point $\omega_0$ on a system with $N_B+1$ bath sites. 
However, it is not necessary to repeat the calculation for each physical quantity of interest. In the linearly corrected current values presented 
in \se~\ref{ssec:resultsjV}, a $\omega_0$ mesh of 200 points was used, whereby this number is likely to be reduced when optimizing the method.

Strictly speaking the coefficient $r$ in \eq{lincorr} should be $1$. However, for cases in which the linear correction is not small, this could produce an ``over-correction''. In order to avoid this, we introduce a smaller ratio $r$ which is determined as follows: We evaluate the corrected self energy at each $\omega$ via \eq{lincorr} and $O=\underline{\Sigma}(\omega)$ with some value of $r<1$ and denote it $\underline{\Sigma}_r(\omega)$. We do the same for the Green's function of the auxiliary system and denote it $\underline{G}_{r}(\omega)$. Using Equations (\ref{dys}) and (\ref{g0}) we now have an estimate  of an effective $r$-dependent auxiliary hybridization function of the linearly corrected system via
\begin{align*}
\und{\Delta}_{\aux,r}(\omega)
\equiv \underline{g}^{-1}_{0}(\omega) - \underline{G}_{r}^{-1}(\omega) -\underline{\Sigma}_r(\omega)\,\mbox{.}
\end{align*}
In principle, for $r=1$ this gives $\und\Delta_{ex}(\omega)$ up to $\mathcal{O}(\und\diffd^2)$. In practice, for finite $\und\diffd(\omega)$, one can introduce a cost function $\chi(r)$ analogous to \eq{eq:SC} to minimize the difference $|\underline{\Delta}_{\aux,r}(\omega) - \underline{\Delta}_{ex}(\omega)|$ as a function of $r$. We checked that for the case in which the linear correction is a good approximation, the minimum occurs at $r=1$. If the minimum of $\chi(r)$ is situated at some value $r_{\text{min}}<1$ then  one corrects also other physical quantities according to \eq{lincorr} with the same $r = r_{\text{min}}$.

Alternatively to the correction \eq{lincorr} discussed above, one can use the numerical functional derivative evaluated via \eq{fder} in order to estimate the sensitivity
of the value of $O$ with respect to variations of $\iim(\Delta^\alpha_\aux(\omega))$ as a function of $\omega$ and $\alpha$. This is of use, in a second step, to adjust the weight function $W^\alpha(\omega)$ in \eq{eq:SC}, so that more sensitive $\omega$ regions acquire a larger weight.

\bibliography{openSiam.bbl}{}

\begin{thebibliography}{166}
\expandafter\ifx\csname natexlab\endcsname\relax\def\natexlab#1{#1}\fi
\expandafter\ifx\csname bibnamefont\endcsname\relax
  \def\bibnamefont#1{#1}\fi
\expandafter\ifx\csname bibfnamefont\endcsname\relax
  \def\bibfnamefont#1{#1}\fi
\expandafter\ifx\csname citenamefont\endcsname\relax
  \def\citenamefont#1{#1}\fi
\expandafter\ifx\csname url\endcsname\relax
  \def\url#1{\texttt{#1}}\fi
\expandafter\ifx\csname urlprefix\endcsname\relax\def\urlprefix{URL }\fi
\providecommand{\bibinfo}[2]{#2}
\providecommand{\eprint}[2][]{\url{#2}}

\bibitem[{\citenamefont{Hartmann et~al.}(2008)\citenamefont{Hartmann,
  Brand{\~a}o, and Plenio}}]{ha.br.08}
\bibinfo{author}{\bibfnamefont{M.}~\bibnamefont{Hartmann}},
  \bibinfo{author}{\bibfnamefont{F.}~\bibnamefont{Brand{\~a}o}},
  \bibnamefont{and} \bibinfo{author}{\bibfnamefont{M.}~\bibnamefont{Plenio}},
  \bibinfo{journal}{Laser {\&} Photonics Rev.} \textbf{\bibinfo{volume}{2}},
  \bibinfo{pages}{527} (\bibinfo{year}{2008}).

\bibitem[{\citenamefont{Raizen et~al.}(1997)\citenamefont{Raizen, Salomon, and
  Niu}}]{ra.sa.97}
\bibinfo{author}{\bibfnamefont{M.}~\bibnamefont{Raizen}},
  \bibinfo{author}{\bibfnamefont{C.}~\bibnamefont{Salomon}}, \bibnamefont{and}
  \bibinfo{author}{\bibfnamefont{Q.}~\bibnamefont{Niu}},
  \bibinfo{journal}{Phys. Today} \textbf{\bibinfo{volume}{50}},
  \bibinfo{pages}{30} (\bibinfo{year}{1997}).

\bibitem[{\citenamefont{Jaksch et~al.}(1998)\citenamefont{Jaksch, Bruder,
  Cirac, Gardiner, and Zoller}}]{ja.br.98}
\bibinfo{author}{\bibfnamefont{D.}~\bibnamefont{Jaksch}},
  \bibinfo{author}{\bibfnamefont{C.}~\bibnamefont{Bruder}},
  \bibinfo{author}{\bibfnamefont{J.~I.} \bibnamefont{Cirac}},
  \bibinfo{author}{\bibfnamefont{C.~W.} \bibnamefont{Gardiner}},
  \bibnamefont{and} \bibinfo{author}{\bibfnamefont{P.}~\bibnamefont{Zoller}},
  \bibinfo{journal}{Phys. Rev. Lett.} \textbf{\bibinfo{volume}{81}},
  \bibinfo{pages}{3108} (\bibinfo{year}{1998}).

\bibitem[{\citenamefont{Greiner et~al.}(2002)\citenamefont{Greiner, Mandel,
  Esslinger, H{\"a}nsch, and Bloch}}]{gr.ma.02}
\bibinfo{author}{\bibfnamefont{M.}~\bibnamefont{Greiner}},
  \bibinfo{author}{\bibfnamefont{O.}~\bibnamefont{Mandel}},
  \bibinfo{author}{\bibfnamefont{T.}~\bibnamefont{Esslinger}},
  \bibinfo{author}{\bibfnamefont{T.~W.} \bibnamefont{H{\"a}nsch}},
  \bibnamefont{and} \bibinfo{author}{\bibfnamefont{I.}~\bibnamefont{Bloch}},
  \bibinfo{journal}{Nature} \textbf{\bibinfo{volume}{415}}, \bibinfo{pages}{39}
  (\bibinfo{year}{2002}).

\bibitem[{\citenamefont{Trotzky et~al.}(2008)\citenamefont{Trotzky, Cheinet,
  F{\"o}lling, Feld, Schnorrberger, Rey, Polkovnikov, Demler, Lukin, and
  Bloch}}]{tr.ch.08}
\bibinfo{author}{\bibfnamefont{S.}~\bibnamefont{Trotzky}},
  \bibinfo{author}{\bibfnamefont{P.}~\bibnamefont{Cheinet}},
  \bibinfo{author}{\bibfnamefont{S.}~\bibnamefont{F{\"o}lling}},
  \bibinfo{author}{\bibfnamefont{M.}~\bibnamefont{Feld}},
  \bibinfo{author}{\bibfnamefont{U.}~\bibnamefont{Schnorrberger}},
  \bibinfo{author}{\bibfnamefont{A.~M.} \bibnamefont{Rey}},
  \bibinfo{author}{\bibfnamefont{A.}~\bibnamefont{Polkovnikov}},
  \bibinfo{author}{\bibfnamefont{E.~A.} \bibnamefont{Demler}},
  \bibinfo{author}{\bibfnamefont{M.~D.} \bibnamefont{Lukin}}, \bibnamefont{and}
  \bibinfo{author}{\bibfnamefont{I.}~\bibnamefont{Bloch}},
  \bibinfo{journal}{Science} \textbf{\bibinfo{volume}{319}},
  \bibinfo{pages}{295} (\bibinfo{year}{2008}).

\bibitem[{\citenamefont{Schneider et~al.}(2012)\citenamefont{Schneider,
  Hackermuller, Ronzheimer, Will, Braun, Best, Bloch, Demler, Mandt, Rasch
  et~al.}}]{sc.ha.12}
\bibinfo{author}{\bibfnamefont{U.}~\bibnamefont{Schneider}},
  \bibinfo{author}{\bibfnamefont{L.}~\bibnamefont{Hackermuller}},
  \bibinfo{author}{\bibfnamefont{J.~P.} \bibnamefont{Ronzheimer}},
  \bibinfo{author}{\bibfnamefont{S.}~\bibnamefont{Will}},
  \bibinfo{author}{\bibfnamefont{S.}~\bibnamefont{Braun}},
  \bibinfo{author}{\bibfnamefont{T.}~\bibnamefont{Best}},
  \bibinfo{author}{\bibfnamefont{I.}~\bibnamefont{Bloch}},
  \bibinfo{author}{\bibfnamefont{E.}~\bibnamefont{Demler}},
  \bibinfo{author}{\bibfnamefont{S.}~\bibnamefont{Mandt}},
  \bibinfo{author}{\bibfnamefont{D.}~\bibnamefont{Rasch}},
  \bibnamefont{et~al.}, \bibinfo{journal}{Nat Phys}
  \textbf{\bibinfo{volume}{8}}, \bibinfo{pages}{213} (\bibinfo{year}{2012}).

\bibitem[{\citenamefont{Iwai et~al.}(2003)\citenamefont{Iwai, Ono, Maeda,
  Matsuzaki, Kishida, Okamoto, and Tokura}}]{iw.on.03}
\bibinfo{author}{\bibfnamefont{S.}~\bibnamefont{Iwai}},
  \bibinfo{author}{\bibfnamefont{M.}~\bibnamefont{Ono}},
  \bibinfo{author}{\bibfnamefont{A.}~\bibnamefont{Maeda}},
  \bibinfo{author}{\bibfnamefont{H.}~\bibnamefont{Matsuzaki}},
  \bibinfo{author}{\bibfnamefont{H.}~\bibnamefont{Kishida}},
  \bibinfo{author}{\bibfnamefont{H.}~\bibnamefont{Okamoto}}, \bibnamefont{and}
  \bibinfo{author}{\bibfnamefont{Y.}~\bibnamefont{Tokura}},
  \bibinfo{journal}{Phys. Rev. Lett.} \textbf{\bibinfo{volume}{91}},
  \bibinfo{pages}{057401} (\bibinfo{year}{2003}).

\bibitem[{\citenamefont{Cavalleri et~al.}(2004)\citenamefont{Cavalleri,
  Dekorsy, Chong, Kieffer, and Schoenlein}}]{ca.de.04}
\bibinfo{author}{\bibfnamefont{A.}~\bibnamefont{Cavalleri}},
  \bibinfo{author}{\bibfnamefont{T.}~\bibnamefont{Dekorsy}},
  \bibinfo{author}{\bibfnamefont{H.~H.~W.} \bibnamefont{Chong}},
  \bibinfo{author}{\bibfnamefont{J.~C.} \bibnamefont{Kieffer}},
  \bibnamefont{and} \bibinfo{author}{\bibfnamefont{R.~W.}
  \bibnamefont{Schoenlein}}, \bibinfo{journal}{Phys. Rev. B}
  \textbf{\bibinfo{volume}{70}}, \bibinfo{pages}{161102}
  (\bibinfo{year}{2004}).

\bibitem[{\citenamefont{Bonilla and Grahn}(2005)}]{bo.gr.05}
\bibinfo{author}{\bibfnamefont{L.~L.} \bibnamefont{Bonilla}} \bibnamefont{and}
  \bibinfo{author}{\bibfnamefont{H.~T.} \bibnamefont{Grahn}},
  \bibinfo{journal}{Rep. Prog. Phys.} \textbf{\bibinfo{volume}{68}},
  \bibinfo{pages}{577} (\bibinfo{year}{2005}).

\bibitem[{\citenamefont{Zutic et~al.}(2004)\citenamefont{Zutic, Fabian, and
  Sarma}}]{zu.fa.04}
\bibinfo{author}{\bibfnamefont{I.}~\bibnamefont{Zutic}},
  \bibinfo{author}{\bibfnamefont{J.}~\bibnamefont{Fabian}}, \bibnamefont{and}
  \bibinfo{author}{\bibfnamefont{S.~D.} \bibnamefont{Sarma}},
  \bibinfo{journal}{Rev. Mod. Phys.} \textbf{\bibinfo{volume}{76}},
  \bibinfo{pages}{323} (\bibinfo{year}{2004}).

\bibitem[{\citenamefont{Cuniberti et~al.}(2005)\citenamefont{Cuniberti, Fagas,
  and Richter}}]{cu.fa.05}
\bibinfo{author}{\bibfnamefont{G.}~\bibnamefont{Cuniberti}},
  \bibinfo{author}{\bibfnamefont{G.}~\bibnamefont{Fagas}}, \bibnamefont{and}
  \bibinfo{author}{\bibfnamefont{K.}~\bibnamefont{Richter}},
  \emph{\bibinfo{title}{Introducing Molecular Electronics}}
  (\bibinfo{publisher}{Springer}, \bibinfo{year}{2005}), ISBN
  \bibinfo{isbn}{3540279946}.

\bibitem[{\citenamefont{Smit et~al.}(2002)\citenamefont{Smit, Noat, Untiedt,
  Lang, van Hemert, and van Ruitenbeek}}]{sm.no.02}
\bibinfo{author}{\bibfnamefont{R.~H.~M.} \bibnamefont{Smit}},
  \bibinfo{author}{\bibfnamefont{Y.}~\bibnamefont{Noat}},
  \bibinfo{author}{\bibfnamefont{C.}~\bibnamefont{Untiedt}},
  \bibinfo{author}{\bibfnamefont{N.~D.} \bibnamefont{Lang}},
  \bibinfo{author}{\bibfnamefont{M.~C.} \bibnamefont{van Hemert}},
  \bibnamefont{and} \bibinfo{author}{\bibfnamefont{J.~M.} \bibnamefont{van
  Ruitenbeek}}, \bibinfo{journal}{Nature} \textbf{\bibinfo{volume}{419}},
  \bibinfo{pages}{906} (\bibinfo{year}{2002}).

\bibitem[{\citenamefont{Park et~al.}(2002)\citenamefont{Park, Pasupathy,
  Goldsmith, Chang, Yaish, Petta, Rinkoski, Sethna, Abruna, {McEuen}
  et~al.}}]{pa.ab.02}
\bibinfo{author}{\bibfnamefont{J.}~\bibnamefont{Park}},
  \bibinfo{author}{\bibfnamefont{A.~N.} \bibnamefont{Pasupathy}},
  \bibinfo{author}{\bibfnamefont{J.~I.} \bibnamefont{Goldsmith}},
  \bibinfo{author}{\bibfnamefont{C.}~\bibnamefont{Chang}},
  \bibinfo{author}{\bibfnamefont{Y.}~\bibnamefont{Yaish}},
  \bibinfo{author}{\bibfnamefont{J.~R.} \bibnamefont{Petta}},
  \bibinfo{author}{\bibfnamefont{M.}~\bibnamefont{Rinkoski}},
  \bibinfo{author}{\bibfnamefont{J.~P.} \bibnamefont{Sethna}},
  \bibinfo{author}{\bibfnamefont{H.~D.} \bibnamefont{Abruna}},
  \bibinfo{author}{\bibfnamefont{P.~L.} \bibnamefont{{McEuen}}},
  \bibnamefont{et~al.}, \bibinfo{journal}{Nature}
  \textbf{\bibinfo{volume}{417}}, \bibinfo{pages}{722} (\bibinfo{year}{2002}).

\bibitem[{\citenamefont{Liang et~al.}(2002)\citenamefont{Liang, Shores,
  Bockrath, Long, and Park}}]{li.sh.02}
\bibinfo{author}{\bibfnamefont{W.}~\bibnamefont{Liang}},
  \bibinfo{author}{\bibfnamefont{M.~P.} \bibnamefont{Shores}},
  \bibinfo{author}{\bibfnamefont{M.}~\bibnamefont{Bockrath}},
  \bibinfo{author}{\bibfnamefont{J.~R.} \bibnamefont{Long}}, \bibnamefont{and}
  \bibinfo{author}{\bibfnamefont{H.}~\bibnamefont{Park}},
  \bibinfo{journal}{Nature} \textbf{\bibinfo{volume}{417}},
  \bibinfo{pages}{725} (\bibinfo{year}{2002}).

\bibitem[{\citenamefont{Agrait et~al.}(2003)\citenamefont{Agrait, Yeyati, and
  van Ruitenbeek}}]{ag.ye.03}
\bibinfo{author}{\bibfnamefont{N.}~\bibnamefont{Agrait}},
  \bibinfo{author}{\bibfnamefont{A.~L.} \bibnamefont{Yeyati}},
  \bibnamefont{and} \bibinfo{author}{\bibfnamefont{J.~M.} \bibnamefont{van
  Ruitenbeek}}, \bibinfo{journal}{Physics Reports}
  \textbf{\bibinfo{volume}{377}}, \bibinfo{pages}{81 } (\bibinfo{year}{2003}).

\bibitem[{\citenamefont{Venkataraman et~al.}(2006)\citenamefont{Venkataraman,
  Klare, Nuckolls, Hybertsen, and Steigerwald}}]{ve.la.06}
\bibinfo{author}{\bibfnamefont{L.}~\bibnamefont{Venkataraman}},
  \bibinfo{author}{\bibfnamefont{J.~E.} \bibnamefont{Klare}},
  \bibinfo{author}{\bibfnamefont{C.}~\bibnamefont{Nuckolls}},
  \bibinfo{author}{\bibfnamefont{M.~S.} \bibnamefont{Hybertsen}},
  \bibnamefont{and} \bibinfo{author}{\bibfnamefont{M.~L.}
  \bibnamefont{Steigerwald}}, \bibinfo{journal}{Nature}
  \textbf{\bibinfo{volume}{442}}, \bibinfo{pages}{904} (\bibinfo{year}{2006}).

\bibitem[{\citenamefont{Goldhaber-Gordon
  et~al.}(1998)\citenamefont{Goldhaber-Gordon, G{\"o}res, Kastner, Shtrikman,
  Mahalu, and Meirav}}]{go.go.98}
\bibinfo{author}{\bibfnamefont{D.}~\bibnamefont{Goldhaber-Gordon}},
  \bibinfo{author}{\bibfnamefont{J.}~\bibnamefont{G{\"o}res}},
  \bibinfo{author}{\bibfnamefont{M.~A.} \bibnamefont{Kastner}},
  \bibinfo{author}{\bibfnamefont{H.}~\bibnamefont{Shtrikman}},
  \bibinfo{author}{\bibfnamefont{D.}~\bibnamefont{Mahalu}}, \bibnamefont{and}
  \bibinfo{author}{\bibfnamefont{U.}~\bibnamefont{Meirav}},
  \bibinfo{journal}{Phys. Rev. Lett.} \textbf{\bibinfo{volume}{81}},
  \bibinfo{pages}{5225} (\bibinfo{year}{1998}).

\bibitem[{\citenamefont{Kretinin et~al.}(2012)\citenamefont{Kretinin,
  Shtrikman, and Mahalu}}]{kr.sh.12}
\bibinfo{author}{\bibfnamefont{A.~V.} \bibnamefont{Kretinin}},
  \bibinfo{author}{\bibfnamefont{H.}~\bibnamefont{Shtrikman}},
  \bibnamefont{and} \bibinfo{author}{\bibfnamefont{D.}~\bibnamefont{Mahalu}},
  \bibinfo{journal}{Phys. Rev. B} \textbf{\bibinfo{volume}{85}},
  \bibinfo{pages}{201301} (\bibinfo{year}{2012}).

\bibitem[{\citenamefont{Mitra et~al.}(2006)\citenamefont{Mitra, Takei, Kim, and
  Millis}}]{mi.ta.06}
\bibinfo{author}{\bibfnamefont{A.}~\bibnamefont{Mitra}},
  \bibinfo{author}{\bibfnamefont{S.}~\bibnamefont{Takei}},
  \bibinfo{author}{\bibfnamefont{Y.~B.} \bibnamefont{Kim}}, \bibnamefont{and}
  \bibinfo{author}{\bibfnamefont{A.~J.} \bibnamefont{Millis}},
  \bibinfo{journal}{Phys. Rev. Lett.} \textbf{\bibinfo{volume}{97}},
  \bibinfo{pages}{236808} (\bibinfo{year}{2006}).

\bibitem[{\citenamefont{Leggett et~al.}(1987)\citenamefont{Leggett,
  Chakravarty, Dorsey, Fisher, Garg, and Zwerger}}]{le.ch.87}
\bibinfo{author}{\bibfnamefont{A.~J.} \bibnamefont{Leggett}},
  \bibinfo{author}{\bibfnamefont{S.}~\bibnamefont{Chakravarty}},
  \bibinfo{author}{\bibfnamefont{A.~T.} \bibnamefont{Dorsey}},
  \bibinfo{author}{\bibfnamefont{M.~P.~A.} \bibnamefont{Fisher}},
  \bibinfo{author}{\bibfnamefont{A.}~\bibnamefont{Garg}}, \bibnamefont{and}
  \bibinfo{author}{\bibfnamefont{W.}~\bibnamefont{Zwerger}},
  \bibinfo{journal}{Rev. Mod. Phys.} \textbf{\bibinfo{volume}{59}},
  \bibinfo{pages}{1} (\bibinfo{year}{1987}).

\bibitem[{\citenamefont{Cazalilla}(2006)}]{caza.06}
\bibinfo{author}{\bibfnamefont{M.~A.} \bibnamefont{Cazalilla}},
  \bibinfo{journal}{Phys. Rev. Lett.} \textbf{\bibinfo{volume}{97}},
  \bibinfo{pages}{156403} (\bibinfo{year}{2006}).

\bibitem[{\citenamefont{Rigol et~al.}(2008)\citenamefont{Rigol, Dunjko, and
  Olshanii}}]{ri.du.08}
\bibinfo{author}{\bibfnamefont{M.}~\bibnamefont{Rigol}},
  \bibinfo{author}{\bibfnamefont{V.}~\bibnamefont{Dunjko}}, \bibnamefont{and}
  \bibinfo{author}{\bibfnamefont{M.}~\bibnamefont{Olshanii}},
  \bibinfo{journal}{Nature} \textbf{\bibinfo{volume}{452}},
  \bibinfo{pages}{854} (\bibinfo{year}{2008}).

\bibitem[{\citenamefont{Nitzan and Ratner}(2003)}]{ni.ra.03}
\bibinfo{author}{\bibfnamefont{A.}~\bibnamefont{Nitzan}} \bibnamefont{and}
  \bibinfo{author}{\bibfnamefont{M.~A.} \bibnamefont{Ratner}},
  \bibinfo{journal}{Science} \textbf{\bibinfo{volume}{300}},
  \bibinfo{pages}{1384} (\bibinfo{year}{2003}).

\bibitem[{\citenamefont{Anderson}(1961)}]{ande.61}
\bibinfo{author}{\bibfnamefont{P.~W.} \bibnamefont{Anderson}},
  \bibinfo{journal}{Phys. Rev.} \textbf{\bibinfo{volume}{124}},
  \bibinfo{pages}{41} (\bibinfo{year}{1961}).

\bibitem[{\citenamefont{Friedel}(1956)}]{frie.56}
\bibinfo{author}{\bibfnamefont{J.}~\bibnamefont{Friedel}},
  \bibinfo{journal}{Can. J. Phys.} \textbf{\bibinfo{volume}{34}},
  \bibinfo{pages}{1190} (\bibinfo{year}{1956}).

\bibitem[{\citenamefont{Clogston et~al.}(1962)\citenamefont{Clogston, Matthias,
  Peter, Williams, Corenzwit, and Sherwood}}]{cl.ma.62}
\bibinfo{author}{\bibfnamefont{A.~M.} \bibnamefont{Clogston}},
  \bibinfo{author}{\bibfnamefont{B.~T.} \bibnamefont{Matthias}},
  \bibinfo{author}{\bibfnamefont{M.}~\bibnamefont{Peter}},
  \bibinfo{author}{\bibfnamefont{H.~J.} \bibnamefont{Williams}},
  \bibinfo{author}{\bibfnamefont{E.}~\bibnamefont{Corenzwit}},
  \bibnamefont{and} \bibinfo{author}{\bibfnamefont{R.~C.}
  \bibnamefont{Sherwood}}, \bibinfo{journal}{Phys. Rev.}
  \textbf{\bibinfo{volume}{125}}, \bibinfo{pages}{541} (\bibinfo{year}{1962}).

\bibitem[{\citenamefont{Brenig and Sch{\"o}nhammer}(1974)}]{br.sc.74}
\bibinfo{author}{\bibfnamefont{W.}~\bibnamefont{Brenig}} \bibnamefont{and}
  \bibinfo{author}{\bibfnamefont{K.}~\bibnamefont{Sch{\"o}nhammer}},
  \bibinfo{journal}{Zeitschrift f{\"u}r Physik} \textbf{\bibinfo{volume}{267}},
  \bibinfo{pages}{201} (\bibinfo{year}{1974}).

\bibitem[{\citenamefont{Georges et~al.}(1996)\citenamefont{Georges, Kotliar,
  Krauth, and Rozenberg}}]{ge.ko.96}
\bibinfo{author}{\bibfnamefont{A.}~\bibnamefont{Georges}},
  \bibinfo{author}{\bibfnamefont{G.}~\bibnamefont{Kotliar}},
  \bibinfo{author}{\bibfnamefont{W.}~\bibnamefont{Krauth}}, \bibnamefont{and}
  \bibinfo{author}{\bibfnamefont{M.~J.} \bibnamefont{Rozenberg}},
  \bibinfo{journal}{Rev. Mod. Phys.} \textbf{\bibinfo{volume}{68}},
  \bibinfo{pages}{13} (\bibinfo{year}{1996}).

\bibitem[{\citenamefont{Cyril}(1997)}]{hews.97}
\bibinfo{author}{\bibfnamefont{A.}~\bibnamefont{Cyril}},
  \emph{\bibinfo{title}{The Kondo Problem to Heavy Fermions}}
  (\bibinfo{publisher}{Cambridge University Press}, \bibinfo{year}{1997}), ISBN
  \bibinfo{isbn}{0521599474}.

\bibitem[{\citenamefont{Vollhardt}(2010)}]{voll.10}
\bibinfo{author}{\bibfnamefont{D.}~\bibnamefont{Vollhardt}}, in
  \emph{\bibinfo{booktitle}{Lecture Notes on the Physics of Strongly Correlated
  Systems}}, edited by \bibinfo{editor}{\bibfnamefont{A.}~\bibnamefont{Avella}}
  \bibnamefont{and} \bibinfo{editor}{\bibfnamefont{F.}~\bibnamefont{Mancini}}
  (\bibinfo{address}{AIP, New York}, \bibinfo{year}{2010}), vol.
  \bibinfo{volume}{1297} of \emph{\bibinfo{series}{AIP Conf. Proc.}}, pp.
  \bibinfo{pages}{339--403}.

\bibitem[{\citenamefont{Metzner and Vollhardt}(1989)}]{me.vo.89}
\bibinfo{author}{\bibfnamefont{W.}~\bibnamefont{Metzner}} \bibnamefont{and}
  \bibinfo{author}{\bibfnamefont{D.}~\bibnamefont{Vollhardt}},
  \bibinfo{journal}{Phys. Rev. Lett.} \textbf{\bibinfo{volume}{62}},
  \bibinfo{pages}{324} (\bibinfo{year}{1989}).

\bibitem[{\citenamefont{Kondo}(1964)}]{kond.64}
\bibinfo{author}{\bibfnamefont{J.}~\bibnamefont{Kondo}},
  \bibinfo{journal}{Progress of Theoretical Physics}
  \textbf{\bibinfo{volume}{32}}, \bibinfo{pages}{37} (\bibinfo{year}{1964}).

\bibitem[{\citenamefont{Anderson}(1970)}]{ande.70}
\bibinfo{author}{\bibfnamefont{P.~W.} \bibnamefont{Anderson}},
  \bibinfo{journal}{Journal of Physics C: Solid State Physics}
  \textbf{\bibinfo{volume}{3}}, \bibinfo{pages}{2436} (\bibinfo{year}{1970}).

\bibitem[{\citenamefont{Yosida and Yamada}(1970)}]{yo.ya.70}
\bibinfo{author}{\bibfnamefont{K.}~\bibnamefont{Yosida}} \bibnamefont{and}
  \bibinfo{author}{\bibfnamefont{K.}~\bibnamefont{Yamada}},
  \bibinfo{journal}{Progress of Theoretical Physics Supplement}
  \textbf{\bibinfo{volume}{46}}, \bibinfo{pages}{244} (\bibinfo{year}{1970}).

\bibitem[{\citenamefont{Yamada}(1975{\natexlab{a}})}]{yama.75}
\bibinfo{author}{\bibfnamefont{K.}~\bibnamefont{Yamada}},
  \bibinfo{journal}{Progress of Theoretical Physics}
  \textbf{\bibinfo{volume}{53}}, \bibinfo{pages}{970}
  (\bibinfo{year}{1975}{\natexlab{a}}).

\bibitem[{\citenamefont{Yosida and Yamada}(1975)}]{yo.ya.75}
\bibinfo{author}{\bibfnamefont{K.}~\bibnamefont{Yosida}} \bibnamefont{and}
  \bibinfo{author}{\bibfnamefont{K.}~\bibnamefont{Yamada}},
  \bibinfo{journal}{Progress of Theoretical Physics}
  \textbf{\bibinfo{volume}{53}}, \bibinfo{pages}{1286} (\bibinfo{year}{1975}).

\bibitem[{\citenamefont{Yamada}(1975{\natexlab{b}})}]{yama.75b}
\bibinfo{author}{\bibfnamefont{K.}~\bibnamefont{Yamada}},
  \bibinfo{journal}{Progress of Theoretical Physics}
  \textbf{\bibinfo{volume}{54}}, \bibinfo{pages}{316}
  (\bibinfo{year}{1975}{\natexlab{b}}).

\bibitem[{\citenamefont{Schrieffer and Wolff}(1966)}]{sc.wo.66}
\bibinfo{author}{\bibfnamefont{J.~R.} \bibnamefont{Schrieffer}}
  \bibnamefont{and} \bibinfo{author}{\bibfnamefont{P.~A.} \bibnamefont{Wolff}},
  \bibinfo{journal}{Phys. Rev.} \textbf{\bibinfo{volume}{149}},
  \bibinfo{pages}{491} (\bibinfo{year}{1966}).

\bibitem[{\citenamefont{M{\"u}ller et~al.}(2013)\citenamefont{M{\"u}ller,
  Pletyukhov, Schuricht, and Andergassen}}]{mu.pl.13}
\bibinfo{author}{\bibfnamefont{S.~Y.} \bibnamefont{M{\"u}ller}},
  \bibinfo{author}{\bibfnamefont{M.}~\bibnamefont{Pletyukhov}},
  \bibinfo{author}{\bibfnamefont{D.}~\bibnamefont{Schuricht}},
  \bibnamefont{and}
  \bibinfo{author}{\bibfnamefont{S.}~\bibnamefont{Andergassen}},
  \bibinfo{journal}{Phys. Rev. B} \textbf{\bibinfo{volume}{87}},
  \bibinfo{pages}{245115} (\bibinfo{year}{2013}).

\bibitem[{\citenamefont{Bohr and Schmitteckert}(2012)}]{bo.sc.12}
\bibinfo{author}{\bibfnamefont{D.}~\bibnamefont{Bohr}} \bibnamefont{and}
  \bibinfo{author}{\bibfnamefont{P.}~\bibnamefont{Schmitteckert}},
  \bibinfo{journal}{Ann. Phys.} \textbf{\bibinfo{volume}{524}},
  \bibinfo{pages}{199} (\bibinfo{year}{2012}).

\bibitem[{\citenamefont{Yu et~al.}(2005)\citenamefont{Yu, Keane, Ciszek, Cheng,
  Tour, Baruah, Pederson, and Natelson}}]{yu.ke.05}
\bibinfo{author}{\bibfnamefont{L.~H.} \bibnamefont{Yu}},
  \bibinfo{author}{\bibfnamefont{Z.~K.} \bibnamefont{Keane}},
  \bibinfo{author}{\bibfnamefont{J.~W.} \bibnamefont{Ciszek}},
  \bibinfo{author}{\bibfnamefont{L.}~\bibnamefont{Cheng}},
  \bibinfo{author}{\bibfnamefont{J.~M.} \bibnamefont{Tour}},
  \bibinfo{author}{\bibfnamefont{T.}~\bibnamefont{Baruah}},
  \bibinfo{author}{\bibfnamefont{M.~R.} \bibnamefont{Pederson}},
  \bibnamefont{and} \bibinfo{author}{\bibfnamefont{D.}~\bibnamefont{Natelson}},
  \bibinfo{journal}{Phys. Rev. Lett.} \textbf{\bibinfo{volume}{95}},
  \bibinfo{pages}{256803} (\bibinfo{year}{2005}).

\bibitem[{\citenamefont{Tosi et~al.}(2012)\citenamefont{Tosi, Roura-Bas, and
  Aligia}}]{to.ro.12}
\bibinfo{author}{\bibfnamefont{L.}~\bibnamefont{Tosi}},
  \bibinfo{author}{\bibfnamefont{P.}~\bibnamefont{Roura-Bas}},
  \bibnamefont{and} \bibinfo{author}{\bibfnamefont{A.~A.}
  \bibnamefont{Aligia}}, \bibinfo{journal}{Journal of Physics: Condensed
  Matter} \textbf{\bibinfo{volume}{24}}, \bibinfo{pages}{365301}
  (\bibinfo{year}{2012}).

\bibitem[{\citenamefont{Pr{\"u}ser et~al.}(2011)\citenamefont{Pr{\"u}ser,
  Wenderoth, Dargel, Weismann, Peters, Pruschke, and Ulbrich}}]{pr.we.11}
\bibinfo{author}{\bibfnamefont{H.}~\bibnamefont{Pr{\"u}ser}},
  \bibinfo{author}{\bibfnamefont{M.}~\bibnamefont{Wenderoth}},
  \bibinfo{author}{\bibfnamefont{P.~E.} \bibnamefont{Dargel}},
  \bibinfo{author}{\bibfnamefont{A.}~\bibnamefont{Weismann}},
  \bibinfo{author}{\bibfnamefont{R.}~\bibnamefont{Peters}},
  \bibinfo{author}{\bibfnamefont{T.}~\bibnamefont{Pruschke}}, \bibnamefont{and}
  \bibinfo{author}{\bibfnamefont{R.~G.} \bibnamefont{Ulbrich}},
  \bibinfo{journal}{Nature Physics} \textbf{\bibinfo{volume}{7}},
  \bibinfo{pages}{203} (\bibinfo{year}{2011}).

\bibitem[{\citenamefont{Aoki et~al.}(2013)\citenamefont{Aoki, Tsuji, Eckstein,
  Kollar, Oka, and Philipp}}]{ao.ts.13u}
\bibinfo{author}{\bibfnamefont{H.}~\bibnamefont{Aoki}},
  \bibinfo{author}{\bibfnamefont{N.}~\bibnamefont{Tsuji}},
  \bibinfo{author}{\bibfnamefont{M.}~\bibnamefont{Eckstein}},
  \bibinfo{author}{\bibfnamefont{M.}~\bibnamefont{Kollar}},
  \bibinfo{author}{\bibfnamefont{T.}~\bibnamefont{Oka}}, \bibnamefont{and}
  \bibinfo{author}{\bibfnamefont{W.}~\bibnamefont{Philipp}}
  (\bibinfo{year}{2013}), \bibinfo{note}{arXiv:1310.5329}.

\bibitem[{\citenamefont{Schmidt and Monien}()}]{sc.mo.02u}
\bibinfo{author}{\bibfnamefont{P.}~\bibnamefont{Schmidt}} \bibnamefont{and}
  \bibinfo{author}{\bibfnamefont{H.}~\bibnamefont{Monien}},
  \bibinfo{note}{arXiv:cond-mat/0202046}.

\bibitem[{\citenamefont{Freericks et~al.}(2006)\citenamefont{Freericks,
  Turkowski, and Zlati{\'{c}}}}]{fr.tu.06}
\bibinfo{author}{\bibfnamefont{J.~K.} \bibnamefont{Freericks}},
  \bibinfo{author}{\bibfnamefont{V.~M.} \bibnamefont{Turkowski}},
  \bibnamefont{and}
  \bibinfo{author}{\bibfnamefont{V.}~\bibnamefont{Zlati{\'{c}}}},
  \bibinfo{journal}{Phys. Rev. Lett.} \textbf{\bibinfo{volume}{97}},
  \bibinfo{pages}{266408} (\bibinfo{year}{2006}).

\bibitem[{\citenamefont{Freericks}(2008)}]{free.08}
\bibinfo{author}{\bibfnamefont{J.~K.} \bibnamefont{Freericks}},
  \bibinfo{journal}{Phys. Rev. B} \textbf{\bibinfo{volume}{77}},
  \bibinfo{pages}{075109} (\bibinfo{year}{2008}).

\bibitem[{\citenamefont{Joura et~al.}(2008)\citenamefont{Joura, Freericks, and
  Pruschke}}]{jo.fr.08}
\bibinfo{author}{\bibfnamefont{A.~V.} \bibnamefont{Joura}},
  \bibinfo{author}{\bibfnamefont{J.~K.} \bibnamefont{Freericks}},
  \bibnamefont{and} \bibinfo{author}{\bibfnamefont{T.}~\bibnamefont{Pruschke}},
  \bibinfo{journal}{Phys. Rev. Lett.} \textbf{\bibinfo{volume}{101}},
  \bibinfo{pages}{196401} (\bibinfo{year}{2008}).

\bibitem[{\citenamefont{Eckstein et~al.}(2009)\citenamefont{Eckstein, Kollar,
  and Werner}}]{ec.ko.09}
\bibinfo{author}{\bibfnamefont{M.}~\bibnamefont{Eckstein}},
  \bibinfo{author}{\bibfnamefont{M.}~\bibnamefont{Kollar}}, \bibnamefont{and}
  \bibinfo{author}{\bibfnamefont{P.}~\bibnamefont{Werner}},
  \bibinfo{journal}{Phys. Rev. Lett.} \textbf{\bibinfo{volume}{103}},
  \bibinfo{pages}{056403} (\bibinfo{year}{2009}).

\bibitem[{\citenamefont{Okamoto}(2007)}]{okam.07}
\bibinfo{author}{\bibfnamefont{S.}~\bibnamefont{Okamoto}},
  \bibinfo{journal}{Phys. Rev. B} \textbf{\bibinfo{volume}{76}},
  \bibinfo{pages}{035105} (\bibinfo{year}{2007}).

\bibitem[{\citenamefont{Arrigoni et~al.}(2013)\citenamefont{Arrigoni, Knap, and
  von~der Linden}}]{ar.kn.13}
\bibinfo{author}{\bibfnamefont{E.}~\bibnamefont{Arrigoni}},
  \bibinfo{author}{\bibfnamefont{M.}~\bibnamefont{Knap}}, \bibnamefont{and}
  \bibinfo{author}{\bibfnamefont{W.}~\bibnamefont{von~der Linden}},
  \bibinfo{journal}{Phys. Rev. Lett.} \textbf{\bibinfo{volume}{110}},
  \bibinfo{pages}{086403} (\bibinfo{year}{2013}).

\bibitem[{\citenamefont{Mehta and Andrei}(2006)}]{me.an.06}
\bibinfo{author}{\bibfnamefont{P.}~\bibnamefont{Mehta}} \bibnamefont{and}
  \bibinfo{author}{\bibfnamefont{N.}~\bibnamefont{Andrei}},
  \bibinfo{journal}{Phys. Rev. Lett.} \textbf{\bibinfo{volume}{96}},
  \bibinfo{pages}{216802} (\bibinfo{year}{2006}).

\bibitem[{\citenamefont{Anders}(2008)}]{ande.08}
\bibinfo{author}{\bibfnamefont{F.~B.} \bibnamefont{Anders}},
  \bibinfo{journal}{Phys. Rev. Lett.} \textbf{\bibinfo{volume}{101}},
  \bibinfo{pages}{066804} (\bibinfo{year}{2008}).

\bibitem[{\citenamefont{Anders and Schmitt}(2010)}]{an.sc.10}
\bibinfo{author}{\bibfnamefont{F.~B.} \bibnamefont{Anders}} \bibnamefont{and}
  \bibinfo{author}{\bibfnamefont{S.}~\bibnamefont{Schmitt}},
  \bibinfo{journal}{Journal of Physics: Conference Series}
  \textbf{\bibinfo{volume}{220}}, \bibinfo{pages}{012021}
  (\bibinfo{year}{2010}).

\bibitem[{\citenamefont{Rosch}(2012)}]{rosc.12}
\bibinfo{author}{\bibfnamefont{A.}~\bibnamefont{Rosch}}, \bibinfo{journal}{Eur.
  Phys. J. B} \textbf{\bibinfo{volume}{85}}, \bibinfo{pages}{6}
  (\bibinfo{year}{2012}).

\bibitem[{\citenamefont{Meir et~al.}(1993)\citenamefont{Meir, Wingreen, and
  Lee}}]{me.wi.93}
\bibinfo{author}{\bibfnamefont{Y.}~\bibnamefont{Meir}},
  \bibinfo{author}{\bibfnamefont{N.~S.} \bibnamefont{Wingreen}},
  \bibnamefont{and} \bibinfo{author}{\bibfnamefont{P.~A.} \bibnamefont{Lee}},
  \bibinfo{journal}{Phys. Rev. Lett.} \textbf{\bibinfo{volume}{70}},
  \bibinfo{pages}{2601} (\bibinfo{year}{1993}).

\bibitem[{\citenamefont{Wingreen and Meir}(1994)}]{wi.me.94}
\bibinfo{author}{\bibfnamefont{N.~S.} \bibnamefont{Wingreen}} \bibnamefont{and}
  \bibinfo{author}{\bibfnamefont{Y.}~\bibnamefont{Meir}},
  \bibinfo{journal}{Phys. Rev. B} \textbf{\bibinfo{volume}{49}},
  \bibinfo{pages}{11040} (\bibinfo{year}{1994}).

\bibitem[{\citenamefont{Fujii and Ueda}(2003)}]{fu.ue.03}
\bibinfo{author}{\bibfnamefont{T.}~\bibnamefont{Fujii}} \bibnamefont{and}
  \bibinfo{author}{\bibfnamefont{K.}~\bibnamefont{Ueda}},
  \bibinfo{journal}{Phys. Rev. B} \textbf{\bibinfo{volume}{68}},
  \bibinfo{pages}{155310} (\bibinfo{year}{2003}).

\bibitem[{\citenamefont{Schoeller and Sch{\"o}n}(1994)}]{sc.sc.94}
\bibinfo{author}{\bibfnamefont{H.}~\bibnamefont{Schoeller}} \bibnamefont{and}
  \bibinfo{author}{\bibfnamefont{G.}~\bibnamefont{Sch{\"o}n}},
  \bibinfo{journal}{Phys. Rev. B} \textbf{\bibinfo{volume}{50}},
  \bibinfo{pages}{18436} (\bibinfo{year}{1994}).

\bibitem[{\citenamefont{Hershfield et~al.}(1991)\citenamefont{Hershfield,
  Davies, and Wilkins}}]{he.da.91}
\bibinfo{author}{\bibfnamefont{S.}~\bibnamefont{Hershfield}},
  \bibinfo{author}{\bibfnamefont{J.~H.} \bibnamefont{Davies}},
  \bibnamefont{and} \bibinfo{author}{\bibfnamefont{J.~W.}
  \bibnamefont{Wilkins}}, \bibinfo{journal}{Phys. Rev. Lett.}
  \textbf{\bibinfo{volume}{67}}, \bibinfo{pages}{3720} (\bibinfo{year}{1991}).

\bibitem[{\citenamefont{Schoeller}(2009)}]{scho.09}
\bibinfo{author}{\bibfnamefont{H.}~\bibnamefont{Schoeller}},
  \bibinfo{journal}{Eur. Phys. J. Special Topics}
  \textbf{\bibinfo{volume}{168}}, \bibinfo{pages}{179} (\bibinfo{year}{2009}).

\bibitem[{\citenamefont{Rosch et~al.}(2005)\citenamefont{Rosch, Paaske, Kroha,
  and W{\"o}lfle}}]{ro.pa.05}
\bibinfo{author}{\bibfnamefont{A.}~\bibnamefont{Rosch}},
  \bibinfo{author}{\bibfnamefont{J.}~\bibnamefont{Paaske}},
  \bibinfo{author}{\bibfnamefont{J.}~\bibnamefont{Kroha}}, \bibnamefont{and}
  \bibinfo{author}{\bibfnamefont{P.}~\bibnamefont{W{\"o}lfle}},
  \bibinfo{journal}{J. Phys. Soc. Jpn.} \textbf{\bibinfo{volume}{74}},
  \bibinfo{pages}{118} (\bibinfo{year}{2005}).

\bibitem[{\citenamefont{Anders and Schiller}(2006)}]{an.sc.06}
\bibinfo{author}{\bibfnamefont{F.~B.} \bibnamefont{Anders}} \bibnamefont{and}
  \bibinfo{author}{\bibfnamefont{A.}~\bibnamefont{Schiller}},
  \bibinfo{journal}{Phys. Rev. B} \textbf{\bibinfo{volume}{74}},
  \bibinfo{pages}{245113} (\bibinfo{year}{2006}).

\bibitem[{\citenamefont{Roosen et~al.}(2008)\citenamefont{Roosen, Wegewijs, and
  Hofstetter}}]{ro.we.08}
\bibinfo{author}{\bibfnamefont{D.}~\bibnamefont{Roosen}},
  \bibinfo{author}{\bibfnamefont{M.~R.} \bibnamefont{Wegewijs}},
  \bibnamefont{and}
  \bibinfo{author}{\bibfnamefont{W.}~\bibnamefont{Hofstetter}},
  \bibinfo{journal}{Phys. Rev. Lett.} \textbf{\bibinfo{volume}{100}},
  \bibinfo{pages}{087201} (\bibinfo{year}{2008}).

\bibitem[{\citenamefont{Doyon and Andrei}(2006)}]{do.an.06}
\bibinfo{author}{\bibfnamefont{B.}~\bibnamefont{Doyon}} \bibnamefont{and}
  \bibinfo{author}{\bibfnamefont{N.}~\bibnamefont{Andrei}},
  \bibinfo{journal}{Phys. Rev. B} \textbf{\bibinfo{volume}{73}},
  \bibinfo{pages}{245326} (\bibinfo{year}{2006}).

\bibitem[{\citenamefont{Weiss et~al.}(2008)\citenamefont{Weiss, Eckel,
  Thorwart, and Egger}}]{we.ec.08}
\bibinfo{author}{\bibfnamefont{S.}~\bibnamefont{Weiss}},
  \bibinfo{author}{\bibfnamefont{J.}~\bibnamefont{Eckel}},
  \bibinfo{author}{\bibfnamefont{M.}~\bibnamefont{Thorwart}}, \bibnamefont{and}
  \bibinfo{author}{\bibfnamefont{R.}~\bibnamefont{Egger}},
  \bibinfo{journal}{Phys. Rev. B} \textbf{\bibinfo{volume}{77}},
  \bibinfo{pages}{195316} (\bibinfo{year}{2008}).

\bibitem[{\citenamefont{Anders and Schiller}(2005)}]{an.sc.05}
\bibinfo{author}{\bibfnamefont{F.~B.} \bibnamefont{Anders}} \bibnamefont{and}
  \bibinfo{author}{\bibfnamefont{A.}~\bibnamefont{Schiller}},
  \bibinfo{journal}{Phys. Rev. Lett.} \textbf{\bibinfo{volume}{95}},
  \bibinfo{pages}{196801} (\bibinfo{year}{2005}).

\bibitem[{\citenamefont{Moeckel and Kehrein}(2008)}]{mo.ke.08}
\bibinfo{author}{\bibfnamefont{M.}~\bibnamefont{Moeckel}} \bibnamefont{and}
  \bibinfo{author}{\bibfnamefont{S.}~\bibnamefont{Kehrein}},
  \bibinfo{journal}{Phys. Rev. Lett.} \textbf{\bibinfo{volume}{100}},
  \bibinfo{pages}{175702} (\bibinfo{year}{2008}).

\bibitem[{\citenamefont{Kehrein}(2005)}]{kehr.05}
\bibinfo{author}{\bibfnamefont{S.}~\bibnamefont{Kehrein}},
  \bibinfo{journal}{Phys. Rev. Lett.} \textbf{\bibinfo{volume}{95}},
  \bibinfo{pages}{056602} (\bibinfo{year}{2005}).

\bibitem[{\citenamefont{Vidal}(2004)}]{vida.04}
\bibinfo{author}{\bibfnamefont{G.}~\bibnamefont{Vidal}},
  \bibinfo{journal}{Phys. Rev. Lett.} \textbf{\bibinfo{volume}{93}},
  \bibinfo{pages}{040502} (\bibinfo{year}{2004}).

\bibitem[{\citenamefont{White}(1993)}]{whit.93}
\bibinfo{author}{\bibfnamefont{S.~R.} \bibnamefont{White}},
  \bibinfo{journal}{Phys. Rev. B} \textbf{\bibinfo{volume}{48}},
  \bibinfo{pages}{10345} (\bibinfo{year}{1993}).

\bibitem[{\citenamefont{Daley et~al.}(2004)\citenamefont{Daley, Kollath,
  Schollw{\"o}ck, and Vidal}}]{da.ko.04}
\bibinfo{author}{\bibfnamefont{A.~J.} \bibnamefont{Daley}},
  \bibinfo{author}{\bibfnamefont{C.}~\bibnamefont{Kollath}},
  \bibinfo{author}{\bibfnamefont{U.}~\bibnamefont{Schollw{\"o}ck}},
  \bibnamefont{and} \bibinfo{author}{\bibfnamefont{G.}~\bibnamefont{Vidal}},
  \bibinfo{journal}{J. Stat. Mech.} \textbf{\bibinfo{volume}{2004}},
  \bibinfo{pages}{P04005} (\bibinfo{year}{2004}).

\bibitem[{\citenamefont{White and Feiguin}(2004)}]{wh.fe.04}
\bibinfo{author}{\bibfnamefont{S.~R.} \bibnamefont{White}} \bibnamefont{and}
  \bibinfo{author}{\bibfnamefont{A.~E.} \bibnamefont{Feiguin}},
  \bibinfo{journal}{Phys. Rev. Lett.} \textbf{\bibinfo{volume}{93}},
  \bibinfo{pages}{076401} (\bibinfo{year}{2004}).

\bibitem[{\citenamefont{Schollwoeck}(2011)}]{scho.11}
\bibinfo{author}{\bibfnamefont{U.}~\bibnamefont{Schollwoeck}},
  \bibinfo{journal}{Annals of Physics} \textbf{\bibinfo{volume}{326}},
  \bibinfo{pages}{96} (\bibinfo{year}{2011}).

\bibitem[{\citenamefont{Schmitteckert}(2004)}]{schm.04}
\bibinfo{author}{\bibfnamefont{P.}~\bibnamefont{Schmitteckert}},
  \bibinfo{journal}{Phys. Rev. B} \textbf{\bibinfo{volume}{70}},
  \bibinfo{pages}{121302} (\bibinfo{year}{2004}).

\bibitem[{\citenamefont{Heidrich-Meisner
  et~al.}(2009)\citenamefont{Heidrich-Meisner, Feiguin, and
  Dagotto}}]{he.fe.09}
\bibinfo{author}{\bibfnamefont{F.}~\bibnamefont{Heidrich-Meisner}},
  \bibinfo{author}{\bibfnamefont{A.~E.} \bibnamefont{Feiguin}},
  \bibnamefont{and} \bibinfo{author}{\bibfnamefont{E.}~\bibnamefont{Dagotto}},
  \bibinfo{journal}{Phys. Rev. B} \textbf{\bibinfo{volume}{79}},
  \bibinfo{pages}{235336} (\bibinfo{year}{2009}).

\bibitem[{\citenamefont{Nuss et~al.}(2013)\citenamefont{Nuss, Ganahl, Evertz,
  Arrigoni, and von~der Linden}}]{nu.ga.13}
\bibinfo{author}{\bibfnamefont{M.}~\bibnamefont{Nuss}},
  \bibinfo{author}{\bibfnamefont{M.}~\bibnamefont{Ganahl}},
  \bibinfo{author}{\bibfnamefont{H.~G.} \bibnamefont{Evertz}},
  \bibinfo{author}{\bibfnamefont{E.}~\bibnamefont{Arrigoni}}, \bibnamefont{and}
  \bibinfo{author}{\bibfnamefont{W.}~\bibnamefont{von~der Linden}},
  \bibinfo{journal}{Phys. Rev. B} \textbf{\bibinfo{volume}{88}},
  \bibinfo{pages}{045132} (\bibinfo{year}{2013}).

\bibitem[{\citenamefont{Nuss et~al.}(2012)\citenamefont{Nuss, Heil, Ganahl,
  Knap, Evertz, Arrigoni, and von~der Linden}}]{nu.he.12}
\bibinfo{author}{\bibfnamefont{M.}~\bibnamefont{Nuss}},
  \bibinfo{author}{\bibfnamefont{C.}~\bibnamefont{Heil}},
  \bibinfo{author}{\bibfnamefont{M.}~\bibnamefont{Ganahl}},
  \bibinfo{author}{\bibfnamefont{M.}~\bibnamefont{Knap}},
  \bibinfo{author}{\bibfnamefont{H.~G.} \bibnamefont{Evertz}},
  \bibinfo{author}{\bibfnamefont{E.}~\bibnamefont{Arrigoni}}, \bibnamefont{and}
  \bibinfo{author}{\bibfnamefont{W.}~\bibnamefont{von~der Linden}},
  \bibinfo{journal}{Phys. Rev. B} \textbf{\bibinfo{volume}{86}},
  \bibinfo{pages}{245119} (\bibinfo{year}{2012}).

\bibitem[{\citenamefont{Knap et~al.}(2011{\natexlab{a}})\citenamefont{Knap,
  {von der Linden}, and Arrigoni}}]{kn.li.11}
\bibinfo{author}{\bibfnamefont{M.}~\bibnamefont{Knap}},
  \bibinfo{author}{\bibfnamefont{W.}~\bibnamefont{{von der Linden}}},
  \bibnamefont{and} \bibinfo{author}{\bibfnamefont{E.}~\bibnamefont{Arrigoni}},
  \bibinfo{journal}{Phys. Rev. B} \textbf{\bibinfo{volume}{84}},
  \bibinfo{pages}{115145} (\bibinfo{year}{2011}{\natexlab{a}}).

\bibitem[{\citenamefont{Hofmann et~al.}(2013)\citenamefont{Hofmann, Eckstein,
  Arrigoni, and Potthoff}}]{ho.ec.13}
\bibinfo{author}{\bibfnamefont{F.}~\bibnamefont{Hofmann}},
  \bibinfo{author}{\bibfnamefont{M.}~\bibnamefont{Eckstein}},
  \bibinfo{author}{\bibfnamefont{E.}~\bibnamefont{Arrigoni}}, \bibnamefont{and}
  \bibinfo{author}{\bibfnamefont{M.}~\bibnamefont{Potthoff}},
  \bibinfo{journal}{Phys. Rev. B} \textbf{\bibinfo{volume}{88}},
  \bibinfo{pages}{165124} (\bibinfo{year}{2013}).

\bibitem[{\citenamefont{Jung et~al.}(2012)\citenamefont{Jung, Lieder, Brener,
  Hafermann, Baxevanis, Chudnovskiy, Rubtsov, Katsnelson, and
  Lichtenstein}}]{ju.li.12}
\bibinfo{author}{\bibfnamefont{C.}~\bibnamefont{Jung}},
  \bibinfo{author}{\bibfnamefont{A.}~\bibnamefont{Lieder}},
  \bibinfo{author}{\bibfnamefont{S.}~\bibnamefont{Brener}},
  \bibinfo{author}{\bibfnamefont{H.}~\bibnamefont{Hafermann}},
  \bibinfo{author}{\bibfnamefont{B.}~\bibnamefont{Baxevanis}},
  \bibinfo{author}{\bibfnamefont{A.}~\bibnamefont{Chudnovskiy}},
  \bibinfo{author}{\bibfnamefont{A.}~\bibnamefont{Rubtsov}},
  \bibinfo{author}{\bibfnamefont{M.}~\bibnamefont{Katsnelson}},
  \bibnamefont{and}
  \bibinfo{author}{\bibfnamefont{A.}~\bibnamefont{Lichtenstein}},
  \bibinfo{journal}{Ann. Phys.} \textbf{\bibinfo{volume}{524}},
  \bibinfo{pages}{49} (\bibinfo{year}{2012}).

\bibitem[{\citenamefont{Gezzi et~al.}(2007)\citenamefont{Gezzi, Pruschke, and
  Meden}}]{ge.pr.07}
\bibinfo{author}{\bibfnamefont{R.}~\bibnamefont{Gezzi}},
  \bibinfo{author}{\bibfnamefont{T.}~\bibnamefont{Pruschke}}, \bibnamefont{and}
  \bibinfo{author}{\bibfnamefont{V.}~\bibnamefont{Meden}},
  \bibinfo{journal}{Phys. Rev. B} \textbf{\bibinfo{volume}{75}},
  \bibinfo{pages}{045324} (\bibinfo{year}{2007}).

\bibitem[{\citenamefont{Jakobs et~al.}(2007)\citenamefont{Jakobs, Meden, and
  Schoeller}}]{ja.me.07}
\bibinfo{author}{\bibfnamefont{S.~G.} \bibnamefont{Jakobs}},
  \bibinfo{author}{\bibfnamefont{V.}~\bibnamefont{Meden}}, \bibnamefont{and}
  \bibinfo{author}{\bibfnamefont{H.}~\bibnamefont{Schoeller}},
  \bibinfo{journal}{Phys. Rev. Lett.} \textbf{\bibinfo{volume}{99}},
  \bibinfo{pages}{150603} (\bibinfo{year}{2007}).

\bibitem[{\citenamefont{Werner et~al.}(2010)\citenamefont{Werner, Oka,
  Eckstein, and Millis}}]{we.ok.10}
\bibinfo{author}{\bibfnamefont{P.}~\bibnamefont{Werner}},
  \bibinfo{author}{\bibfnamefont{T.}~\bibnamefont{Oka}},
  \bibinfo{author}{\bibfnamefont{M.}~\bibnamefont{Eckstein}}, \bibnamefont{and}
  \bibinfo{author}{\bibfnamefont{A.~J.} \bibnamefont{Millis}},
  \bibinfo{journal}{Phys. Rev. B} \textbf{\bibinfo{volume}{81}},
  \bibinfo{pages}{035108} (\bibinfo{year}{2010}).

\bibitem[{\citenamefont{Cohen et~al.}(2013)\citenamefont{Cohen, Gull, Reichman,
  and Millis}}]{co.gu.13u}
\bibinfo{author}{\bibfnamefont{G.}~\bibnamefont{Cohen}},
  \bibinfo{author}{\bibfnamefont{E.}~\bibnamefont{Gull}},
  \bibinfo{author}{\bibfnamefont{D.~R.} \bibnamefont{Reichman}},
  \bibnamefont{and} \bibinfo{author}{\bibfnamefont{A.~J.} \bibnamefont{Millis}}
  (\bibinfo{year}{2013}), \bibinfo{note}{arXiv:1310.4151}.

\bibitem[{\citenamefont{Han}(2006)}]{han.06}
\bibinfo{author}{\bibfnamefont{J.~E.} \bibnamefont{Han}},
  \bibinfo{journal}{Phys. Rev. B} \textbf{\bibinfo{volume}{73}},
  \bibinfo{pages}{125319} (\bibinfo{year}{2006}).

\bibitem[{\citenamefont{Han and Heary}(2007)}]{ha.he.07}
\bibinfo{author}{\bibfnamefont{J.~E.} \bibnamefont{Han}} \bibnamefont{and}
  \bibinfo{author}{\bibfnamefont{R.~J.} \bibnamefont{Heary}},
  \bibinfo{journal}{Phys. Rev. Lett.} \textbf{\bibinfo{volume}{99}},
  \bibinfo{pages}{236808} (\bibinfo{year}{2007}).

\bibitem[{\citenamefont{Dirks et~al.}(2010)\citenamefont{Dirks, Werner,
  Jarrell, and Pruschke}}]{di.we.10}
\bibinfo{author}{\bibfnamefont{A.}~\bibnamefont{Dirks}},
  \bibinfo{author}{\bibfnamefont{P.}~\bibnamefont{Werner}},
  \bibinfo{author}{\bibfnamefont{M.}~\bibnamefont{Jarrell}}, \bibnamefont{and}
  \bibinfo{author}{\bibfnamefont{T.}~\bibnamefont{Pruschke}},
  \bibinfo{journal}{Phys. Rev. E} \textbf{\bibinfo{volume}{82}},
  \bibinfo{pages}{026701} (\bibinfo{year}{2010}).

\bibitem[{\citenamefont{Han et~al.}(2012)\citenamefont{Han, Dirks, and
  Pruschke}}]{ha.di.12}
\bibinfo{author}{\bibfnamefont{J.~E.} \bibnamefont{Han}},
  \bibinfo{author}{\bibfnamefont{A.}~\bibnamefont{Dirks}}, \bibnamefont{and}
  \bibinfo{author}{\bibfnamefont{T.}~\bibnamefont{Pruschke}},
  \bibinfo{journal}{Phys. Rev. B} \textbf{\bibinfo{volume}{86}},
  \bibinfo{pages}{155130} (\bibinfo{year}{2012}).

\bibitem[{\citenamefont{Dirks et~al.}(2013{\natexlab{a}})\citenamefont{Dirks,
  Han, Jarrell, and Pruschke}}]{ha.di.12b}
\bibinfo{author}{\bibfnamefont{A.}~\bibnamefont{Dirks}},
  \bibinfo{author}{\bibfnamefont{J.~E.} \bibnamefont{Han}},
  \bibinfo{author}{\bibfnamefont{M.}~\bibnamefont{Jarrell}}, \bibnamefont{and}
  \bibinfo{author}{\bibfnamefont{T.}~\bibnamefont{Pruschke}},
  \bibinfo{journal}{Phys. Rev. B} \textbf{\bibinfo{volume}{87}},
  \bibinfo{pages}{235140} (\bibinfo{year}{2013}{\natexlab{a}}).

\bibitem[{\citenamefont{Dutt et~al.}(2011)\citenamefont{Dutt, Koch, Han, and
  {Le Hur}}}]{du.ko.11}
\bibinfo{author}{\bibfnamefont{P.}~\bibnamefont{Dutt}},
  \bibinfo{author}{\bibfnamefont{J.}~\bibnamefont{Koch}},
  \bibinfo{author}{\bibfnamefont{J.}~\bibnamefont{Han}}, \bibnamefont{and}
  \bibinfo{author}{\bibfnamefont{K.}~\bibnamefont{{Le Hur}}},
  \bibinfo{journal}{Annals of Physics} \textbf{\bibinfo{volume}{326}},
  \bibinfo{pages}{2963} (\bibinfo{year}{2011}).

\bibitem[{\citenamefont{Mu\~noz et~al.}(2013)\citenamefont{Mu\~noz, Bolech, and
  Kirchner}}]{mu.bo.13}
\bibinfo{author}{\bibfnamefont{E.}~\bibnamefont{Mu\~noz}},
  \bibinfo{author}{\bibfnamefont{C.~J.} \bibnamefont{Bolech}},
  \bibnamefont{and} \bibinfo{author}{\bibfnamefont{S.}~\bibnamefont{Kirchner}},
  \bibinfo{journal}{Phys. Rev. Lett.} \textbf{\bibinfo{volume}{110}},
  \bibinfo{pages}{016601} (\bibinfo{year}{2013}).

\bibitem[{\citenamefont{Uimonen et~al.}(2011)\citenamefont{Uimonen, Khosravi,
  Stan, Stefanucci, Kurth, van Leeuwen, and Gross}}]{ui.kh.11}
\bibinfo{author}{\bibfnamefont{A.~M.} \bibnamefont{Uimonen}},
  \bibinfo{author}{\bibfnamefont{E.}~\bibnamefont{Khosravi}},
  \bibinfo{author}{\bibfnamefont{A.}~\bibnamefont{Stan}},
  \bibinfo{author}{\bibfnamefont{G.}~\bibnamefont{Stefanucci}},
  \bibinfo{author}{\bibfnamefont{S.}~\bibnamefont{Kurth}},
  \bibinfo{author}{\bibfnamefont{R.}~\bibnamefont{van Leeuwen}},
  \bibnamefont{and} \bibinfo{author}{\bibfnamefont{E.~K.~U.}
  \bibnamefont{Gross}}, \bibinfo{journal}{Phys. Rev. B}
  \textbf{\bibinfo{volume}{84}}, \bibinfo{pages}{115103}
  (\bibinfo{year}{2011}).

\bibitem[{\citenamefont{Smirnov and Grifoni}(2011)}]{sm.gr.11}
\bibinfo{author}{\bibfnamefont{S.}~\bibnamefont{Smirnov}} \bibnamefont{and}
  \bibinfo{author}{\bibfnamefont{M.}~\bibnamefont{Grifoni}},
  \bibinfo{journal}{Phys. Rev. B} \textbf{\bibinfo{volume}{84}},
  \bibinfo{pages}{125303} (\bibinfo{year}{2011}).

\bibitem[{\citenamefont{Schoeller and K{\"o}nig}(2000)}]{so.ko.00}
\bibinfo{author}{\bibfnamefont{H.}~\bibnamefont{Schoeller}} \bibnamefont{and}
  \bibinfo{author}{\bibfnamefont{J.}~\bibnamefont{K{\"o}nig}},
  \bibinfo{journal}{Phys. Rev. Lett.} \textbf{\bibinfo{volume}{84}},
  \bibinfo{pages}{3686} (\bibinfo{year}{2000}).

\bibitem[{\citenamefont{Schiro and Fabrizio}(2010)}]{sc.fa.10}
\bibinfo{author}{\bibfnamefont{M.}~\bibnamefont{Schiro}} \bibnamefont{and}
  \bibinfo{author}{\bibfnamefont{M.}~\bibnamefont{Fabrizio}},
  \bibinfo{journal}{Phys. Rev. Lett.} \textbf{\bibinfo{volume}{105}},
  \bibinfo{pages}{076401} (\bibinfo{year}{2010}).

\bibitem[{\citenamefont{Timm}(2008)}]{timm.08}
\bibinfo{author}{\bibfnamefont{C.}~\bibnamefont{Timm}}, \bibinfo{journal}{Phys.
  Rev. B} \textbf{\bibinfo{volume}{77}}, \bibinfo{pages}{195416}
  (\bibinfo{year}{2008}).

\bibitem[{\citenamefont{Eckel et~al.}(2010)\citenamefont{Eckel,
  Heidrich-Meisner, Jakobs, Thorwart, Pletyukhov, and Egger}}]{ec.he.10}
\bibinfo{author}{\bibfnamefont{J.}~\bibnamefont{Eckel}},
  \bibinfo{author}{\bibfnamefont{F.}~\bibnamefont{Heidrich-Meisner}},
  \bibinfo{author}{\bibfnamefont{S.~G.} \bibnamefont{Jakobs}},
  \bibinfo{author}{\bibfnamefont{M.}~\bibnamefont{Thorwart}},
  \bibinfo{author}{\bibfnamefont{M.}~\bibnamefont{Pletyukhov}},
  \bibnamefont{and} \bibinfo{author}{\bibfnamefont{R.}~\bibnamefont{Egger}},
  \bibinfo{journal}{New. J. Phys.} \textbf{\bibinfo{volume}{12}},
  \bibinfo{pages}{043042} (\bibinfo{year}{2010}).

\bibitem[{\citenamefont{Andergassen et~al.}(2010)\citenamefont{Andergassen,
  Meden, Schoeller, Splettstoesser, and Wegewijs}}]{an.me.10}
\bibinfo{author}{\bibfnamefont{S.}~\bibnamefont{Andergassen}},
  \bibinfo{author}{\bibfnamefont{V.}~\bibnamefont{Meden}},
  \bibinfo{author}{\bibfnamefont{H.}~\bibnamefont{Schoeller}},
  \bibinfo{author}{\bibfnamefont{J.}~\bibnamefont{Splettstoesser}},
  \bibnamefont{and} \bibinfo{author}{\bibfnamefont{M.~R.}
  \bibnamefont{Wegewijs}}, \bibinfo{journal}{Nanotechnology}
  \textbf{\bibinfo{volume}{21}}, \bibinfo{pages}{272001}
  (\bibinfo{year}{2010}).

\bibitem[{\citenamefont{Contreras-Pulido
  et~al.}(2012)\citenamefont{Contreras-Pulido, Splettstoesser, Governale,
  K{\"o}nig, and B{\"u}ttiker}}]{pu.sp.12}
\bibinfo{author}{\bibfnamefont{L.~D.} \bibnamefont{Contreras-Pulido}},
  \bibinfo{author}{\bibfnamefont{J.}~\bibnamefont{Splettstoesser}},
  \bibinfo{author}{\bibfnamefont{M.}~\bibnamefont{Governale}},
  \bibinfo{author}{\bibfnamefont{J.}~\bibnamefont{K{\"o}nig}},
  \bibnamefont{and}
  \bibinfo{author}{\bibfnamefont{M.}~\bibnamefont{B{\"u}ttiker}},
  \bibinfo{journal}{Phys. Rev. B} \textbf{\bibinfo{volume}{85}},
  \bibinfo{pages}{075301} (\bibinfo{year}{2012}).

\bibitem[{\citenamefont{Falicov and Kimball}(1969)}]{fa.ki.69}
\bibinfo{author}{\bibfnamefont{L.~M.} \bibnamefont{Falicov}} \bibnamefont{and}
  \bibinfo{author}{\bibfnamefont{J.~C.} \bibnamefont{Kimball}},
  \bibinfo{journal}{Phys. Rev. Lett.} \textbf{\bibinfo{volume}{22}},
  \bibinfo{pages}{997} (\bibinfo{year}{1969}).

\bibitem[{\citenamefont{Eckstein and Kollar}(2008)}]{ec.ko.08}
\bibinfo{author}{\bibfnamefont{M.}~\bibnamefont{Eckstein}} \bibnamefont{and}
  \bibinfo{author}{\bibfnamefont{M.}~\bibnamefont{Kollar}},
  \bibinfo{journal}{Phys. Rev. Lett.} \textbf{\bibinfo{volume}{100}},
  \bibinfo{pages}{120404} (\bibinfo{year}{2008}).

\bibitem[{\citenamefont{Eckstein et~al.}(2010)\citenamefont{Eckstein, Kollar,
  and Werner}}]{ec.ko.10}
\bibinfo{author}{\bibfnamefont{M.}~\bibnamefont{Eckstein}},
  \bibinfo{author}{\bibfnamefont{M.}~\bibnamefont{Kollar}}, \bibnamefont{and}
  \bibinfo{author}{\bibfnamefont{P.}~\bibnamefont{Werner}},
  \bibinfo{journal}{Phys. Rev. B} \textbf{\bibinfo{volume}{81}},
  \bibinfo{pages}{115131} (\bibinfo{year}{2010}).

\bibitem[{\citenamefont{Okamoto}(2008)}]{okam.08}
\bibinfo{author}{\bibfnamefont{S.}~\bibnamefont{Okamoto}},
  \bibinfo{journal}{Phys. Rev. Lett.} \textbf{\bibinfo{volume}{101}},
  \bibinfo{pages}{116807} (\bibinfo{year}{2008}).

\bibitem[{\citenamefont{Aron et~al.}(2012)\citenamefont{Aron, Kotliar, and
  Weber}}]{ar.ko.12}
\bibinfo{author}{\bibfnamefont{C.}~\bibnamefont{Aron}},
  \bibinfo{author}{\bibfnamefont{G.}~\bibnamefont{Kotliar}}, \bibnamefont{and}
  \bibinfo{author}{\bibfnamefont{C.}~\bibnamefont{Weber}},
  \bibinfo{journal}{Phys. Rev. Lett.} \textbf{\bibinfo{volume}{108}},
  \bibinfo{pages}{086401} (\bibinfo{year}{2012}).

\bibitem[{\citenamefont{Gramsch et~al.}(2013)\citenamefont{Gramsch, Balzer,
  Eckstein, and Kollar}}]{gr.ba.13}
\bibinfo{author}{\bibfnamefont{C.}~\bibnamefont{Gramsch}},
  \bibinfo{author}{\bibfnamefont{K.}~\bibnamefont{Balzer}},
  \bibinfo{author}{\bibfnamefont{M.}~\bibnamefont{Eckstein}}, \bibnamefont{and}
  \bibinfo{author}{\bibfnamefont{M.}~\bibnamefont{Kollar}},
  \bibinfo{journal}{Phys. Rev. B} \textbf{\bibinfo{volume}{88}},
  \bibinfo{pages}{235106} (\bibinfo{year}{2013}).

\bibitem[{\citenamefont{Carmichael}(2002)}]{carmichael1}
\bibinfo{author}{\bibfnamefont{H.~J.} \bibnamefont{Carmichael}},
  \emph{\bibinfo{title}{Statistical Methods in Quantum Optics: Master Equations
  and Fokker-Planck Equations}}, vol.~\bibinfo{volume}{1} of
  \emph{\bibinfo{series}{Texts and monographs in physics}}
  (\bibinfo{publisher}{Springer}, \bibinfo{address}{Singapore},
  \bibinfo{year}{2002}).

\bibitem[{foo({\natexlab{a}})}]{footnote1}
\bibinfo{note}{In our convention, lowercase $g$ denote Green's functions of the
  system where the impurity is disconnected from the reservoirs, while capital
  $G$ denote Green's functions of the connected system.}

\bibitem[{foo({\natexlab{b}})}]{footnote14}
\bibinfo{note}{Conventions for branch cuts are such that $g^R$ is causal.}

\bibitem[{\citenamefont{Economou}(2006)}]{economou}
\bibinfo{author}{\bibfnamefont{E.~N.} \bibnamefont{Economou}},
  \emph{\bibinfo{title}{Green~s Functions in Quantum Physics}}
  (\bibinfo{publisher}{Springer}, \bibinfo{address}{Heidelberg},
  \bibinfo{year}{2006}).

\bibitem[{foo({\natexlab{c}})}]{footnote13}
\bibinfo{note}{Note that in the present formalism, temperature would enter
  through the hybridization function $\Delta^K(\omega)$ only.}

\bibitem[{foo({\natexlab{d}})}]{footnote2}
\bibinfo{note}{This is in general true unless the system has bound states.}

\bibitem[{\citenamefont{Kadanoff and Baym}(1962)}]{kad.baym}
\bibinfo{author}{\bibfnamefont{L.~P.} \bibnamefont{Kadanoff}} \bibnamefont{and}
  \bibinfo{author}{\bibfnamefont{G.}~\bibnamefont{Baym}},
  \emph{\bibinfo{title}{Quantum Statistical Mechanics: Green's Function Methods
  in Equilibrium and Nonequilibrium Problems}}
  (\bibinfo{publisher}{Addison-Wesley}, \bibinfo{address}{Redwood City, CA},
  \bibinfo{year}{1962}).

\bibitem[{\citenamefont{Schwinger}(1961)}]{schw.61}
\bibinfo{author}{\bibfnamefont{J.}~\bibnamefont{Schwinger}},
  \bibinfo{journal}{J. Math. Phys.} \textbf{\bibinfo{volume}{2}},
  \bibinfo{pages}{407} (\bibinfo{year}{1961}).

\bibitem[{\citenamefont{Keldysh}(1965)}]{keld.65}
\bibinfo{author}{\bibfnamefont{L.~V.} \bibnamefont{Keldysh}},
  \bibinfo{journal}{Sov. Phys. JETP} \textbf{\bibinfo{volume}{20}},
  \bibinfo{pages}{1018} (\bibinfo{year}{1965}).

\bibitem[{\citenamefont{Haug and Jauho}(1998)}]{ha.ja}
\bibinfo{author}{\bibfnamefont{H.}~\bibnamefont{Haug}} \bibnamefont{and}
  \bibinfo{author}{\bibfnamefont{A.-P.} \bibnamefont{Jauho}},
  \emph{\bibinfo{title}{Quantum Kinetics in Transport and Optics of
  Semiconductors}} (\bibinfo{publisher}{Springer},
  \bibinfo{address}{Heidelberg}, \bibinfo{year}{1998}).

\bibitem[{\citenamefont{Rammer and Smith}(1986)}]{ra.sm.86}
\bibinfo{author}{\bibfnamefont{J.}~\bibnamefont{Rammer}} \bibnamefont{and}
  \bibinfo{author}{\bibfnamefont{H.}~\bibnamefont{Smith}},
  \bibinfo{journal}{Rev. Mod. Phys.} \textbf{\bibinfo{volume}{58}},
  \bibinfo{pages}{323} (\bibinfo{year}{1986}).

\bibitem[{\citenamefont{Kamenev}(2011)}]{kame.11}
\bibinfo{author}{\bibfnamefont{A.}~\bibnamefont{Kamenev}},
  \emph{\bibinfo{title}{Field Theory of Non-Equilibrium Systems}}
  (\bibinfo{publisher}{Cambridge University Press},
  \bibinfo{address}{Cambridge}, \bibinfo{year}{2011}), ISBN
  \bibinfo{isbn}{0521760828}.

\bibitem[{\citenamefont{Caffarel and Krauth}(1994)}]{ca.kr.94}
\bibinfo{author}{\bibfnamefont{M.}~\bibnamefont{Caffarel}} \bibnamefont{and}
  \bibinfo{author}{\bibfnamefont{W.}~\bibnamefont{Krauth}},
  \bibinfo{journal}{Phys. Rev. Lett.} \textbf{\bibinfo{volume}{72}},
  \bibinfo{pages}{1545} (\bibinfo{year}{1994}).

\bibitem[{\citenamefont{Lanczos}(1951)}]{lanc.51}
\bibinfo{author}{\bibfnamefont{C.}~\bibnamefont{Lanczos}},
  \bibinfo{journal}{Journal of research of the National Bureau of Standards}
  \textbf{\bibinfo{volume}{45}}, \bibinfo{pages}{255} (\bibinfo{year}{1951}).

\bibitem[{foo({\natexlab{e}})}]{footnote3}
\bibinfo{note}{But see \tcite{ha.he.07}.}

\bibitem[{\citenamefont{Breuer and Petruccione}(2009)}]{br.pe}
\bibinfo{author}{\bibfnamefont{H.-P.} \bibnamefont{Breuer}} \bibnamefont{and}
  \bibinfo{author}{\bibfnamefont{F.}~\bibnamefont{Petruccione}},
  \emph{\bibinfo{title}{The Theory of Open Quantum Systems}}
  (\bibinfo{publisher}{Oxford University Press}, \bibinfo{address}{Oxford,
  England}, \bibinfo{year}{2009}).

\bibitem[{foo({\natexlab{f}})}]{footnote4}
\bibinfo{note}{Operators are denoted by a hat: $\hat{o}$, while super-operators
  acting on operators are denoted by a double hat $\hat{\hat{o}}$. For
  elementary fermionic creation/annihilation operators we omit the hat.
  Finally, we use boldface for matrices and vectors in orbital indices.}

\bibitem[{foo({\natexlab{g}})}]{footnote5}
\bibinfo{note}{Alternatively, one could use the ``star'' representation, in
  which only diagonal and $E_{f,\nu}$ terms are nonzero.}

\bibitem[{\citenamefont{Prosen}(2008)}]{pros.08}
\bibinfo{author}{\bibfnamefont{T.}~\bibnamefont{Prosen}}, \bibinfo{journal}{New
  J. Phys.} \textbf{\bibinfo{volume}{10}}, \bibinfo{pages}{043026}
  (\bibinfo{year}{2008}).

\bibitem[{\citenamefont{Dzhioev and Kosov}(2011)}]{dz.ko.11}
\bibinfo{author}{\bibfnamefont{A.~A.} \bibnamefont{Dzhioev}} \bibnamefont{and}
  \bibinfo{author}{\bibfnamefont{D.~S.} \bibnamefont{Kosov}},
  \bibinfo{journal}{J. Chem. Phys.} \textbf{\bibinfo{volume}{134}},
  \bibinfo{pages}{044121} (\bibinfo{year}{2011}).

\bibitem[{\citenamefont{Schmutz}(1978)}]{schm.78}
\bibinfo{author}{\bibfnamefont{M.}~\bibnamefont{Schmutz}}, \bibinfo{journal}{Z.
  Phys. B.} \textbf{\bibinfo{volume}{30}}, \bibinfo{pages}{97}
  (\bibinfo{year}{1978}).

\bibitem[{\citenamefont{Harbola and Mukamel}(2008)}]{ha.mu.08}
\bibinfo{author}{\bibfnamefont{U.}~\bibnamefont{Harbola}} \bibnamefont{and}
  \bibinfo{author}{\bibfnamefont{S.}~\bibnamefont{Mukamel}},
  \bibinfo{journal}{Physics Reports} \textbf{\bibinfo{volume}{465}},
  \bibinfo{pages}{191} (\bibinfo{year}{2008}).

\bibitem[{foo({\natexlab{h}})}]{footnote12}
\bibinfo{note}{From now on we will omit the spin index, unless necessary.}

\bibitem[{foo({\natexlab{i}})}]{footnote9}
\bibinfo{note}{In our convention, $G_{\mu\nu}^{>+}(t)$ and $G_{\mu\nu}^{<-}(t)$
  are zero for $t<0$, and vice-versa.}

\bibitem[{foo({\natexlab{j}})}]{footnote7}
\bibinfo{note}{Notice that $\vv D$ commutes with $\vv \lambda$, so $\frac{\vv
  D}{\omega - i \vv \lambda}$ is well defined.}

\bibitem[{\citenamefont{Shanno}(1970)}]{shan.70}
\bibinfo{author}{\bibfnamefont{D.~F.} \bibnamefont{Shanno}},
  \bibinfo{journal}{Math. Comp.} \textbf{\bibinfo{volume}{24}},
  \bibinfo{pages}{647} (\bibinfo{year}{1970}).

\bibitem[{\citenamefont{Press et~al.}(2007)\citenamefont{Press, Teukolsky,
  Vetterling, and Flannery}}]{pr.te.07}
\bibinfo{author}{\bibfnamefont{W.~H.} \bibnamefont{Press}},
  \bibinfo{author}{\bibfnamefont{S.~A.} \bibnamefont{Teukolsky}},
  \bibinfo{author}{\bibfnamefont{W.~T.} \bibnamefont{Vetterling}},
  \bibnamefont{and} \bibinfo{author}{\bibfnamefont{B.~P.}
  \bibnamefont{Flannery}}, \emph{\bibinfo{title}{Numerical Recipes 3rd Edition:
  The Art of Scientific Computing}} (\bibinfo{publisher}{Cambridge University
  Press}, \bibinfo{year}{2007}), \bibinfo{edition}{3rd} ed., ISBN
  \bibinfo{isbn}{0521880688}, \urlprefix\url{http://www.nr.com/}.

\bibitem[{\citenamefont{Jackson}(1975)}]{jack.75}
\bibinfo{author}{\bibfnamefont{J.~D.} \bibnamefont{Jackson}},
  \emph{\bibinfo{title}{Classical Electrodynamics}} (\bibinfo{publisher}{New
  York: Wiley}, \bibinfo{year}{1975}), \bibinfo{edition}{2nd} ed., ISBN
  \bibinfo{isbn}{0-471-43132-X}.

\bibitem[{foo({\natexlab{k}})}]{footnote8}
\bibinfo{note}{For the particle-hole symmetric model, the auxiliary system
  on-site energies are restricted to $E_{ff} = -\frac{U}{2}$ and $E_{\mu\mu} =
  -E_{N_B+1-\mu,N_B+1-\mu}$ for $\mu\neq f$ as well as nearest-neighbor $<>$
  hopping to $E_{<\mu\nu>} = (-1)^{\mu+\nu+1}E_{<N_B+1-\mu,N_B+1-\nu>}$ while
  the dissipation matrices have to fulfill $\Gamma^{(1)}_{\mu\nu} =
  (-1)^{\mu+\nu}\Gamma^{(2)}_{N_B+1-\mu,N_B+1-\nu}$.}

\bibitem[{\citenamefont{Meir and Wingreen}(1992)}]{me.wi.92}
\bibinfo{author}{\bibfnamefont{Y.}~\bibnamefont{Meir}} \bibnamefont{and}
  \bibinfo{author}{\bibfnamefont{N.~S.} \bibnamefont{Wingreen}},
  \bibinfo{journal}{Phys. Rev. Lett.} \textbf{\bibinfo{volume}{68}},
  \bibinfo{pages}{2512} (\bibinfo{year}{1992}).

\bibitem[{\citenamefont{Jauho}(2006)}]{jauh}
\bibinfo{author}{\bibfnamefont{A.-P.} \bibnamefont{Jauho}}
  (\bibinfo{year}{2006}), \bibinfo{note}{preprint}.

\bibitem[{\citenamefont{Negele and Orland}(1988)}]{negele.orland}
\bibinfo{author}{\bibfnamefont{J.~W.} \bibnamefont{Negele}} \bibnamefont{and}
  \bibinfo{author}{\bibfnamefont{H.}~\bibnamefont{Orland}},
  \emph{\bibinfo{title}{Quantum many-particle systems}},
  vol.~\bibinfo{volume}{68} of \emph{\bibinfo{series}{Frontiers in physics}}
  (\bibinfo{publisher}{Addison-Wesley}, \bibinfo{address}{Redwood City,
  Calif.}, \bibinfo{year}{1988}).

\bibitem[{\citenamefont{De~Franceschi et~al.}(2002)\citenamefont{De~Franceschi,
  Hanson, van~der Wiel, Elzerman, Wijpkema, Fujisawa, Tarucha, and
  Kouwenhoven}}]{fr.ha.02}
\bibinfo{author}{\bibfnamefont{S.}~\bibnamefont{De~Franceschi}},
  \bibinfo{author}{\bibfnamefont{R.}~\bibnamefont{Hanson}},
  \bibinfo{author}{\bibfnamefont{W.~G.} \bibnamefont{van~der Wiel}},
  \bibinfo{author}{\bibfnamefont{J.~M.} \bibnamefont{Elzerman}},
  \bibinfo{author}{\bibfnamefont{J.~J.} \bibnamefont{Wijpkema}},
  \bibinfo{author}{\bibfnamefont{T.}~\bibnamefont{Fujisawa}},
  \bibinfo{author}{\bibfnamefont{S.}~\bibnamefont{Tarucha}}, \bibnamefont{and}
  \bibinfo{author}{\bibfnamefont{L.~P.} \bibnamefont{Kouwenhoven}},
  \bibinfo{journal}{Phys. Rev. Lett.} \textbf{\bibinfo{volume}{89}},
  \bibinfo{pages}{156801} (\bibinfo{year}{2002}).

\bibitem[{\citenamefont{Leturcq et~al.}(2005)\citenamefont{Leturcq, Schmid,
  Ensslin, Meir, Driscoll, and Gossard}}]{le.sc.05}
\bibinfo{author}{\bibfnamefont{R.}~\bibnamefont{Leturcq}},
  \bibinfo{author}{\bibfnamefont{L.}~\bibnamefont{Schmid}},
  \bibinfo{author}{\bibfnamefont{K.}~\bibnamefont{Ensslin}},
  \bibinfo{author}{\bibfnamefont{Y.}~\bibnamefont{Meir}},
  \bibinfo{author}{\bibfnamefont{D.~C.} \bibnamefont{Driscoll}},
  \bibnamefont{and} \bibinfo{author}{\bibfnamefont{A.~C.}
  \bibnamefont{Gossard}}, \bibinfo{journal}{Phys. Rev. Lett.}
  \textbf{\bibinfo{volume}{95}}, \bibinfo{pages}{126603}
  (\bibinfo{year}{2005}).

\bibitem[{\citenamefont{Langer and Ambegaokar}(1961)}]{lang.61}
\bibinfo{author}{\bibfnamefont{J.~S.} \bibnamefont{Langer}} \bibnamefont{and}
  \bibinfo{author}{\bibfnamefont{V.}~\bibnamefont{Ambegaokar}},
  \bibinfo{journal}{Phys. Rev.} \textbf{\bibinfo{volume}{121}},
  \bibinfo{pages}{1090} (\bibinfo{year}{1961}).

\bibitem[{\citenamefont{Langreth}(1966)}]{lang.66}
\bibinfo{author}{\bibfnamefont{D.~C.} \bibnamefont{Langreth}},
  \bibinfo{journal}{Phys. Rev.} \textbf{\bibinfo{volume}{150}},
  \bibinfo{pages}{516} (\bibinfo{year}{1966}).

\bibitem[{\citenamefont{M{\"u}hlbacher
  et~al.}(2011)\citenamefont{M{\"u}hlbacher, Urban, and Komnik}}]{mu.ur.11}
\bibinfo{author}{\bibfnamefont{L.}~\bibnamefont{M{\"u}hlbacher}},
  \bibinfo{author}{\bibfnamefont{D.~F.} \bibnamefont{Urban}}, \bibnamefont{and}
  \bibinfo{author}{\bibfnamefont{A.}~\bibnamefont{Komnik}},
  \bibinfo{journal}{Phys. Rev. B} \textbf{\bibinfo{volume}{83}},
  \bibinfo{pages}{075107} (\bibinfo{year}{2011}).

\bibitem[{\citenamefont{K{\"o}nig et~al.}(1996)\citenamefont{K{\"o}nig, Schmid,
  Schoeller, and Sch{\"o}n}}]{ko.ju.96}
\bibinfo{author}{\bibfnamefont{J.}~\bibnamefont{K{\"o}nig}},
  \bibinfo{author}{\bibfnamefont{J.}~\bibnamefont{Schmid}},
  \bibinfo{author}{\bibfnamefont{H.}~\bibnamefont{Schoeller}},
  \bibnamefont{and}
  \bibinfo{author}{\bibfnamefont{G.}~\bibnamefont{Sch{\"o}n}},
  \bibinfo{journal}{Phys. Rev. B} \textbf{\bibinfo{volume}{54}},
  \bibinfo{pages}{16820} (\bibinfo{year}{1996}).

\bibitem[{\citenamefont{Dirks et~al.}(2013{\natexlab{b}})\citenamefont{Dirks,
  Han, Jarrell, and Pruschke}}]{di.ha.13}
\bibinfo{author}{\bibfnamefont{A.}~\bibnamefont{Dirks}},
  \bibinfo{author}{\bibfnamefont{J.~E.} \bibnamefont{Han}},
  \bibinfo{author}{\bibfnamefont{M.}~\bibnamefont{Jarrell}}, \bibnamefont{and}
  \bibinfo{author}{\bibfnamefont{T.}~\bibnamefont{Pruschke}},
  \bibinfo{journal}{Phys. Rev. B} \textbf{\bibinfo{volume}{87}},
  \bibinfo{pages}{235140} (\bibinfo{year}{2013}{\natexlab{b}}).

\bibitem[{\citenamefont{Rosch et~al.}(2003)\citenamefont{Rosch, Paaske, Kroha,
  and W{\"o}lfle}}]{ro.pa.03}
\bibinfo{author}{\bibfnamefont{A.}~\bibnamefont{Rosch}},
  \bibinfo{author}{\bibfnamefont{J.}~\bibnamefont{Paaske}},
  \bibinfo{author}{\bibfnamefont{J.}~\bibnamefont{Kroha}}, \bibnamefont{and}
  \bibinfo{author}{\bibfnamefont{P.}~\bibnamefont{W{\"o}lfle}},
  \bibinfo{journal}{Phys. Rev. Lett.} \textbf{\bibinfo{volume}{90}},
  \bibinfo{pages}{076804} (\bibinfo{year}{2003}).

\bibitem[{\citenamefont{Dalibard et~al.}(1992)\citenamefont{Dalibard, Castin,
  and M\o{}lmer}}]{da.ca.92}
\bibinfo{author}{\bibfnamefont{J.}~\bibnamefont{Dalibard}},
  \bibinfo{author}{\bibfnamefont{Y.}~\bibnamefont{Castin}}, \bibnamefont{and}
  \bibinfo{author}{\bibfnamefont{K.}~\bibnamefont{M\o{}lmer}},
  \bibinfo{journal}{Phys. Rev. Lett.} \textbf{\bibinfo{volume}{68}},
  \bibinfo{pages}{580} (\bibinfo{year}{1992}).

\bibitem[{\citenamefont{Daley et~al.}(2009)\citenamefont{Daley, Taylor, Diehl,
  Baranov, and Zoller}}]{da.ta.09}
\bibinfo{author}{\bibfnamefont{A.~J.} \bibnamefont{Daley}},
  \bibinfo{author}{\bibfnamefont{J.~M.} \bibnamefont{Taylor}},
  \bibinfo{author}{\bibfnamefont{S.}~\bibnamefont{Diehl}},
  \bibinfo{author}{\bibfnamefont{M.}~\bibnamefont{Baranov}}, \bibnamefont{and}
  \bibinfo{author}{\bibfnamefont{P.}~\bibnamefont{Zoller}},
  \bibinfo{journal}{Phys. Rev. Lett.} \textbf{\bibinfo{volume}{102}},
  \bibinfo{pages}{040402} (\bibinfo{year}{2009}).

\bibitem[{\citenamefont{Prosen and Znidaric}(2009)}]{pr.zn.09}
\bibinfo{author}{\bibfnamefont{T.}~\bibnamefont{Prosen}} \bibnamefont{and}
  \bibinfo{author}{\bibfnamefont{M.}~\bibnamefont{Znidaric}},
  \bibinfo{journal}{J. Stat. Mech.} \textbf{\bibinfo{volume}{2009}},
  \bibinfo{pages}{P02035} (\bibinfo{year}{2009}).

\bibitem[{\citenamefont{Alvermann and Fehske}(2009)}]{al.fe.09}
\bibinfo{author}{\bibfnamefont{A.}~\bibnamefont{Alvermann}} \bibnamefont{and}
  \bibinfo{author}{\bibfnamefont{H.}~\bibnamefont{Fehske}},
  \bibinfo{journal}{Phys. Rev. Lett.} \textbf{\bibinfo{volume}{102}},
  \bibinfo{pages}{150601} (\bibinfo{year}{2009}).

\bibitem[{\citenamefont{Weisse et~al.}(2006)\citenamefont{Weisse, Wellein,
  Alvermann, and Fehske}}]{we.we.06}
\bibinfo{author}{\bibfnamefont{A.}~\bibnamefont{Weisse}},
  \bibinfo{author}{\bibfnamefont{G.}~\bibnamefont{Wellein}},
  \bibinfo{author}{\bibfnamefont{A.}~\bibnamefont{Alvermann}},
  \bibnamefont{and} \bibinfo{author}{\bibfnamefont{H.}~\bibnamefont{Fehske}},
  \bibinfo{journal}{Rev. Mod. Phys.} \textbf{\bibinfo{volume}{78}},
  \bibinfo{pages}{275} (\bibinfo{year}{2006}).

\bibitem[{\citenamefont{Sherrill and III}(1999)}]{lo.sa.99}
\bibinfo{author}{\bibfnamefont{C.~D.} \bibnamefont{Sherrill}} \bibnamefont{and}
  \bibinfo{author}{\bibfnamefont{H.~F.~S.} \bibnamefont{III}},
  \textbf{\bibinfo{volume}{34}}, \bibinfo{pages}{143 } (\bibinfo{year}{1999}),
  ISSN \bibinfo{issn}{0065-3276}.

\bibitem[{\citenamefont{Saad}(2011)}]{saad.11}
\bibinfo{author}{\bibfnamefont{Y.}~\bibnamefont{Saad}},
  \emph{\bibinfo{title}{Numerical Methods for Large Eigenvalue Problems,
  Revised Edition}} (\bibinfo{publisher}{Society for Industrial and Applied
  Mathematics}, \bibinfo{year}{2011}), ISBN \bibinfo{isbn}{9781611970722}.

\bibitem[{\citenamefont{Arbenz}(2012)}]{arbe.12u}
\bibinfo{author}{\bibfnamefont{P.}~\bibnamefont{Arbenz}}
  (\bibinfo{year}{2012}), \bibinfo{note}{[Online; accessed 10-November-2013]}.

\bibitem[{\citenamefont{Bai et~al.}(1987)\citenamefont{Bai, Demmel, Dongarra,
  Ruhe, and van~der Vorst}}]{ba.de.87}
\bibinfo{author}{\bibfnamefont{Z.}~\bibnamefont{Bai}},
  \bibinfo{author}{\bibfnamefont{J.}~\bibnamefont{Demmel}},
  \bibinfo{author}{\bibfnamefont{J.}~\bibnamefont{Dongarra}},
  \bibinfo{author}{\bibfnamefont{A.}~\bibnamefont{Ruhe}}, \bibnamefont{and}
  \bibinfo{author}{\bibfnamefont{H.}~\bibnamefont{van~der Vorst}},
  \emph{\bibinfo{title}{Templates for the Solution of Algebraic Eigenvalue
  Problems: A Practical Guide (Software, Environments and Tools)}}
  (\bibinfo{publisher}{Society for Industrial and Applied Mathematics},
  \bibinfo{year}{1987}), ISBN \bibinfo{isbn}{0898714710}.

\bibitem[{\citenamefont{Knap et~al.}(2011{\natexlab{b}})\citenamefont{Knap,
  Arrigoni, {von der Linden}, and Cole}}]{kn.ar.11.ec}
\bibinfo{author}{\bibfnamefont{M.}~\bibnamefont{Knap}},
  \bibinfo{author}{\bibfnamefont{E.}~\bibnamefont{Arrigoni}},
  \bibinfo{author}{\bibfnamefont{W.}~\bibnamefont{{von der Linden}}},
  \bibnamefont{and} \bibinfo{author}{\bibfnamefont{J.~H.} \bibnamefont{Cole}},
  \bibinfo{journal}{Phys. Rev. A} \textbf{\bibinfo{volume}{83}},
  \bibinfo{pages}{023821} (\bibinfo{year}{2011}{\natexlab{b}}).

\bibitem[{\citenamefont{Saad}(2003)}]{saad.03}
\bibinfo{author}{\bibfnamefont{Y.}~\bibnamefont{Saad}},
  \emph{\bibinfo{title}{Iterative Methods for Sparse Linear Systems, Second
  Edition}} (\bibinfo{publisher}{Society for Industrial and Applied
  Mathematics}, \bibinfo{year}{2003}), ISBN \bibinfo{isbn}{9780898715347}.

\bibitem[{\citenamefont{Barrett et~al.}(1994)\citenamefont{Barrett, Berry,
  Chan, Demmel, Donato, Dongarra, Eijkhout, Pozo, Romine, and der
  Vorst}}]{ba.be.94}
\bibinfo{author}{\bibfnamefont{R.}~\bibnamefont{Barrett}},
  \bibinfo{author}{\bibfnamefont{M.}~\bibnamefont{Berry}},
  \bibinfo{author}{\bibfnamefont{T.~F.} \bibnamefont{Chan}},
  \bibinfo{author}{\bibfnamefont{J.}~\bibnamefont{Demmel}},
  \bibinfo{author}{\bibfnamefont{J.}~\bibnamefont{Donato}},
  \bibinfo{author}{\bibfnamefont{J.}~\bibnamefont{Dongarra}},
  \bibinfo{author}{\bibfnamefont{V.}~\bibnamefont{Eijkhout}},
  \bibinfo{author}{\bibfnamefont{R.}~\bibnamefont{Pozo}},
  \bibinfo{author}{\bibfnamefont{C.}~\bibnamefont{Romine}}, \bibnamefont{and}
  \bibinfo{author}{\bibfnamefont{H.~V.} \bibnamefont{der Vorst}},
  \emph{\bibinfo{title}{Templates for the Solution of Linear Systems: Building
  Blocks for Iterative Methods, 2nd Edition}} (\bibinfo{publisher}{SIAM},
  \bibinfo{address}{Philadelphia, PA}, \bibinfo{year}{1994}).

\bibitem[{\citenamefont{Park and Light}(1986)}]{pa.ju.86}
\bibinfo{author}{\bibfnamefont{T.~J.} \bibnamefont{Park}} \bibnamefont{and}
  \bibinfo{author}{\bibfnamefont{J.~C.} \bibnamefont{Light}},
  \bibinfo{journal}{The Journal of Chemical Physics}
  \textbf{\bibinfo{volume}{85}}, \bibinfo{pages}{5870} (\bibinfo{year}{1986}).

\bibitem[{\citenamefont{Bazaliy et~al.}(1997)\citenamefont{Bazaliy, Demler, and
  Zhang}}]{ba.de.97}
\bibinfo{author}{\bibfnamefont{Y.~B.} \bibnamefont{Bazaliy}},
  \bibinfo{author}{\bibfnamefont{E.}~\bibnamefont{Demler}}, \bibnamefont{and}
  \bibinfo{author}{\bibfnamefont{S.-C.} \bibnamefont{Zhang}},
  \bibinfo{journal}{Phys. Rev. Lett.} \textbf{\bibinfo{volume}{79}},
  \bibinfo{pages}{1921} (\bibinfo{year}{1997}).

\bibitem[{\citenamefont{Gutknecht}(1990)}]{gutk.90}
\bibinfo{author}{\bibfnamefont{M.~H.} \bibnamefont{Gutknecht}},
  \emph{\bibinfo{title}{The unsymmetric lanczos algorithms and their relations
  to pade approximation, continued fractions, and the qd algorithm}},
  \bibinfo{howpublished}{\url{http://www.math.ethz.ch/~mhg/}}
  (\bibinfo{year}{1990}).

\bibitem[{\citenamefont{Gutknecht}(1999)}]{gutk.99}
\bibinfo{author}{\bibfnamefont{M.~H.} \bibnamefont{Gutknecht}},
  \emph{\bibinfo{title}{Lanczos-type solvers for non-hermitian linear
  systems}}, \bibinfo{howpublished}{\url{http://www.math.ethz.ch/~mhg/}}
  (\bibinfo{year}{1999}).

\bibitem[{\citenamefont{Freund et~al.}(1993)\citenamefont{Freund, Gutknecht,
  and Nachtigal}}]{fe.gu.93}
\bibinfo{author}{\bibfnamefont{R.}~\bibnamefont{Freund}},
  \bibinfo{author}{\bibfnamefont{M.}~\bibnamefont{Gutknecht}},
  \bibnamefont{and}
  \bibinfo{author}{\bibfnamefont{N.}~\bibnamefont{Nachtigal}},
  \bibinfo{journal}{SIAM Journal on Scientific Computing}
  \textbf{\bibinfo{volume}{14}}, \bibinfo{pages}{137} (\bibinfo{year}{1993}).

\bibitem[{\citenamefont{Parlett et~al.}(1985)\citenamefont{Parlett, Taylor, and
  Liu}}]{be.pa.85}
\bibinfo{author}{\bibfnamefont{B.~N.} \bibnamefont{Parlett}},
  \bibinfo{author}{\bibfnamefont{D.~R.} \bibnamefont{Taylor}},
  \bibnamefont{and} \bibinfo{author}{\bibfnamefont{Z.~A.} \bibnamefont{Liu}},
  \bibinfo{journal}{Mathematics of Computation} \textbf{\bibinfo{volume}{44}},
  \bibinfo{pages}{pp. 105} (\bibinfo{year}{1985}).

\bibitem[{\citenamefont{Weikert et~al.}(1996)\citenamefont{Weikert, Meyer,
  Cederbaum, and Tarantelli}}]{we.me.96}
\bibinfo{author}{\bibfnamefont{H.-G.} \bibnamefont{Weikert}},
  \bibinfo{author}{\bibfnamefont{H.-D.} \bibnamefont{Meyer}},
  \bibinfo{author}{\bibfnamefont{L.~S.} \bibnamefont{Cederbaum}},
  \bibnamefont{and}
  \bibinfo{author}{\bibfnamefont{F.}~\bibnamefont{Tarantelli}},
  \bibinfo{journal}{The Journal of Chemical Physics}
  \textbf{\bibinfo{volume}{104}}, \bibinfo{pages}{7122} (\bibinfo{year}{1996}).

\bibitem[{foo({\natexlab{l}})}]{footnote11}
\bibinfo{note}{For the sake of clarity, we specifically introduce the subscript
  $_{ex}$ to denote the exact hybridisation function $\und \Delta_{ex}$. This
  will be used only in this section.}

\end{thebibliography}

\end{document}